\documentclass[a4paper,fleqn,usenatbib]{mnras}

\usepackage[T1]{fontenc}
\usepackage{ae,aecompl}

\usepackage{graphicx}	% Including figure files
\usepackage{amsmath}	% Advanced maths commands
\usepackage{amssymb}	% Extra maths symbols
\usepackage{epstopdf}
\usepackage{color}
\usepackage{float}
\usepackage{pdflscape}
\usepackage{longtable}
\usepackage{booktabs}
\title[Spatial resonant periodic orbits in the RTBP]{Spatial resonant periodic orbits in the restricted three-body problem} 

\author[K. I. Antoniadou and A.-S. Libert]{Kyriaki I. Antoniadou\thanks{E-mail: kyriaki.antoniadou@unamur.be (KIA)}
and Anne-Sophie Libert
\\
NaXys, Department of Mathematics, University of Namur, 8 Rempart de la Vierge, 5000 Namur, Belgium\\
}

\date{Accepted XXX. Received YYY; in original form ZZZ}

\pubyear{2018}

\begin{document}
\label{firstpage}
\pagerange{\pageref{firstpage}--\pageref{lastpage}}
\maketitle

\begin{abstract}
The quest of exo-Earths has become a prominent field. In this work, we study the stability of non-coplanar planetary configurations consisting of an inclined inner terrestrial planet in mean-motion resonance with an outer giant planet. We examine the families of circular and elliptic symmetric periodic orbits with respect to the vertical stability, and identify the vertical critical orbits from which the spatial families emanate. We showcase that stable spatial periodic orbits can exist for both prograde and retrograde motion in 3/2, 2/1, 5/2, 3/1, 4/1 and 5/1 resonances for broad ranges of inclinations, when the giant evolves on a circular orbit. When the orbit of the giant is elliptic, only the 2/1 resonance has stable periodic orbits up to high inclinations, while the 3/1, 4/1 and 5/1 resonances possess segments of stability for low inclinations. Furthermore, we show that regular motion can also take place in the vicinity of both horizontally and vertically stable planar periodic orbits, even for very high inclinations. Finally, the results are discussed in the context of asteroid dynamics.
\end{abstract}


%%%%%%%%%%%%%%%%%%%%%%%%%%%%%%%%%%%%%%%%%%%%%%%%%%

%%%%%%%%%%%%%%%%% BODY OF PAPER %%%%%%%%%%%%%%%%%%

\section{Introduction}
To-date, ground-based telescopes and space missions have revealed thousands of extrasolar planets. Many celestial architectures different than the one observed in the Solar system pose questions regarding the formation, the dynamical evolution and the long-term stability of planetary systems. In particular, the quest of exo-Earths within the region of habitability, where water can be found in liquid form, receives constant attention. Dynamical mechanisms like the mean-motion resonance (MMR) can act to maintain the long-term stability of terrestrial planets.  

Stability regions can be revealed by constructing maps of dynamical stability (DS-maps) by means of a chaotic indicator. For instance, \citet{sandor07} studied the hypothetical existence of a terrestrial planet in the habitable zone of 15 single-giant exoplanetary systems. \citet{funk09} explored the stability of inclined terrestrial planets evolving in the habitable zone for configurations of terrestrial planets in binary systems, with an inner gas giant, with an outer gas giant, and on a Trojan orbit. Later on, \citet{funkK} investigated the influence of the Kozai mechanism induced by an eccentric giant planet on the long-term stability of inclined Earth-like planets in the habitable zone. 

Recently, \citet{kiaasl} extensively studied the dynamical stability of an inner terrestrial planet in MMR or not with an outer giant planet as the eccentricity of the inner body increases. They showed that stable domains exist for the complete range of eccentricities of the outer body. In particular, the potential existence of a terrestrial planet was discussed for 22 RV-detected planetary systems with high eccentricities. Families of stable periodic orbits are responsible for the appearance of stable domains, as shown by \citet{kiaasl} for the 2/1 MMR and by \citet{spis} for the 3/2, 5/2, 3/1, 4/1 and 5/1 MMRs in the planar circular and elliptic restricted three-body problem (2D-CRTBP and 2D-ERTBP).

In this work, the question we aim to answer relates the potential existence of spatial resonant configurations with Earth-like planets. Spatial resonant configurations have mainly been studied in the spatial general three-body problem (3D-GTBP), namely for the 2/1 MMR by \citet{av12} and for 4/3, 3/2, 5/2, 3/1 and 4/1 MMRs by \citet{av13}. Inclination excitation of two-planet systems has been observed via differential migration during the protoplanetary disc phase \citep{thommes03,leetho09,litsi09b} and it is associated with the vertical critical orbits (v.c.o.) that exist along the planar families. Indeed, assuming initially quasi-coplanar and quasi-circular orbits, \citet{vat14} showed for the 2/1 and 3/1 MMRs, that after resonant capture, the two planets migrate along a planar family of elliptic periodic orbits and get caught in \textit{inclination resonance}, when they meet a v.c.o. Then, they follow the spatial family that bifurcates from it, i.e. they reach a spatial stable configuration. Same holds for the spatial families which emanate from the circular family, as shown for the 5/2 and 7/3 MMRs by \citet{av17}. Recently, simulations by \citet{lsa18} additionally showed that spatial families of periodic orbits act as attractors that trap the planets after instability events during their migration.

Regarding the RTBP and more specifically the exterior MMRs, \citet{kot05}, \citet{kotvoy05} and \citet{vk05} studied the inclined symmetric periodic orbits of trans-Neptunian objects in the planar and spatial RTBP. Recently, \citet{vta} studied inclined asymmetric periodic solutions of the same system.

We herein extend our previous contributions to the 2D-RTBPs \citep{kiaasl,spis} to the 3D-RTBPs by computing spatial families for the restricted cases, and particularly, for the 3/2, 2/1, 5/2, 3/1, 4/1 and 5/1 (interior) MMRs. The study is organised as follows. In Sect. \ref{model}, we introduce the models of the 3D-RTBPs (circular and elliptic cases), define the spatial symmetric periodic orbits in each case and discuss the linear horizontal and vertical stability of the periodic orbits. In Sect. \ref{cont}, we mention the methods of continuation of periodic orbits for the computation of families. In Sect. \ref{res}, we report our results ordered by MMR, i.e. in the 3/2, 2/1, 5/2, 3/1, 4/1 and 5/1 MMRs for each problem. We showcase the neighbourhood of spatial configurations through DS-maps in Sect. \ref{appi} and finally conclude in Sect. \ref{con}.

We should note that the results shown herein can be applied to any celestial bodies which can be effectively modelled by the 3D-RTBPs. For instance, we may cite star-asteroid-giant planet, planet-spacecraft-satellite and binary star-circumprimary planet. As an application, the highly inclined orbits in asteroid dynamics are discussed in Sect. \ref{aster}.

\section{Model set-up}\label{model}
\subsection{The spatial restricted three-body problem (3D-RTBP)}
We consider a system, composed by a star, an inner terrestrial planet and an outer giant planet, of masses $m_0$, $m_1=0$ and $m_2=m_J$, respectively. Therefore, the terrestrial planet is assumed as a massless body. The masses are normalised to unity (hence, $m_0=1-m_2=0.999$), and the mass parameter, $\mu$, equals $\mu=\frac{m_2}{m_0+m_2}=0.001$. The primaries, star and giant planet, revolve around their centre of mass in a Keplerian orbit, under their mutual gravitational attraction (in our case, the gravitational constant is $G=1$) with the terrestrial planet being in interior MMR with the giant planet.

We introduce a suitable rotating frame of reference, $Oxyz$ \citep[see e.g.][]{sze,murray}. The $Oxy$-plane contains always the primaries, the $Ox$-axis is directed from the star to the giant planet, and its origin, $O$, is their centre of mass. The $Oz$-axis is perpendicular to the plane $Oxy$.

The Lagrangian of the small massless body in the rotating frame $Oxyz$ is
\begin{equation} \label{lagr}
\begin{array}{l}
{\cal{L}}= 0.5[(\dot x -\dot \theta y)^2+(\dot y +\dot\theta x)^2+{\dot z}^2]+\frac{1-\mu}{r_1}+\frac{\mu}{r_2},\\ \vspace{0.1cm}
{\textnormal {where}}\\ \vspace{0.1cm}
r_1=\sqrt{(x+\mu r)^2+y^2+z^2},\\ \vspace{0.1cm} {\textnormal {and}}\\ \vspace{0.1cm} r_2=\sqrt{[x-(1-\mu) r]^2+y^2+z^2}.
\end{array}
\end{equation}
The distance, $r=r(t)$, is the distance between the primaries and the angle $\theta=\theta(t)$ is the true anomaly of the giant planet.

In our study, we additionally refer to the following orbital elements: $a_i$ (semi-major axis), $e_i$ (eccentricity), $i_i$ (inclination), $\omega_i$ (argument of pericentre), $M_i$ (mean anomaly) and $\Delta\Omega$ (longitude of ascending node) with $i=1, 2$. For terrestrial planets, we are mainly interested in \textit{prograde} or \textit{direct} ($i_i<90^\circ$) orbits, even if the continuation of the families of periodic orbits is made up to $180^{\circ}$ (\textit{retrograde} orbits) for reasons of completeness. The mutual inclination will be denoted by $\Delta i$ and is  given by the cosine rule
\begin{equation}\cos\Delta i=\cos i_1 \cos i_2+\sin i_1 \sin i_2\cos(\Omega_1-\Omega_2).
\end{equation}  
For the interior MMRs studied herein, $a_1<a_2$ holds and we set $a_2=1$. Therefore, subscript 1 (2) shall always refer to the inner (outer) body.

For a circular orbit of the giant (3D-CRTBP), i.e. $e_2=0$, the positions of the primaries are fixed, namely, $(-\mu,0)$ and $(1-\mu,0)$ for the star and the giant, respectively. We have $r=1$ and $\dot\theta=1$ and this problem is of three degrees of freedom. 

When the orbit of the giant planet is elliptic (3D-ERTBP), i.e. $e_2\neq 0$, the distance between the primaries is not constant and the $Oxyz$ frame is not uniformly rotating. Therefore, we have a non-autonomous system, i.e. of four degrees of freedom.

The Lagrangian of the corresponding planar problems is derived simply by setting $z=0$ and $\dot z=0$.

In this work, we compute spatial symmetric periodic orbits, which exist inside the orbit (either circular or elliptic) of the giant planet ($a_2=1$), and whose period in our normalised units equals to $T_0=2\pi$.

\subsection{Resonant periodic orbits}\label{rpo}
As the periodic orbit describes the motion of $P_1$ (massless body), we can explore three cases with regards to its orbits \citep[see][for more details on their origin and continuation]{spis}:
\begin{itemize}
	\item[--] The circular family, where both the primaries (star and giant planet) and the massless body (terrestrial planet) evolve on circular orbits. The periodic orbits are circular and symmetric, while the MMR varies along this family.
	\item[--] The CRTBP, where the primaries evolve on circular orbits and the orbits of $P_1$ are elliptic. These orbits are periodic in the rotating frame only if they are resonant.
	\item[--] The ERTBP, where the primaries evolve on elliptic orbits and the orbits of $P_1$ are elliptic. These orbits can be either symmetric or asymmetric and periodic in the rotating frame as long as they are resonant.
\end{itemize}

Mean-motion resonances are associated with the rational ratio of mean motions for which $\frac{n_1}{n_2}\approx\frac{p+q}{p}$, where $q,\,p\neq0$ are integers with $q$ defining the \textit{resonance order}  and $n_i$ ($i=1,2$) denoting the mean motion of the inner or outer body.  In this case, we can introduce the \textit{resonant angles}, $\theta_i$, previously used by \citet{spis}, for each order of the MMR studied hereafter:
\begin{itemize}
	\item when the order of the resonance, $q$, is odd (in 3/2, 2/1, 5/2 and 4/1 MMRs in our study) we use the pair $(\theta_1,\theta_2)$,
	\item when $q=2$ (in 3/1 MMR) we use the pair $(\theta_3,\theta_1)$ and
	\item when $q=4$ (in 5/1 MMR) we use the pair $(\theta_4,\theta_1)$,
\end{itemize}
with
\begin{equation}\label{th}\begin{array}{l}
\theta_1=p\lambda_1-(p+q)\lambda_2+q\varpi_1, \\
\theta_2=p\lambda_1-(p+q)\lambda_2+q\varpi_2, \\
\theta_3=\lambda_1-3\lambda_2+\varpi_1+\varpi_2,\\
\theta_4=\lambda_1-5\lambda_2+3\varpi_2+\varpi_1,
\end{array}
%\nonumber
\end{equation}
where $\lambda_i=M_i+\varpi_i$ is the mean longitude. The stationary solutions where $\dot\theta_i=0$ ($i=1,2$) are also called apsidal corotation resonances (ACRs, e.g. in \citet{femibe06}).  A periodic orbit in the rotating frame showcases the exact location of an MMR in phase space and an ACR corresponds to a periodic orbit. 

When the bodies are initially at conjunction in the state of apsidal alignment we refer also to the apsidal difference $\Delta\varpi=\varpi_2-\varpi_1$. Therefore, based on the above pairs of resonant angles, we can have four different symmetric configurations: $(0,0), (0,\pi), (\pi,0)$ and $(\pi,\pi)$.

When the two bodies are not coplanar, we may additionally introduce the resonant angles that define the \textit{inclination resonance} for at least second order resonances  
\begin{equation}\label{phi}
\begin{array}{l}
\varphi _{11}=p\lambda _{1}-(p+q)\lambda _{2}+q\Omega _{1},\\ 
\varphi _{22}=p\lambda _{1}-(p+q)\lambda _{2}+q\Omega _{2},\\
\varphi _{12}=(\varphi_{11}+\varphi_{22})/q,\\
\end{array}
\end{equation}
as well as the zeroth-order secular resonance angle
\begin{equation}
\begin{array}{l}
\varphi _{\Omega}=\Omega _{1}-\Omega_2=(\varphi_{11}-\varphi_{22})/q.
\end{array}
\end{equation}

\subsection{Symmetries of spatial periodic orbits in the 3D-RTBPs}
Let us consider the rotating frame, $OXYZ$, and define the Poincar\'e section plane in phase space $\hat\pi=\{y=0,\dot y>0\}$. Initial conditions shall always be chosen on this plane. Then, the periodic orbits are the fixed or periodic points of the Poincar\'e map \citep{pnc} and satisfy the conditions
\begin{equation} \label{pos}
\begin{array}{lll}
x'(0)=x'(T),& \dot x'(0)=\dot x'(T),&\\
x(0)=x(T), & \dot x(0)=\dot x(T),& \dot y(0)=\dot y(T)\\
z(0)=z(T),& \dot z(0)=\dot z(T), &
\end{array}
\end{equation}
provided that $y(0)=y(T)$ and $T$ is the period. Primed quantities refer to the giant planet. The multiplicity of the periodic orbit is defined as the number of successive intersections with the plane $y=0$ and in the same direction ($\dot y>0$ or $\dot y<0$) during one period. 

According to the definition of the rotating frame, two symmetries can be defined for the restricted problems:

\begin{itemize}
	\item \textbf{Symmetry with respect to the $xz$-plane }\\
The periodic orbit obeying this symmetry has two perpendicular crossings with the $xz$-plane and the initial conditions at $t=0$ of such a periodic orbit should be 
\begin{equation}
\begin{array}{llll}
x'(0)=x'_{0}, &\quad x(0)=x_{0}, &\quad y(0)=0 & \quad z(0)=z_{0},  \\
\dot x'(0)=0, &\quad \dot x(0)=0, &\quad \dot y(0)=\dot y_{0}, &\quad \dot z(0)=0.
\label{xzsym}
\end{array}
\end{equation}
Therefore, an $\textit{xz}$-\textit{symmetric} periodic orbit can be represented by a point in the four-dimensional space of initial conditions: $\{(x'_{0},x_{0},z_{0},\dot{y}_{0})\}$. Given the geometry of the orbit, at $t=0$ it holds that $\Omega_1=90^{\circ}$ and $\varpi_1=0^{\circ}$ or $180^{\circ}$.\vspace{0.2cm}

\item \textbf{Symmetry with respect to the $x$-axis }\\
When the orbit has two perpendicular crossings with the $x$-axis, then it is symmetric with respect to this symmetry. Starting $(t=0)$ perpendicularly from the $x$-axis, then, the periodic orbit has initial conditions
\begin{equation}
\begin{array}{llll}
x'(0)=x'_{0}, &\quad x(0)=x_{0}, &\quad y(0)=0, &\quad z(0)=0,\\
\dot x'(0)=0, &\quad \dot x(0)=0, &\quad \dot y(0)=\dot y_{0}, &\quad \dot z(0)=\dot z_{0}.
\label{xsym}
\end{array}
\end{equation}
Hence, an $\textit{x}$-\textit{symmetric} periodic orbit can be represented by one point in the four-dimensional space of initial conditions$\{(x'_{0},x_{0},\dot{y}_{0},\dot{z}_{0})\}$. The geometry of the orbit shows that at $t=0$, $\Omega_1=0^{\circ}$ and $\varpi_1=0^{\circ}$ or $180^{\circ}$.
\end{itemize}
 
The above conditions hold for the periodic orbits of the 3D-ERTBP.
In the 3D-CRTBP, $x'$ is constant defined by the normalisation adopted for our system and in the computations it equates to $1-\frac{m_2}{m_0+m_2}$. Thus, the $xz$-symmetric periodic orbits can be represented by a point in the three-dimensional space of initial conditions
$\{(x_{0},z_{0},\dot{y}_{0})\}$ and the $x$-symmetric ones by $\{(x_{0},\dot{y}_{0},\dot{z}_{0})\}$.

\begin{figure}
\begin{center}
$\begin{array}{c}
\includegraphics[width=.815\columnwidth]{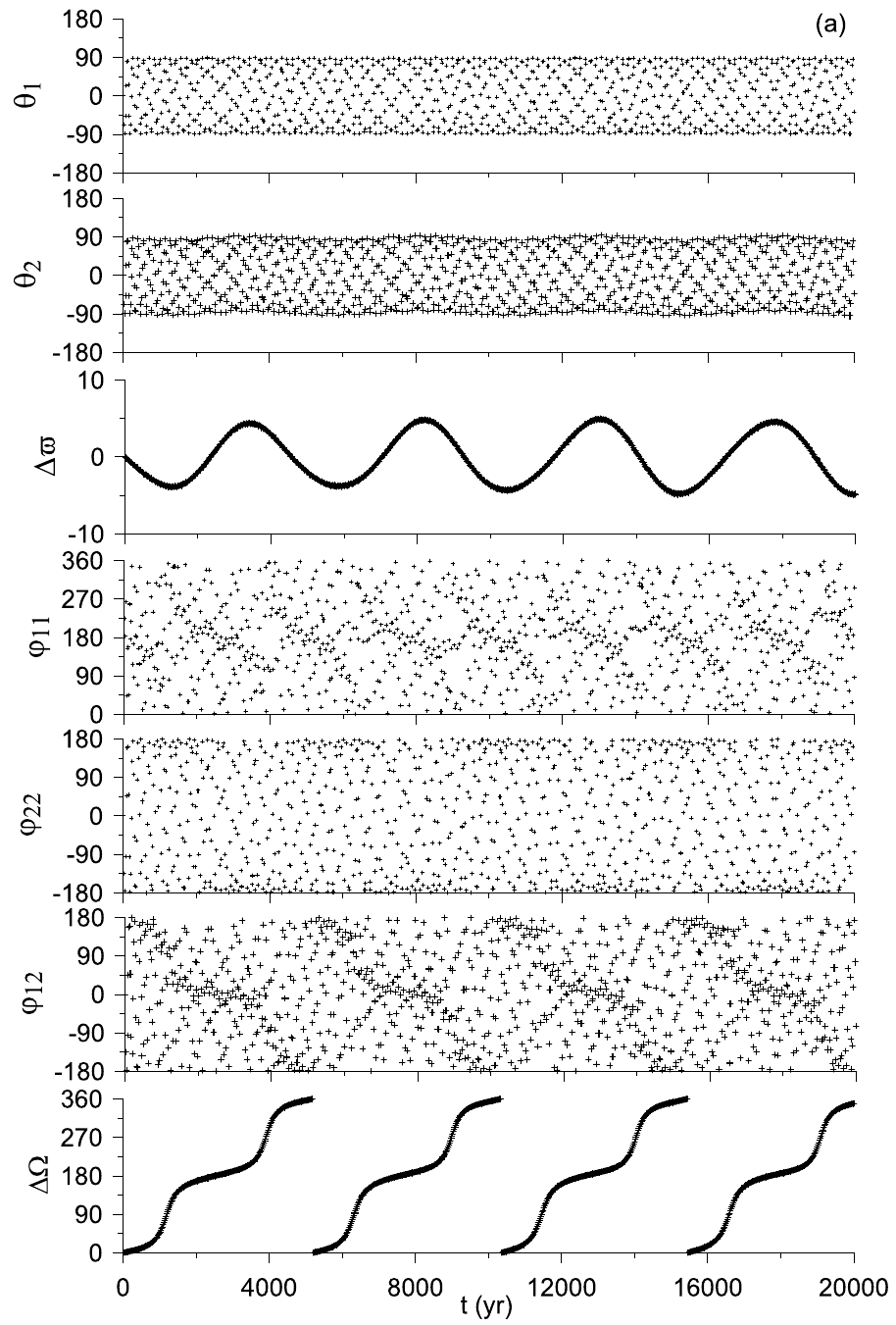}\vspace{-0.1cm} \\ \includegraphics[width=.815\columnwidth]{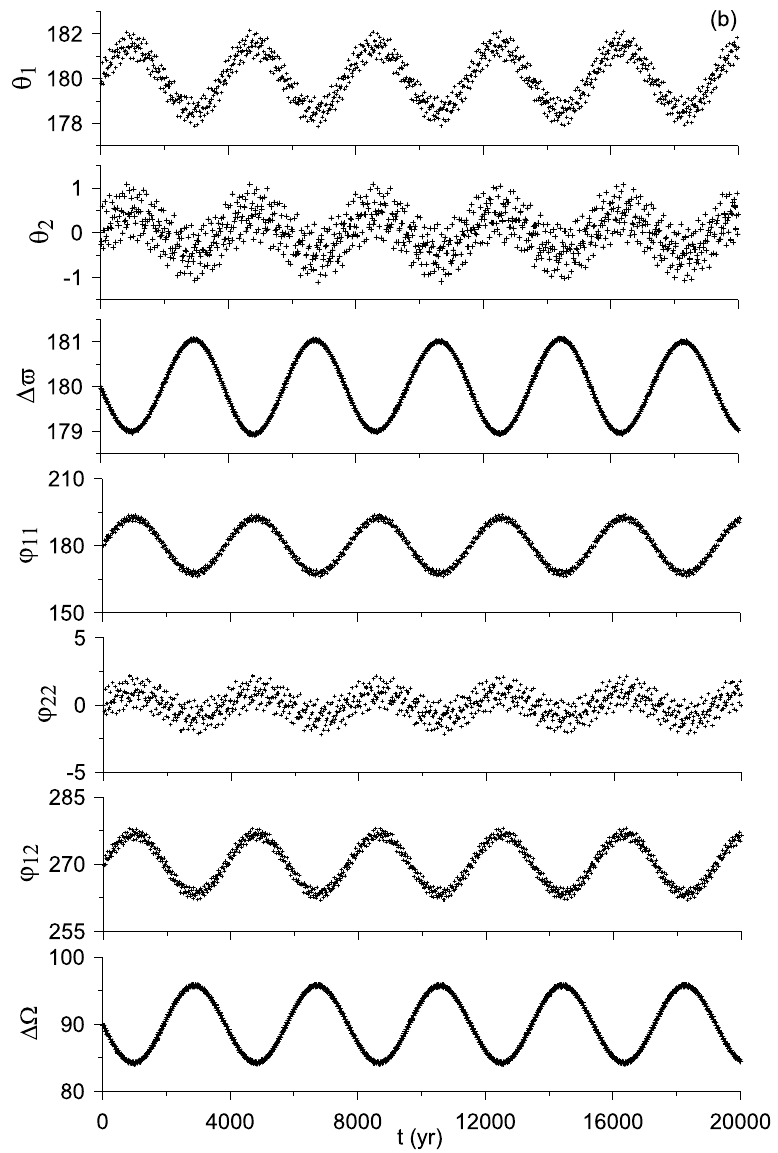}\vspace{-0.4cm}
\end{array} $
\end{center}
\caption{Typical angular evolutions observed in the 2D- and 3D-ERTBPs for the 2/1 MMR. (a) In the neighbourhood of a planar periodic orbit, we observe the libration of the resonant angles $\theta_i (i=1,2)$ and $\Delta\varpi$ linked with the MMR and the rotation of the inclination resonant angles $\varphi_{ii}$ and $\Delta\Omega$ associated with the inclination resonance. (b) In the vicinity of a spatial periodic orbit, the inclination resonance takes place in tandem with the MMR; the libration of all the resonant angles, apsidal difference and nodal difference is observed. Evolution (a) emanates from the configuration $(0,0)$, shown in Fig. \ref{21evco}, and evolution (b) from the configuration $(\pi,0)$, presented in Fig. \ref{21e3d}.}
\label{rang}
\end{figure}

\subsection{Linear stability}
The linear stability of the spatial periodic orbits is derived through the computation of the eigenvalues (conjugate ones in reciprocal pairs) of the monodromy matrix of the variational equations, the number of which depends on the degrees of freedom of each problem. The periodic orbit is called linearly stable iff all the eigenvalues lie on the unit circle. The different types of stability of periodic orbits of high dimensional Hamiltonian systems have been studied e.g. by \citet{marchal90} and \citet{sk01}.

An estimate of the accuracy of the computation of the eigenvalues is retrieved via the accuracy of the computation of the pair of the unit eigenvalues, which exists, due to the energy integral. This accuracy is about 10-11 digits. Therefore, we additionally compute the evolution of stability for at least $t=3$ Myr by checking the detrended Fast Lyapunov Indicator (DFLI) \citep{froe97,voyatzis08}. Orbits with $\log\rm{DFLI}\leq 2$ are classified as stable.

As will be further detailed in Sect. \ref{cont}, one of the vital steps for finding bifurcations of spatial periodic orbits is the computation of the vertical stability index, $a_v$, \citep{hen} of the planar periodic orbits, which belong to the 2D-CRTBP and 2D-ERTBP. Therefore, the planar periodic orbits, apart from being horizontally stable or unstable, can be vertically stable or unstable. The points between the latter transition, where $|a_v|=1$, are called \textit{vertical critical orbits} (v.c.o.). 
Therefore, four combinations of horizontal and vertical stability can exist and can bring to light more spatial configurations. 

The above information is depicted in our plots in the following way:
\begin{itemize}
	\item Blue (red) colour: Horizontal linear stability (instability) of the planar periodic orbits and linear stability (instability) of the spatial periodic orbits;
	\item Solid (dashed) lines: Vertical stability (instability) of the planar periodic orbits;
	\item Magenta dots: Vertical critical orbits (v.c.o.) which generate $xz$-symmetric spatial families of periodic orbits; 
	\item Green dots: v.c.o. which generate $x$-symmetric spatial families of periodic orbits.
\end{itemize}

In Hamiltonian systems, the stable periodic orbits are surrounded by invariant tori, where the motion is regular and quasi-periodic, exhibited by the libration of the resonant angles $\theta_i$ and the apsidal difference, $\Delta \varpi$, whereas in the vicinity of unstable periodic orbits homoclinic webs are formed, chaos emerges and these angles rotate. 

In the neighbourhood of planar stable periodic orbits, the resonant angles $\theta_i$ (Eq. \eqref{th}) and the apsidal difference, $\Delta\varpi$, librate, but no inclination resonance is achieved for the inclined systems. A typical evolution is shown in Fig. \ref{rang}a. In the neighbourhood of spatial stable periodic orbits, the bodies evolve in MMR along with the inclination resonance, as showcased by the libration of the angles $\varphi_{ii}$ (Eq. \eqref{phi}) and the nodal difference, $\Delta\Omega$, in Fig. \ref{rang}b.

\section{Continuation of families of periodic orbits in the 3D-RTBPs}\label{cont}
Changing the value of $z$ ($xz$-symmetric) or $\dot z$ ($x$-symmetric) yields a \textit{monoparametric family} of spatial symmetric periodic orbits and there exist two methods to generate them in the 3D-ERTBP: \vspace{0.2cm}

\noindent 
\textbf{Method I:}\\ 
\noindent
Examine the circular family and the planar families of the 2D-CRTBP and 2D-ERTBP with respect to the vertical stability. The v.c.o. are the bifurcation points that can potentially generate spatial periodic orbits continued analytically to the 3D-CRTBP and 3D-ERTBP. \vspace{0.2cm}

\noindent
\textbf{Method II:}\\
\noindent		
Examine the spatial families of 3D-CRTBP for periodic orbits whose period gets equal to $T=k T_0/m$, where $m$ is the multiplicity of the generated periodic orbit. These orbits constitute bifurcation points that generate spatial periodic orbits in the 3D-ERTBP. Two families, in general, emanate from them: one corresponding to the location of the outer giant planet to pericentre and another to its location at apocentre, since this body was initially on a circular orbit ($e_2=0$) and now it is allowed to evolve on an elliptic one ($e_2\neq 0$).\vspace{0.2cm}

Hereafter, we follow the notation of \citet{av12,av13}, and denote the spatial families of $xz$-symmetric periodic orbits by $F^{\frac{p+q}{p}}_{N}$ and the ones of $x$-symmetric periodic orbits by $G^{\frac{p+q}{p}}_{N}$, where $\frac{p+q}{p}$ is the MMR. $N$ stands either for the configuration ($(\theta_1,\theta_2)$, $(\theta_3,\theta_1)$ or $(\theta_4,\theta_1)$) of the planar family in the 2D-ERTBP or for the name of the planar family, either the circular family or the family of the 2D-CRTBP from which they emanate, accordingly. When the families are generated in the 3D-ERTBP from bifurcation points of the 3D-CRTBP, $N$ is followed by $p$ or $a$ denoting the location of the giant at pericentre or apocentre, respectively. 

Finally, the v.c.o., provided in Sect. \ref{res}, are accordingly denoted, but $F$ or $G$ has a circumflex accent, in order for the bifurcation point to be distinguished from the spatial family. If there are two v.c.o. and hence, two spatial families originating from the same planar family of a particular configuration, the latter ones (points or families) are primed.

\section{Results}\label{res}
Based on Method I, we examine the circular family and the planar families in 3/2, 2/1, 5/2, 3/1, 4/1 and 5/1 MMRs of the 2D-CRTBP and 2D-ERTBP presented in \citet{kiaasl,spis} with respect to the vertical stability and identify the existence of v.c.o. We also examine the spatial families of the 3D-CRTBP according to Method II, in order to find periodic orbits of period multiple of the period of the primary. Finally, we provide the spatial periodic orbits that exist in the 3D-CRTBP and 3D-ERTBP along with their linear stability. The results are classified by MMR.

\subsection{The circular family}\label{cf}
\citet{av17} have shown for the 2D-GTBP that the more the multiplicity of the circular periodic orbits of the circular family increases, the more v.c.o. become apparent. In Appendix \ref{appendix}, we show that this property holds for the restricted case, as well.

As showcased in \citet{spis}, the circular family ``breaks'' when the order of the resonance is $q=1$, namely at 3/2, 2/1, etc., therefore no v.c.o. are apparent at these MMRs. Given the MMRs studied herein, we focus on the segment $\frac{p+q}{p}\in(2,5.1]$ of the circular family and compute the vertical stability for different multiplicity values.

More specifically, in Fig. \ref{circmul}, we present the circular family, denoted by $C_i$, where $i\in[1,9]$ stands for the multiplicity value, with respect to the vertical stability index, $a_v$, and in Table \ref{multab}, we show the MMRs (possessing a v.c.o. at $|a_v|=1$) that emerge in a systematic manner, as the multiplicity of the circular periodic orbits increases.

At these v.c.o. of the circular family, we have rational values of the mean-motion ratio, $\frac{p+q}{p}$ and the periodic orbits are of period $T=2\pi(a_1^{-3/2}-1)^{-1}$ (given that $a_2=1.0$). Therefore, we have bifurcation points from which elliptic symmetric periodic orbits will be generated in the 3D-CRTBP directly.

\subsection{3/2 MMR}

\subsubsection{3D-CRTBP}
As mentioned in Sect. \ref{cf}, there is a gap at 3/2 MMR along the circular family. Therefore, no v.c.o. exist that could generate spatial families according to Method I.

\begin{figure}
\begin{center}
$\begin{array}{c}
%\textnormal{3/2 MMR}  & \textnormal{2/1 MMR} \\
\includegraphics[width=0.9\columnwidth,keepaspectratio]{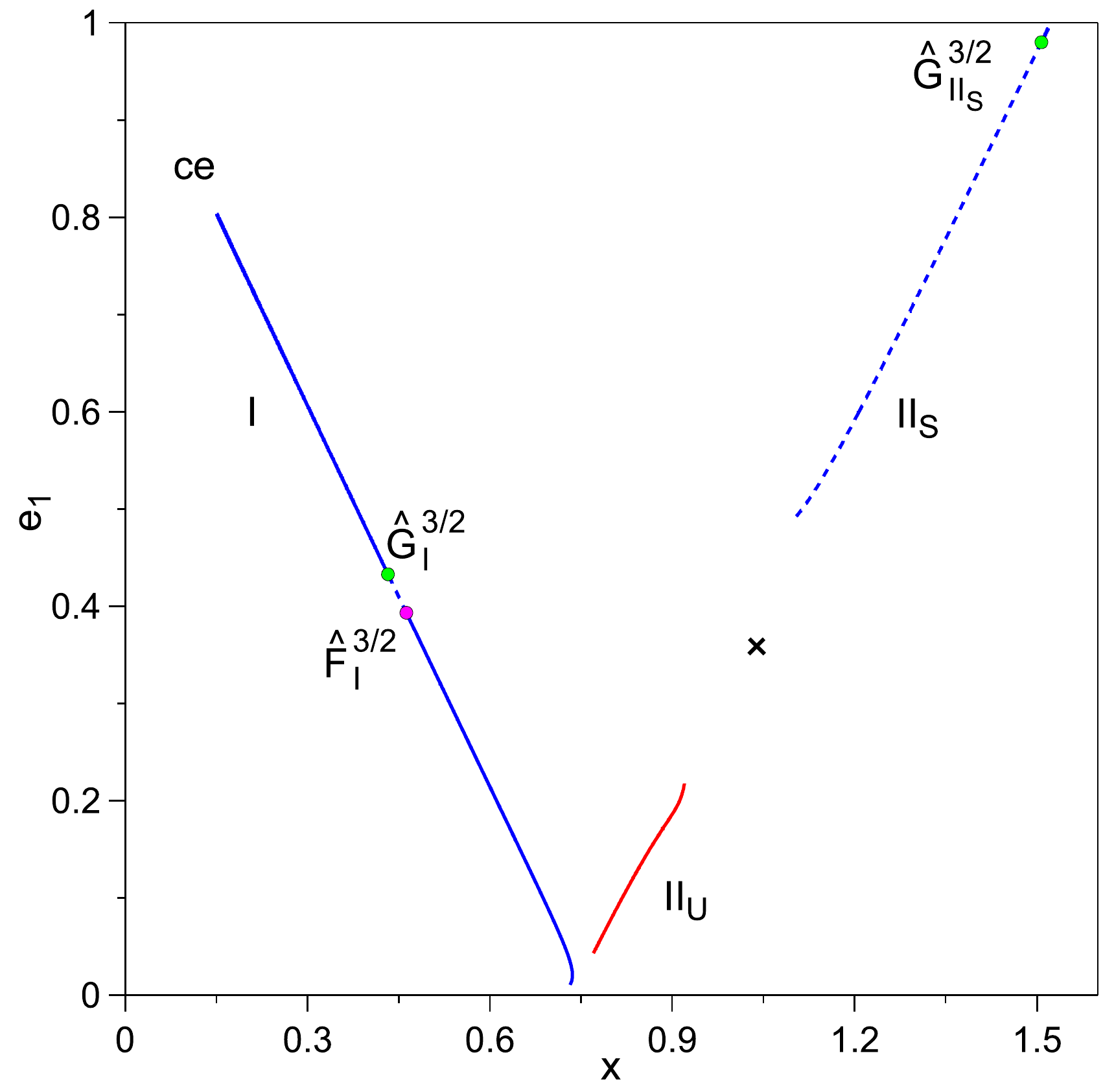} 
\end{array} $
\end{center}
\caption{Families of periodic orbits in the 2D-CRTBP for  3/2 MMR examined with respect to Method I, i.e. scanned with regards to the vertical stability. The inner body is located at pericentre at family $I$ and at apocentre at the families $II$. Blue (red) coloured lines refers to the horizontal stability (instability), while the solid (dashed) coloured lines to the vertical stability (instability) of the periodic orbits. The magenta (green) dots correspond to the vertical critical orbits which generate $xz$- ($x$-) symmetric spatial periodic orbits in the 3D-CRTBP. The cross symbol corresponds to collision between the bodies, while ``ce'' to close encounters.}
\label{32cvco}
\end{figure}

In Fig. \ref{32cvco}, we present the families of the 3/2 MMR in the 2D-CRTBP, i.e. when $e_2=0$. We observe that family $I$ is horizontally stable (blue) and vertically stable (solid line) except the region between the two v.c.o. where it gets vertically unstable (dashed line). These v.c.o., $\hat{F}^{3/2}_I$ and $\hat{G}^{3/2}_I$, with eccentricity values $e_1=0.39$ and $e_1=0.43$, respectively, are called pairs\footnote{Pairs of v.c.o. are neighbouring v.c.o. belonging to the same family, one of which yields $xz$-symmetric spatial periodic orbits and the other $x$-symmetric orbits.} in \citet{av13} and we will keep this terminology. The family $II_S$ is whole horizontally stable (blue), but essentially totally vertically unstable (dashed line), since the v.c.o., $\hat{G}^{3/2}_{II_S}$, and hence, the transition of stability is found at very high eccentricity values and particularly at $e_1=0.98$. Family $II_U$ is horizontally unstable (red), but vertically stable (solid line). 

\begin{figure}
\begin{center}
$\begin{array}{c}
\includegraphics[width=0.9\columnwidth]{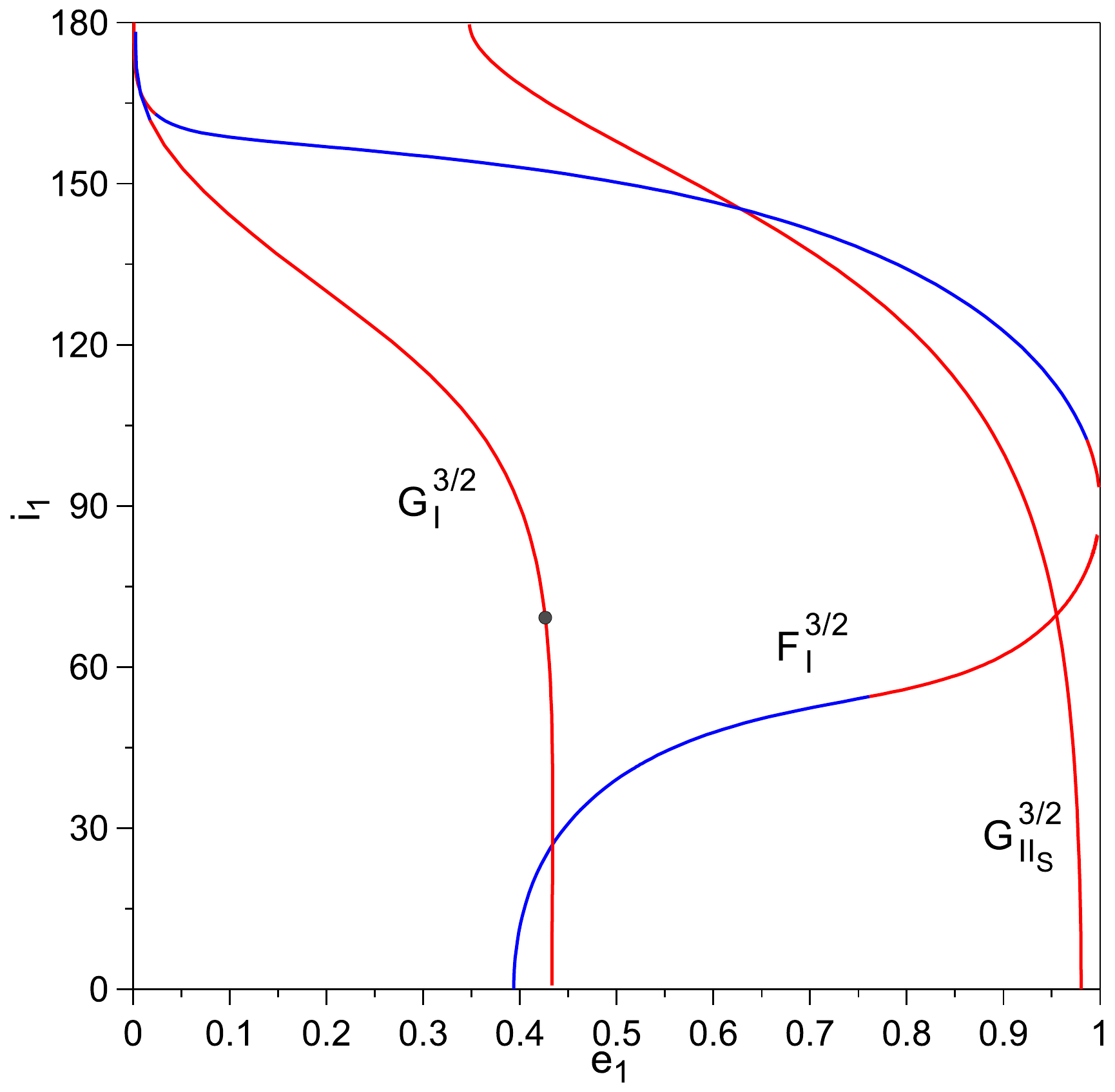}
\end{array} $
\end{center}
\caption{Spatial families of periodic orbits in the 3D-CRTB in the 3/2 MMR bifurcating from the 2D-CRTBP projected on the plane ($e_1,i_1$). $G^{3/2}_I$ and $F^{3/2}_I$ bifurcate from the v.c.o. at $e_1=0.39$ and $e_1=0.43$ of the family $I$. $G^{3/2}_I$ and $F^{3/2}_I$ have segments of stability (blue). $G^{3/2}_{II_S}$ bifurcates from the v.c.o. at $e_1=0.98$ of the family $II_S$ and is whole unstable (red). The grey dot at $(e_1,i_1)=(0.42,69^{\circ})$ corresponds to a bifurcation point from the 3D-CRTBP to the 3D-ERTBP.} %$54.7^{\circ}$
\label{32c3d}
\end{figure}

In Fig. \ref{32c3d}, we present the spatial families in the 3D-CRTBP projected on the plane $(e_1,i_1)$. The spatial family $F^{3/2}_I$ is stable up to $\sim55^{\circ}$ and in the segment between $103^{\circ}$ and $165^{\circ}$ and consists of $xz$-symmetric periodic orbits. At $90^{\circ}$, $e_1$ tends to 1 and the family could not be computed locally\footnote{In the 3D-CRTBP, whenever $P_1$ approaches $P_0$ ($e_1\rightarrow 1$ and $i_1=90^{\circ}$), the stability type might not be accurate, as the integration stops. A study of this region is beyond the scope of the present work.}. The spatial families of $x$-symmetric periodic orbits, $G^{3/2}_I$ and $G^{3/2}_{II_S}$, are unstable with $G^{3/2}_I$ having a stable segment that starts at $162^{\circ}$ and ends at $178^{\circ}$. All of the families terminate at planar retrograde orbits ($i_1=180^{\circ}$).
These spatial families are examined via Method II and only the family $G^{3/2}_I$ possesses a bifurcation point to the 3D-ERTBP at $(e_1,i_1)=(0.42,69^{\circ})$ (grey dot).

\subsubsection{3D-ERTBP}
\begin{figure}
\begin{center}
$\begin{array}{cc}
\includegraphics[width=0.9\columnwidth,keepaspectratio]{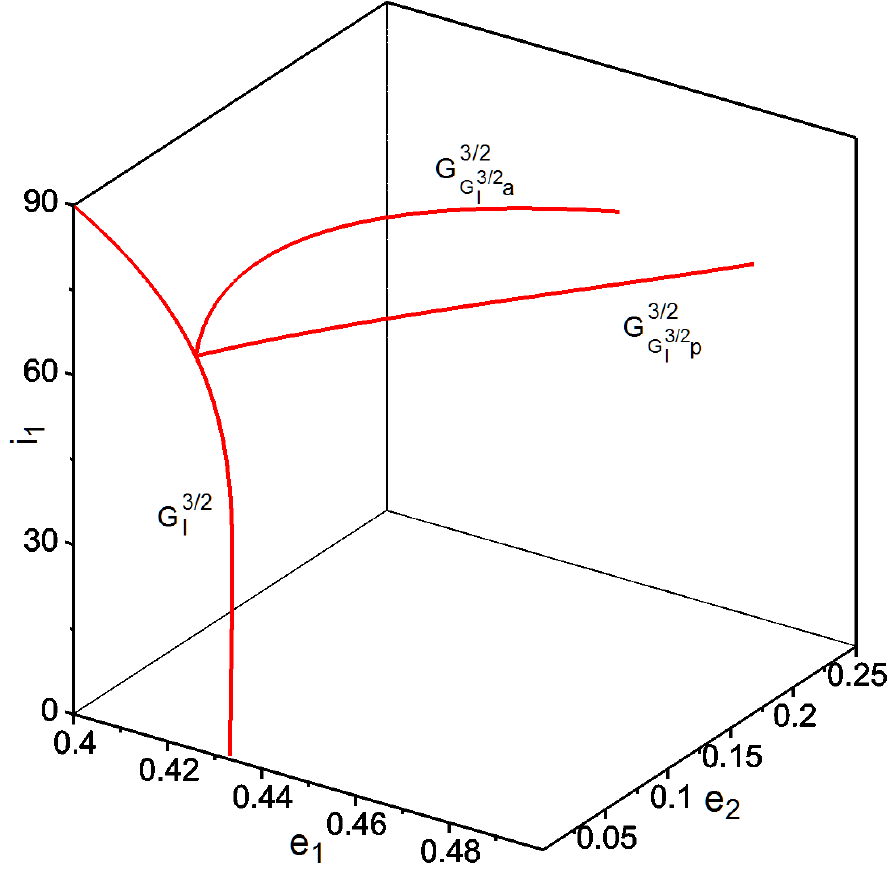} 
\end{array} $
\end{center}
\caption{The two unstable families of periodic orbits, $G^{3/2}_{G^{3/2}_Ip}$ and $G^{3/2}_{G^{3/2}_Ia}$, in the 3D-ERTBP for 3/2 MMR, which bifurcate, according to Method II, from the family $G^{3/2}_I$ (replotted) at the point $(e_1,i_1)=(0.42,69^{\circ})$, of the 3D-CRTBP shown in Fig. \ref{32c3d}.}
\label{32el}
\end{figure}

The bifurcation point of the spatial family $G^{3/2}_I$ of the 3D-CRTBP generates two families in the 3D-ERTBP, $G^{3/2}_{G^{3/2}_Ip}$ and $G^{3/2}_{G^{3/2}_Ia}$, one corresponding to the location of the giant at pericentre and the other at apocentre. They are whole unstable and are presented in Fig \ref{32el}.

\begin{figure}
\begin{center}
$\begin{array}{c}%\textnormal{3/2 MMR}  & \textnormal{2/1 MMR} \\
\includegraphics[width=0.9\columnwidth,keepaspectratio]{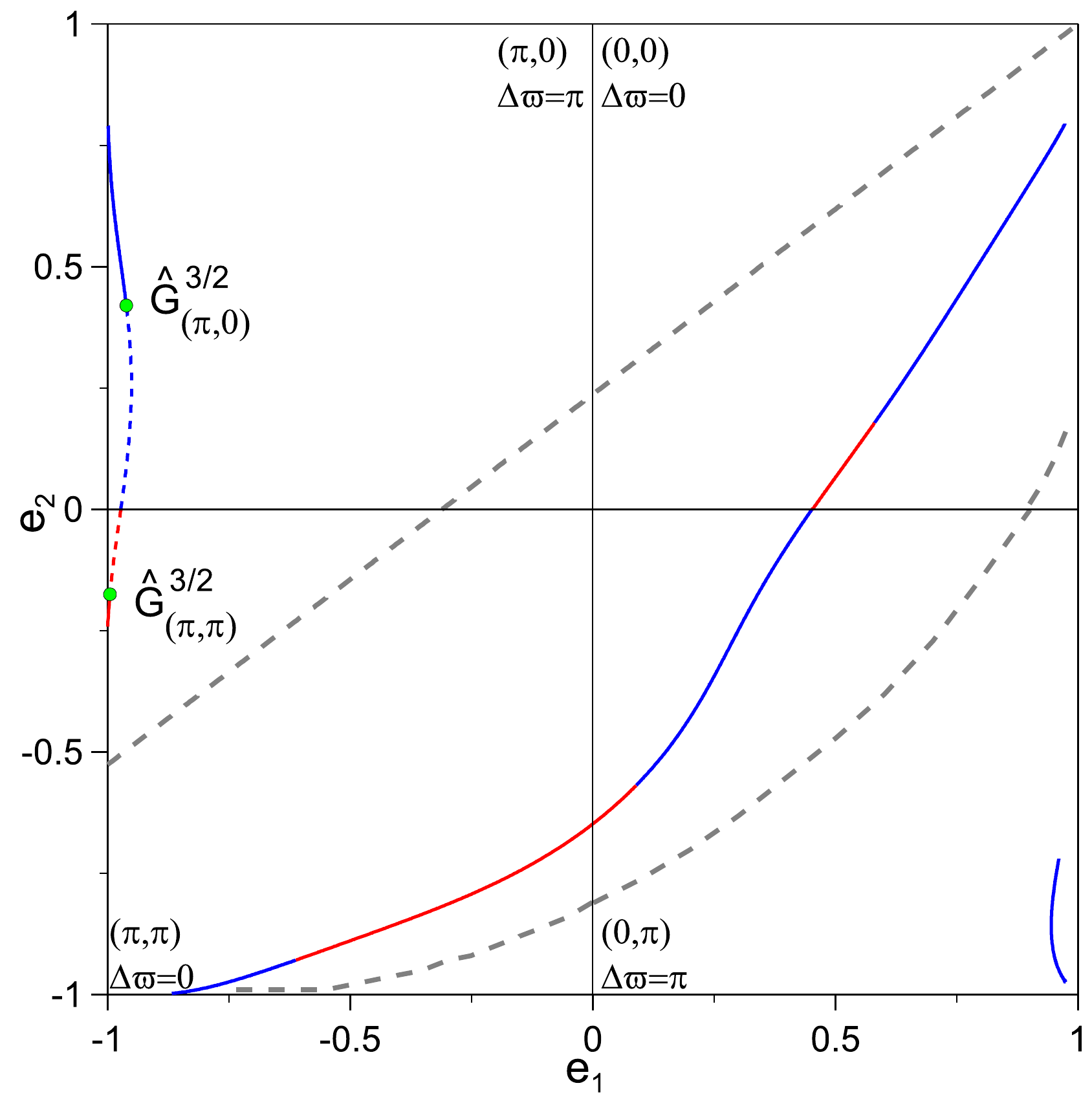} 
\end{array} $
\end{center}
\caption{Families of periodic orbits in the 2D-ERTBP for 3/2 MMR examined with respect to vertical stability (Method I). The v.c.o. (coloured dots) generate spatial families in the 3D-ERTBP. The bold dashed pale grey line and curve correspond to the collision and the close encounters between the primary and the secondary, respectively. Colours and lines as in Fig. \ref{32cvco}.}
\label{32evco}
\end{figure} 

In Fig. \ref{32evco}, we present the families of the 3/2 MMR in the 2D-ERTBP examined with regards to the vertical stability. We observe that the majority of the families is horizontally and vertically stable (blue solid lines). The spatial dynamical neighbourhood of the family of the configuration $(0,\pi)$ with low to moderate values of $e_1$ is examined in Sect. \ref{app}. There are two v.c.o., $\hat{G}^{3/2}_{(\pi,0)}$ located at $(e_1,e_2)=(0.96,0.12)$ and $\hat{G}^{3/2}_{(\pi,\pi)}$ located at $(e_1,e_2)=(0.99,0.17)$, which generate $x$-symmetric spatial periodic orbits. The planar families are vertically unstable (dashed line) between the two v.c.o. and vertically stable (solid line) after the v.c.o. as $e_2$ increases.

\begin{figure}
\begin{center}
$\begin{array}{c}
\includegraphics[width=0.8\columnwidth]{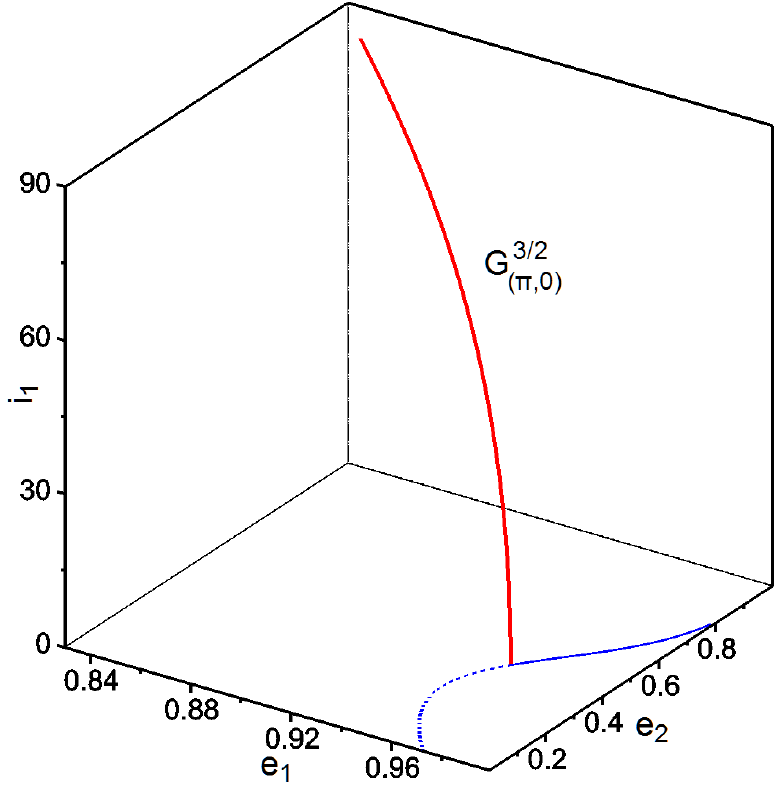}\\ \includegraphics[width=0.8\columnwidth]{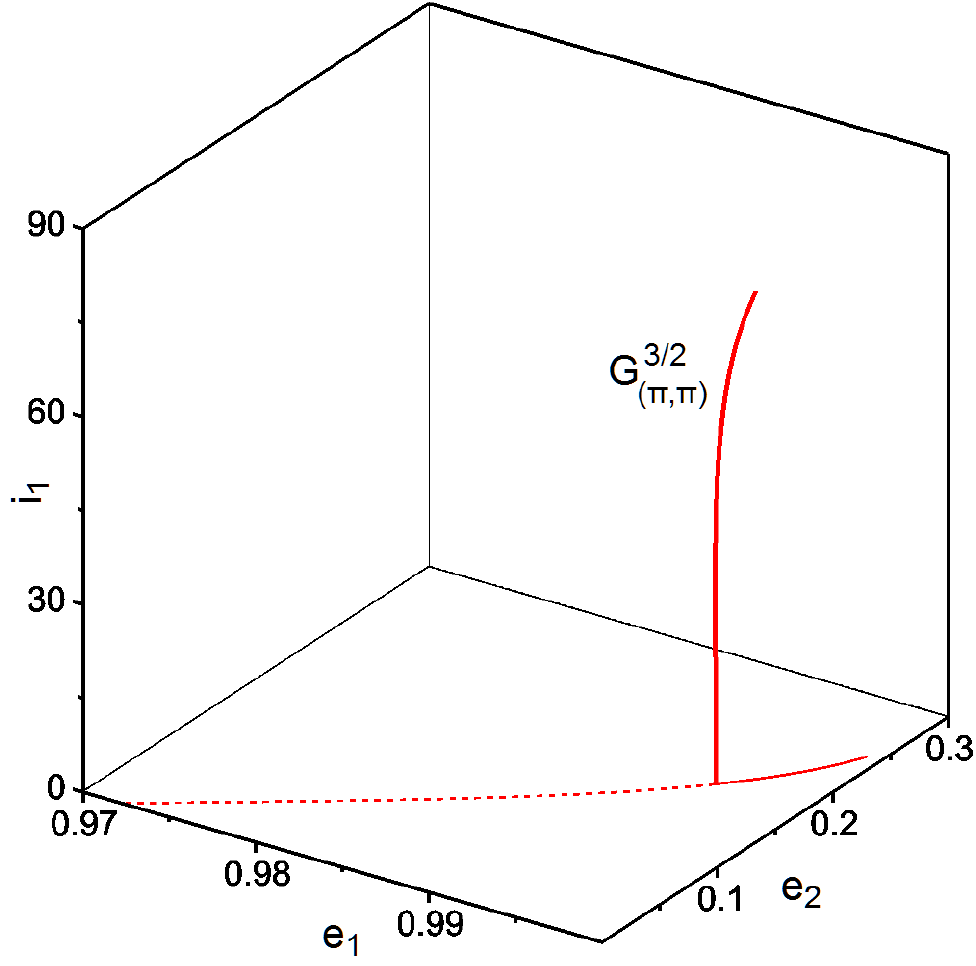}\\
\end{array} $
\end{center}
\caption{Spatial families of unstable periodic orbits in the 3D-ERTBP in the 3/2 MMR. $G^{3/2}_{(\pi,0)}$ bifurcates from the v.c.o. at $(e_1,e_2)=(0.96,0.42)$ of the configuration $(\pi,0)$. $G^{3/2}_{(\pi,\pi)}$ bifurcates from the v.c.o. at $(e_1,e_2)=(0.99,0.17)$ of the configuration $(\pi,\pi)$. The families of the 2D-ERTBP are replotted.}
\label{32e3d}
\end{figure}

In Fig. \ref{32e3d}, we present the spatial families $G^{3/2}_{(\pi,0)}$ and $G^{3/2}_{(\pi,\pi)}$ of $x$-symmetric periodic orbits, which extend up to high inclination values, but are whole unstable.

\subsection{2/1 MMR}

\subsubsection{3D-CRTBP}
Likewise 3/2 MMR, there is a gap at 2/1 MMR along the circular family. Therefore, no spatial families can be generated by this family based on Method I.

\begin{figure}
\begin{center}
$\begin{array}{c}
%\textnormal{3/2 MMR}  & \textnormal{2/1 MMR} \\
\includegraphics[width=0.9\columnwidth,keepaspectratio]{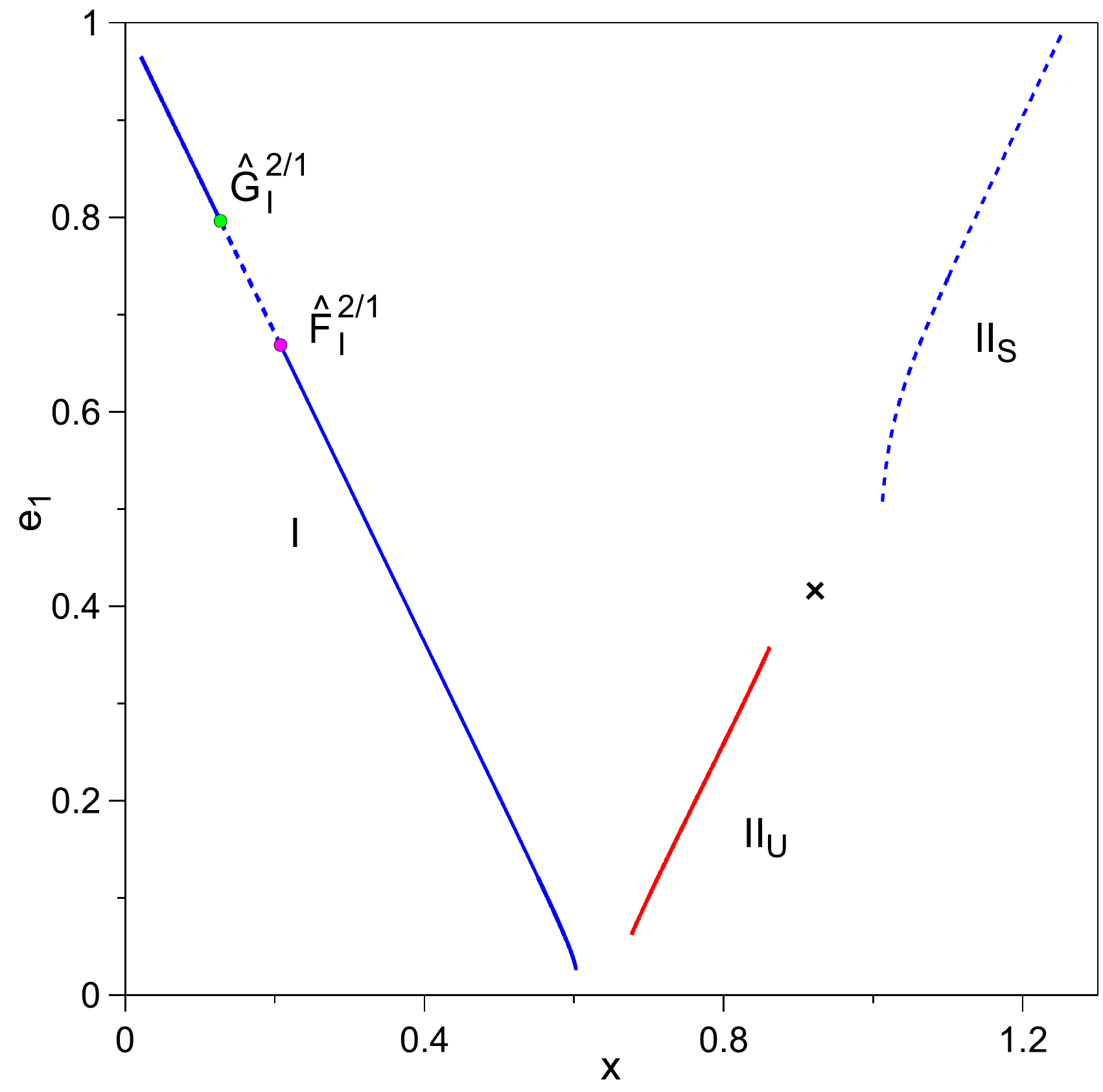} 
\end{array} $
\end{center}
\caption{Families of periodic orbits in the 2D-CRTBP for 2/1 MMR examined with respect to Method I. The  v.c.o. generate spatial symmetric periodic orbits in the 3D-CRTBP. Colours and lines as in Fig. \ref{32cvco}.
}
\label{21cvco}
\end{figure}

In Fig. \ref{21cvco}, we present the planar families of the 2/1 MMR in the 2D-CRTBP examined with regards to the vertical stability (Method I), which exhibit a similar trend to the 3/2 MMR. Precisely, the horizontally stable family $I$ has a pair of v.c.o., $\hat{F}^{2/1}_I$ and $\hat{G}^{2/1}_I$, with eccentricity values $e_1=0.67$ and $e_1=0.79$, respectively, and gets vertically unstable only between them. The family $II_U$ is horizontally unstable, but vertically stable, while the family $II_S$ is horizontally stable, but vertically unstable, since we did not find any v.c.o. until the eccentricity value $e_1=0.993$ up to which it is continued.

\begin{figure}
\begin{center}
$\begin{array}{c}
\includegraphics[width=0.9\columnwidth]{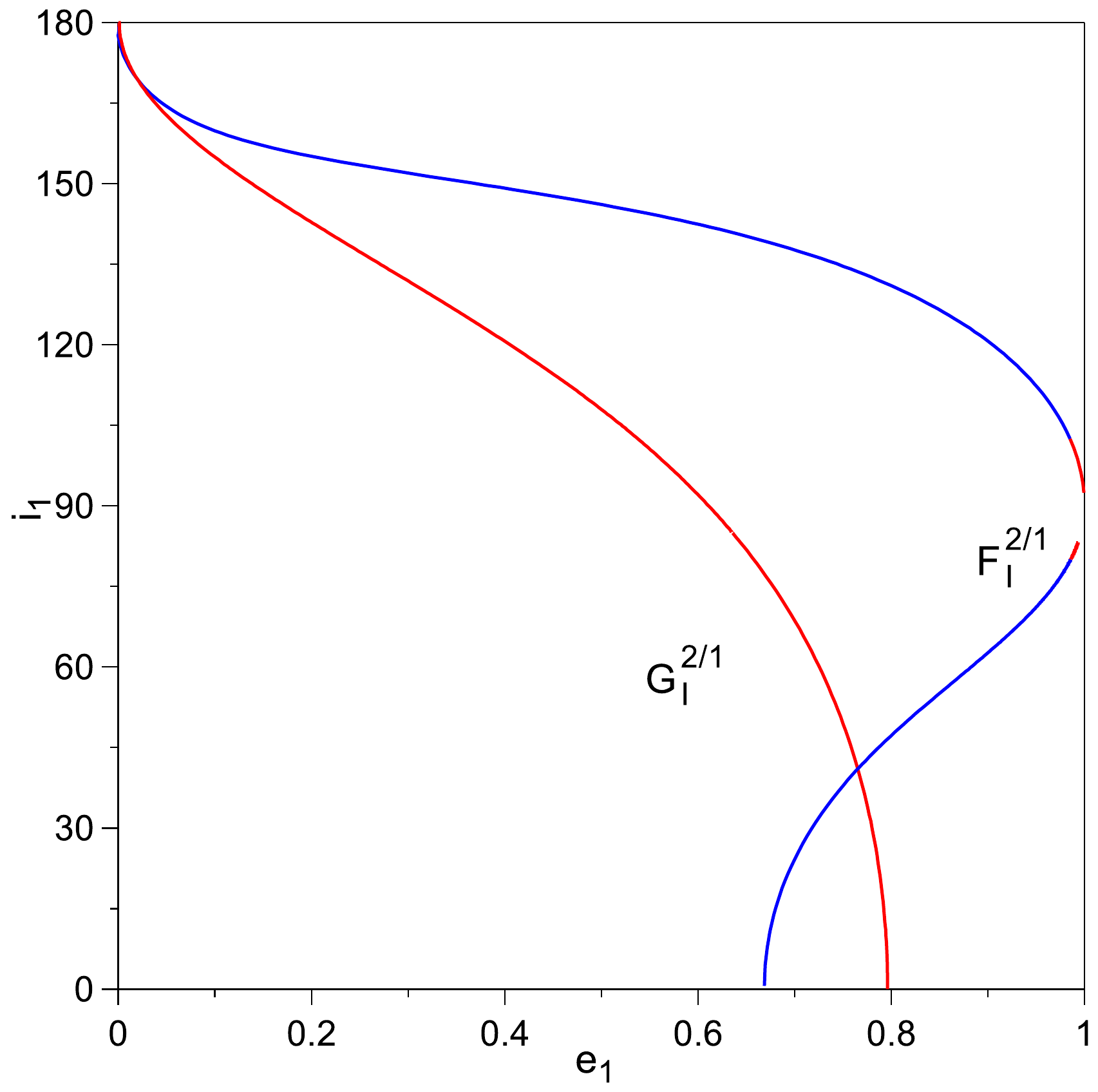}
\end{array} $
\end{center}
\caption{Spatial families of periodic orbits in the 3D-CRTB in the 2/1 MMR bifurcating from the 2D-CRTBP projected on the plane ($e_1,i_1$). $G^{2/1}_I$ and $F^{2/1}_I$ bifurcate from the v.c.o. at $e_1=0.79$ and $e_1=0.66$ of the family $I$. Stability (instability) is depicted by blue (red).} 
\label{21c3d}
\end{figure}

In Fig. \ref{21c3d}, we present the two spatial families in the 3D-CRTBP projected on the plane $(e_1,i_1)$. The spatial family $F^{2/1}_I$ is stable for both prograde and retrograde orbits except of the small region close to $90^{\circ}$ as $e_1\rightarrow 1$ and consists of $xz$-symmetric periodic orbits, whereas $G^{2/1}_I$, the spatial family of $x$-symmetric periodic orbits, is unstable. Both of the families terminate at planar circular retrograde orbits, while the continuation of $F^{2/1}_I$ is interrupted at $90^{\circ}$. These families were also computed by \citet{av12}, but for lower values of the inclination.
Examination of the spatial families in the 3D-CRTBP with respect to Method II showed that none of them possesses a bifurcation point to the 3D-ERTBP.

\subsubsection{3D-ERTBP}
\begin{figure}
\begin{center}
$\begin{array}{c}%\textnormal{3/2 MMR}  & \textnormal{2/1 MMR} \\
\includegraphics[width=0.9\columnwidth,keepaspectratio]{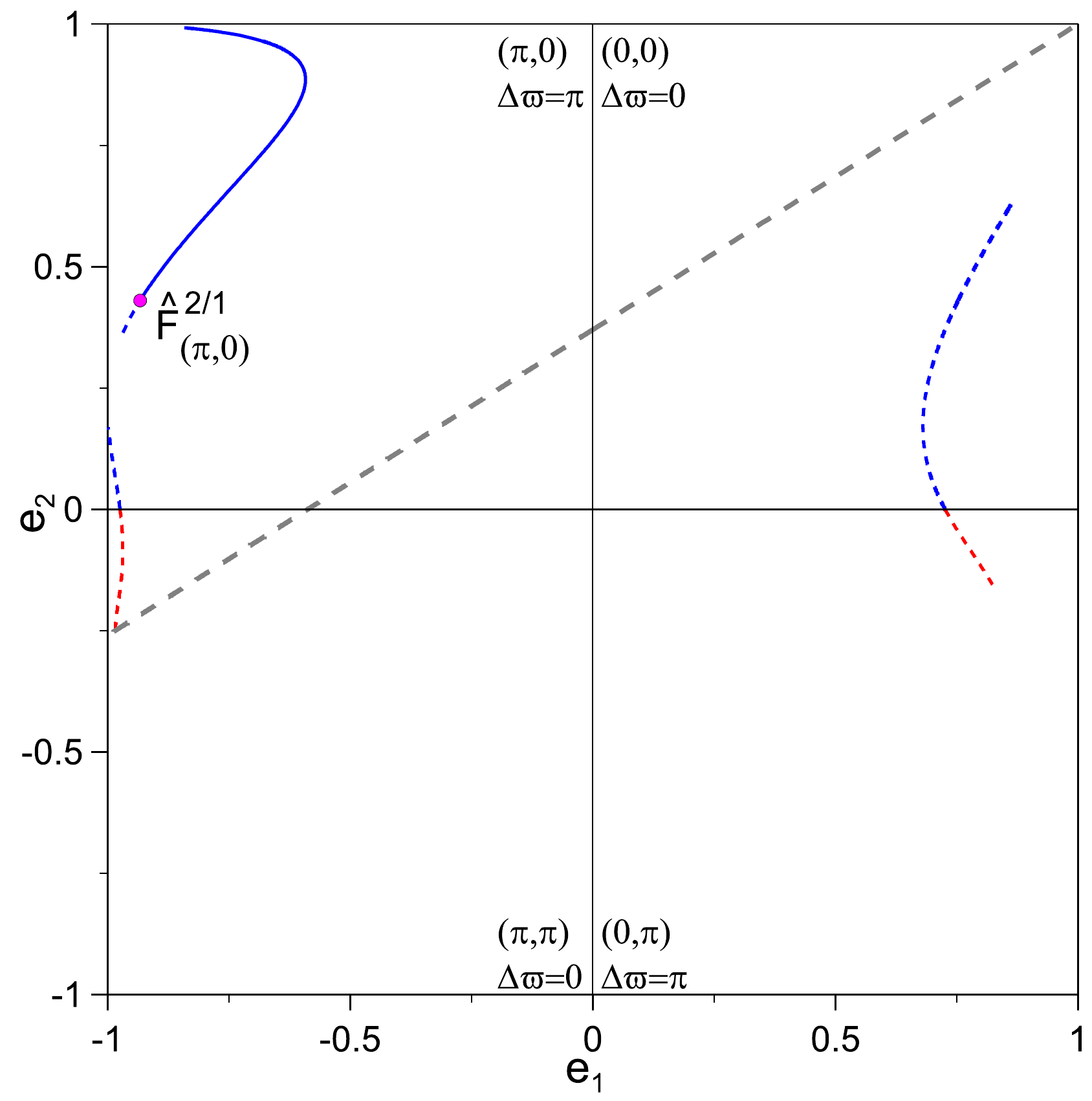} 
\end{array} $
\end{center}
\caption{Families of periodic orbits in the 2D-ERTBP for 2/1 MMR examined with respect to Method I. The v.c.o. generate spatial families in the 3D-ERTBP. Presented as in Fig. \ref{32evco}.}
\label{21evco}
\end{figure}

In Fig. \ref{21evco}, we present the families of the 2/1 MMR in the 2D-ERTBP examined with regards to the vertical stability. The majority of the families is vertically unstable (dashed lines), in tandem with being horizontally stable or unstable. There is only one large segment at high eccentricity values $e_2$ of the horizontally stable family in the configuration $(\pi,0)$ which is also vertically stable. This segment is examined in Sect. \ref{app} regarding the spatial neighbourhood of the horizontally and vertically stable planar periodic orbits. This family possesses one v.c.o. (hence the transition of the vertical stability along it), $\hat{F}^{2/1}_{(\pi,0)}$, located at $(e_1,e_2)=(0.93,0.43)$,  which generates $xz$-symmetric spatial periodic orbits in the 3D-ERTBP. 

\begin{figure}
\begin{center}
$\begin{array}{c}
\includegraphics[width=0.8\columnwidth]{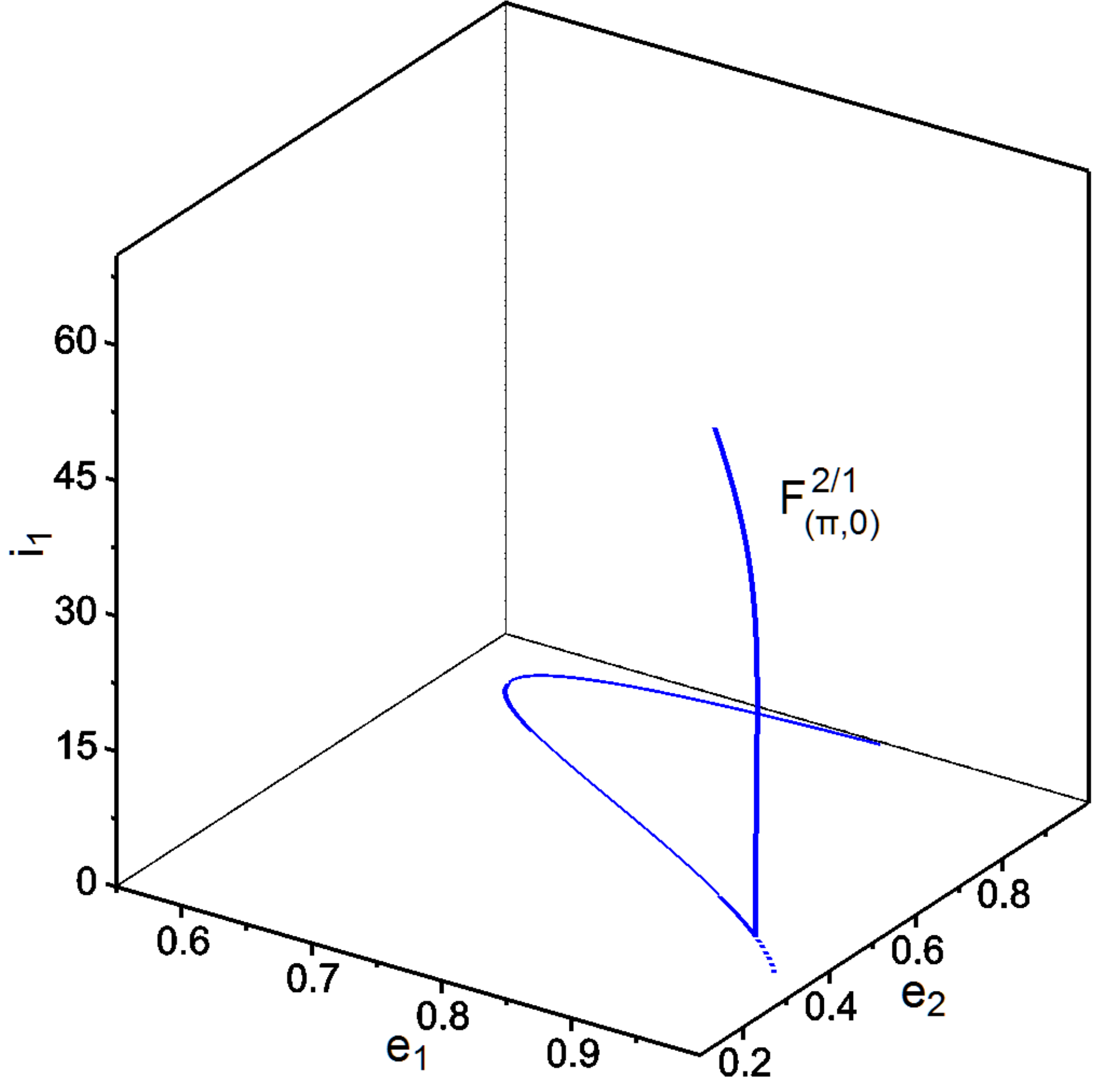}
\end{array} $
\end{center}
\caption{Spatial family of stable periodic orbits in the 3D-ERTBP in the 2/1 MMR of the configuration $(\pi,0)$ together with the generating orbits of the families of the 2D-ERTBP. The family $F^{2/1}_{(\pi,0)}$ bifurcates from the v.c.o. at $(e_1,e_2)=(0.93,0.43)$.}
\label{21e3d}
\end{figure}

In Fig. \ref{21e3d}, we present the spatial family $F^{2/1}_{(\pi,0)}$ of $xz$-symmetric periodic orbits, which is stable.

\subsection{5/2 MMR}

\begin{figure}
\begin{center}
$\begin{array}{c}
\includegraphics[width=0.9\columnwidth]{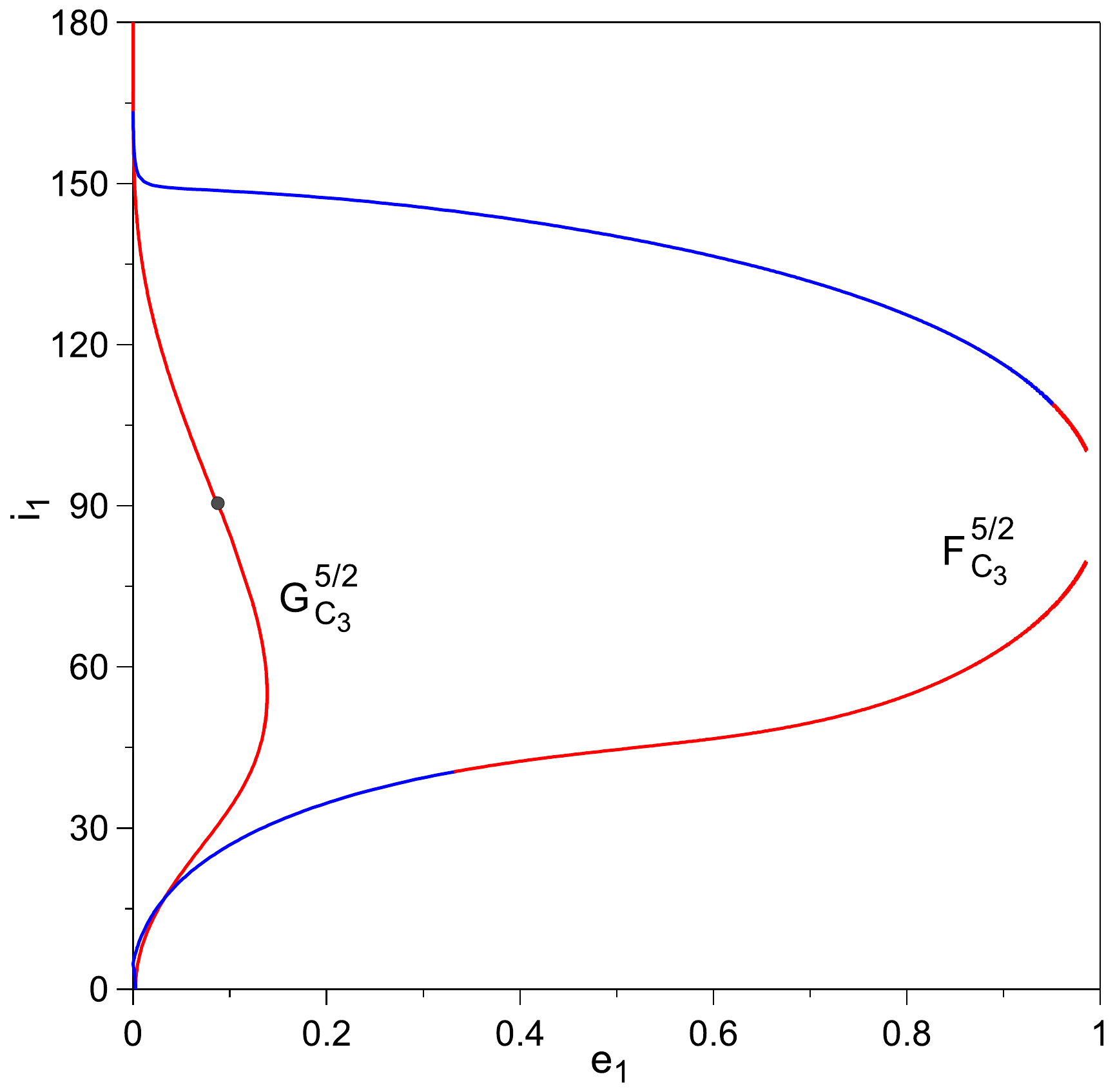}
\end{array} $
\end{center}
\caption{The spatial families, $F_{C_3}^{5/2}$ and $G_{C_3}^{5/2}$, in the 5/2 MMR bifurcating from the circular family, $C_3$, projected on the plane ($e_1,i_1$). The grey dot at $(e_1,i_1)=(0.089,90^{\circ})$ corresponds to a bifurcation point from the 3D-CRTBP to the 3D-ERTBP. Stable (unstable) periodic orbits are blue (red) coloured.}
\label{52circ}
\end{figure}

\subsubsection{3D-CRTBP}

When the multiplicity of the circular periodic orbits becomes equal to 3, we have the first occurrence of v.c.o. at 5/2 MMR along the circular family (see $C_3$, $C_6$ and $C_9$ in Fig. \ref{circmul} and Table \ref{multab}).

In Fig. \ref{52circ}, we present the spatial family $F_{C_3}^{5/2}$ of $xz$-symmetric periodic orbits, which is stable up to $40^{\circ}$ for prograde orbits and for $109^{\circ}<i_1<164^{\circ}$ for the retrograde ones, and the spatial family $G_{C_3}^{5/2}$ of $x$-symmetric periodic orbits which is unstable. Both families terminate at planar circular retrograde orbits. However, at $90^{\circ}$ the continuation of $F_{C_3}^{5/2}$ stopped as $e_1\rightarrow 1$. When examined with respect to Method II, $G_{C_3}^{5/2}$ has a bifurcation point to the 3D-ERTBP at $(e_1,i_1)=(0.0891812,90^{\circ})$. The families emanating from this bifurcation point will not be computed, since we only continue prograde orbits in the 3D-ERTBP in the following.

\begin{figure}
\begin{center}
$\begin{array}{cc}
\includegraphics[width=0.9\columnwidth,keepaspectratio]{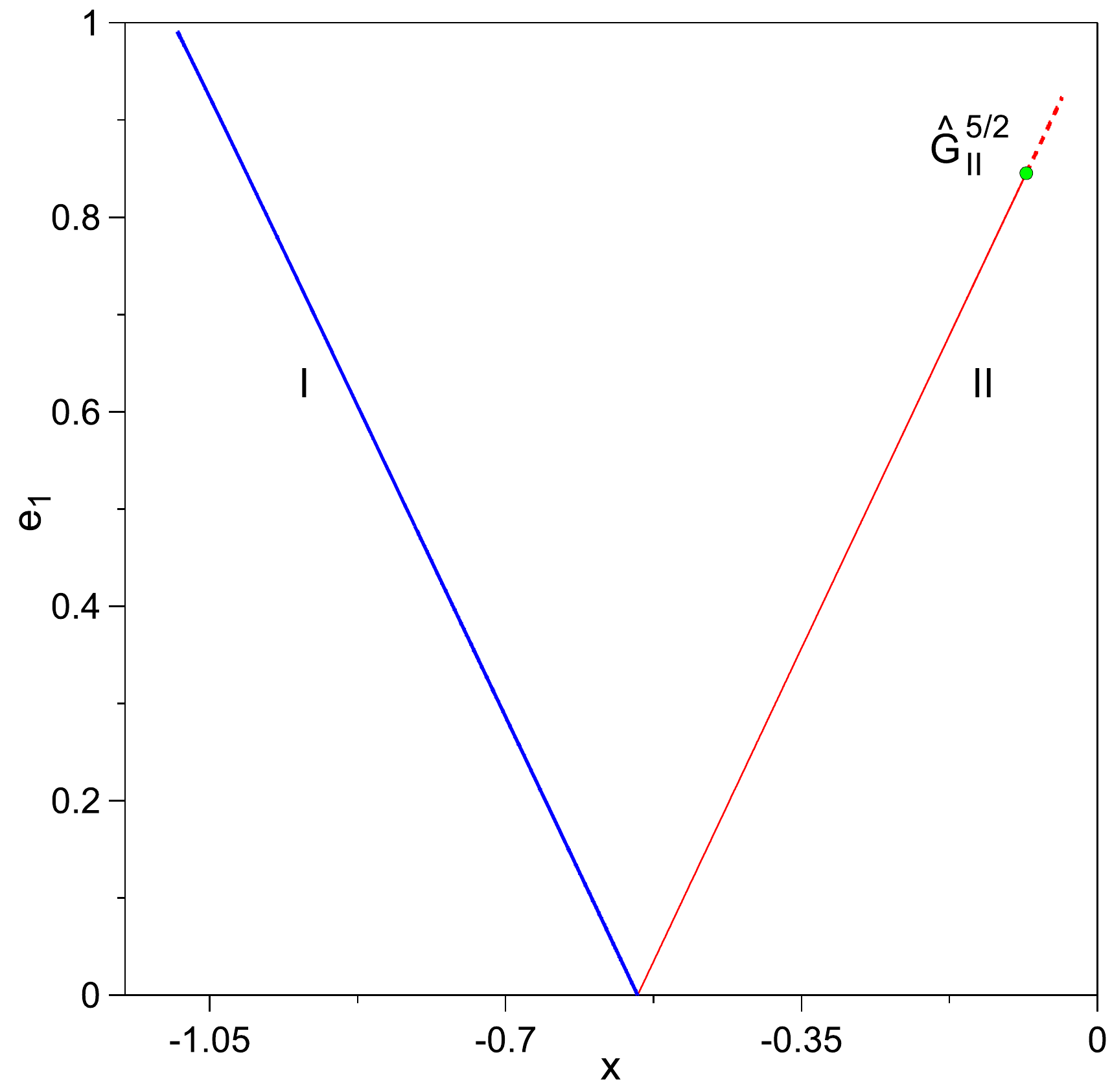} 
\end{array} $
\end{center}
\caption{Families of periodic orbits in the 2D-CRTBP for 5/2 MMR examined with respect to Method I. The  v.c.o. (green dot) generates spatial symmetric periodic orbits in the 3D-CRTBP. Colours and lines as in Fig. \ref{32cvco}.
}
\label{52cvco}
\end{figure}

In Fig. \ref{52cvco}, we present the families of 5/2 MMR in the 2D-CRTBP with respect to both the horizontal and vertical (Method I) stability. The family $I$ is both horizontally and vertically stable, while the family $II$ is horizontally unstable and vertically stable up to the v.c.o. $\hat{G}^{5/2}_{II}$, with eccentricity value $e_1=0.84$. Afterwards, this family is both horizontally and vertically unstable.

\begin{figure}
\begin{center}
$\begin{array}{c}
\includegraphics[width=0.9\columnwidth]{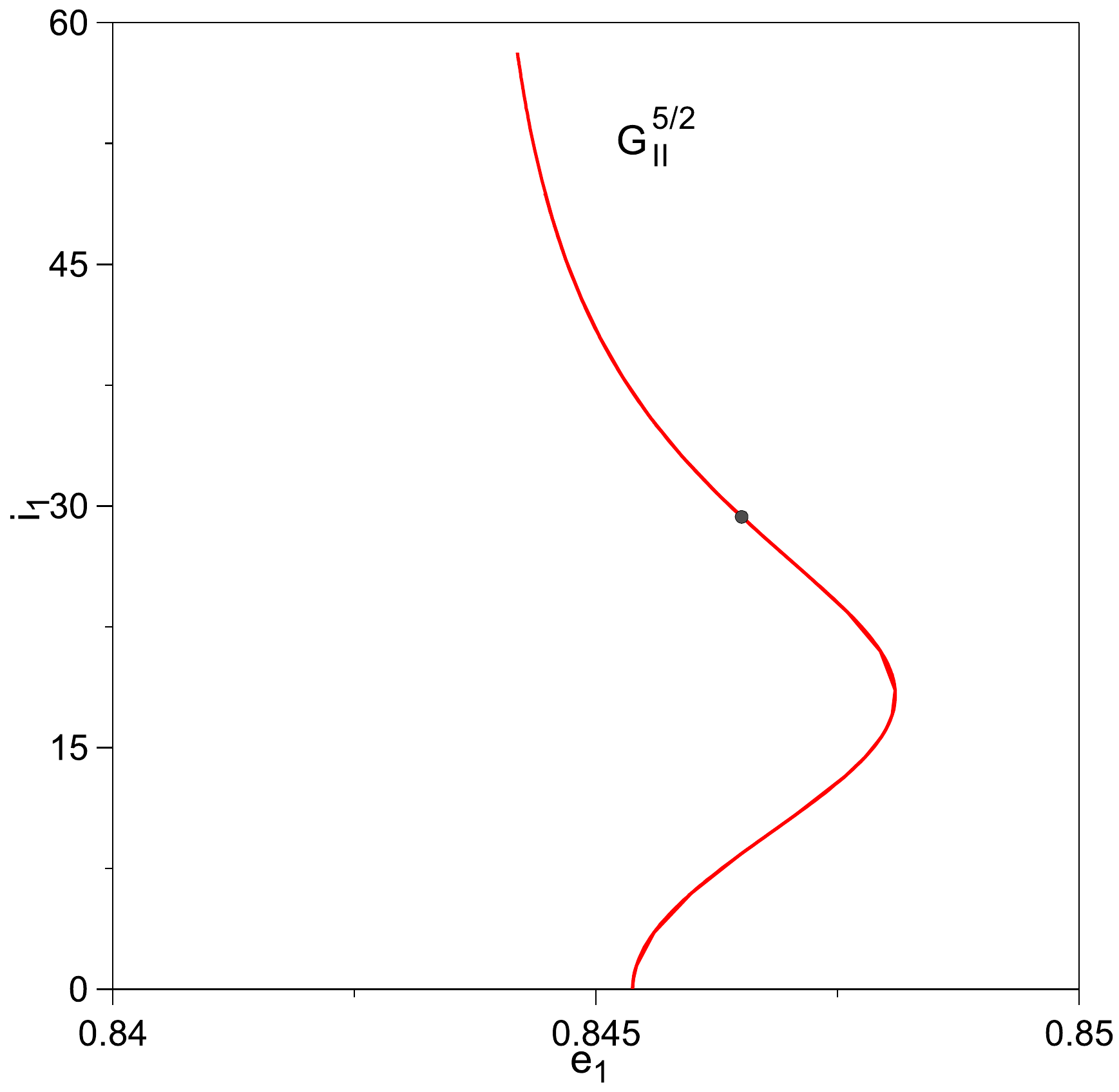}
\end{array} $
\end{center}
\caption{Spatial family of periodic orbits in the 3D-CRTB in the 5/2 MMR bifurcating from the 2D-CRTBP projected on the plane ($e_1,i_1$). $G^{5/2}_{II}$ bifurcates from the v.c.o. at $e_1=0.84$ of the family $II$. The grey dot at $(e_1,i_1)=(0.84,29^{\circ})$ corresponds to a bifurcation point from the 3D-CRTBP to the 3D-ERTBP.}
\label{52c3d}
\end{figure}

In Fig. \ref{52c3d}, we present the spatial family, $G^{5/2}_{II}$, of $x$-symmetric periodic orbits, which is unstable up to the inclination value $\sim 60^{\circ}$ it was continued. 
With regards to Method II, this family possesses one bifurcation point from the 3D-CRTBP to the 3D-ERTBP at $(e_1,i_1)=(0.84,29^{\circ})$ (grey dot). 

\subsubsection{3D-ERTBP}

\begin{figure}
\begin{center}
$\begin{array}{cc}
\includegraphics[width=0.9\columnwidth,keepaspectratio]{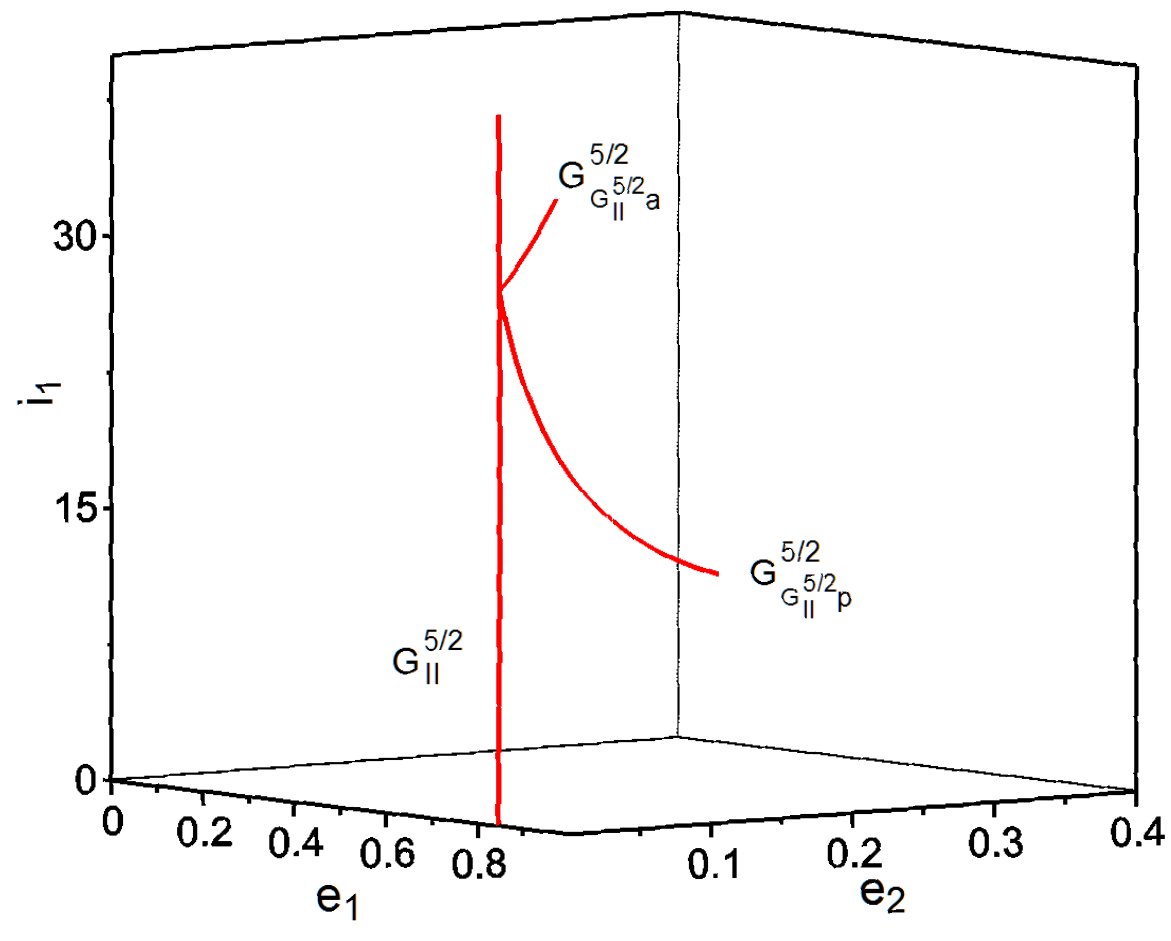} 
\end{array} $
\end{center}
\caption{The two unstable families, $G^{5/2}_{G^{5/2}_{II}p}$ and $G^{5/2}_{G^{5/2}_{II}a}$ of periodic orbits in the 3D-ERTBP for 3/2 MMR, which bifurcate, according to Method II, from the family $G^{5/2}_{II}$ (replotted) at the point $(e_1,i_1)=(0.84,29^{\circ})$, of the 3D-CRTBP shown in Fig. \ref{52c3d}.}
\label{52el}
\end{figure}

The bifurcation point of the spatial family $G^{5/2}_{II}$ of the 3D-CRTBP generates two families in the 3D-ERTBP, $G^{5/2}_{G^{5/2}_{II}p}$ and $G^{5/2}_{G^{5/2}_{II}a}$, one corresponding to the location of the giant at pericentre and the other at apocentre. They are whole unstable and are presented in Fig \ref{52el}.

\begin{figure}
\begin{center}
$\begin{array}{cc}%\textnormal{3/2 MMR}  & \textnormal{2/1 MMR} \\
\includegraphics[width=0.9\columnwidth,keepaspectratio]{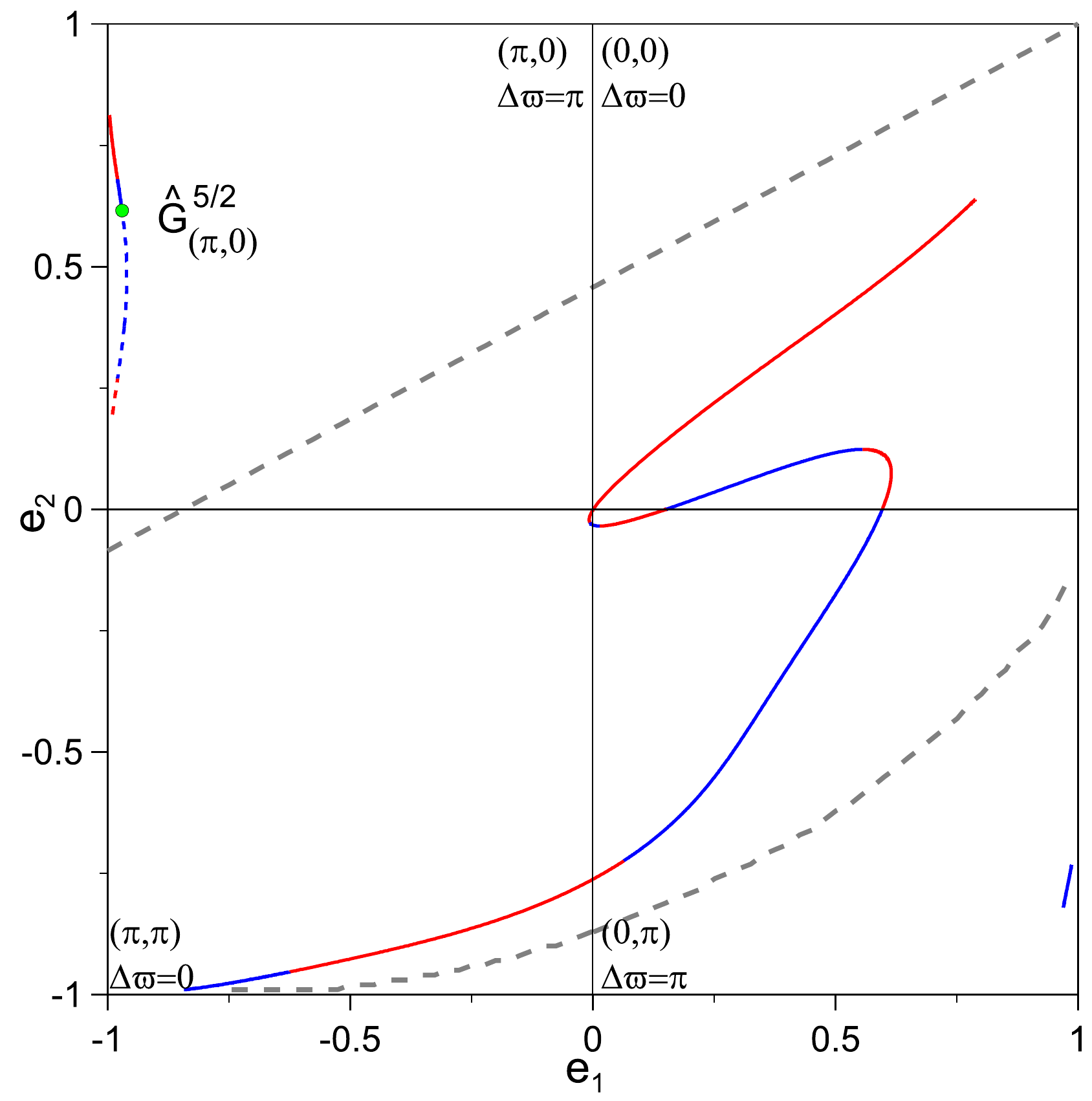} 
\end{array} $
\end{center}
\caption{Families of periodic orbits in the 2D-ERTBP for 5/2 MMR examined with respect to Method I. The v.c.o. generate spatial families in the 3D-ERTBP. Presented as in Fig. \ref{32evco}.}
\label{52evco}
\end{figure}

In Fig. \ref{52evco}, we examine the families of the 5/2 MMR in the 2D-ERTBP with respect to Method I. The vertical stability behaves similarly to the one observed in 3/2 MMR (Fig. \ref{32evco}). Particularly, all of the families in the configurations $(0,0)$, $(0,\pi)$ and $(\pi,\pi)$ are vertically stable (solid lines), while being either horizontally stable (blue) or unstable (red). The family in the configuration $(\pi,0)$ possesses one v.c.o.,  $\hat{G}^{5/2}_{(\pi,0)}$, and is vertically unstable (dashed line) up to this point located at $(e_1,e_2)=(0.96,0.61)$. As a consequence, the vertical instability along this family overlaps with the horizontal stability to a great extent.

\begin{figure}
\begin{center}
$\begin{array}{c}
 \includegraphics[width=0.8\columnwidth]{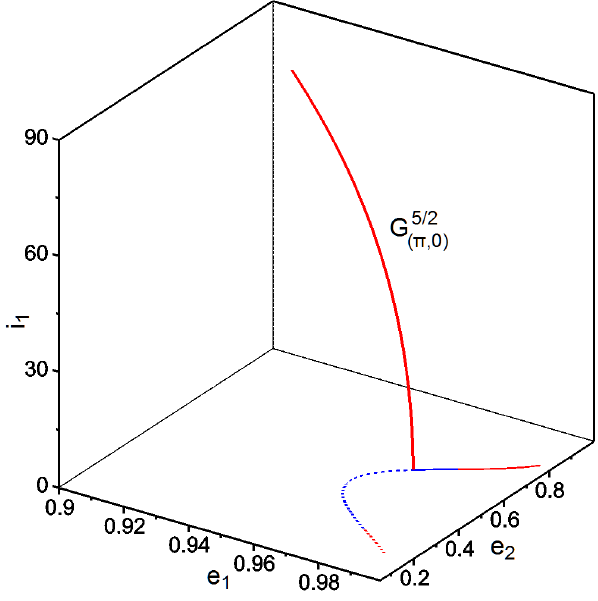}
\end{array} $
\end{center}
\caption{Spatial family of unstable periodic orbits in the 3D-ERTBP in the 5/2 MMR of the configuration $(\pi,0)$ together with the generating orbits of the families of the 2D-ERTBP. The family $G^{5/2}_{(\pi,0)}$ bifurcates from the v.c.o. at $(e_1,e_2)=(0.96,0.61)$.}
\label{52e3d}
\end{figure}

In Fig. \ref{52e3d}, we present the spatial family $G^{5/2}_{(\pi,0)}$ consisting of unstable $x$-symmetric periodic orbits.

\subsection{3/1 MMR}

\begin{figure}
\begin{center}
$\begin{array}{c}
\includegraphics[width=0.9\columnwidth]{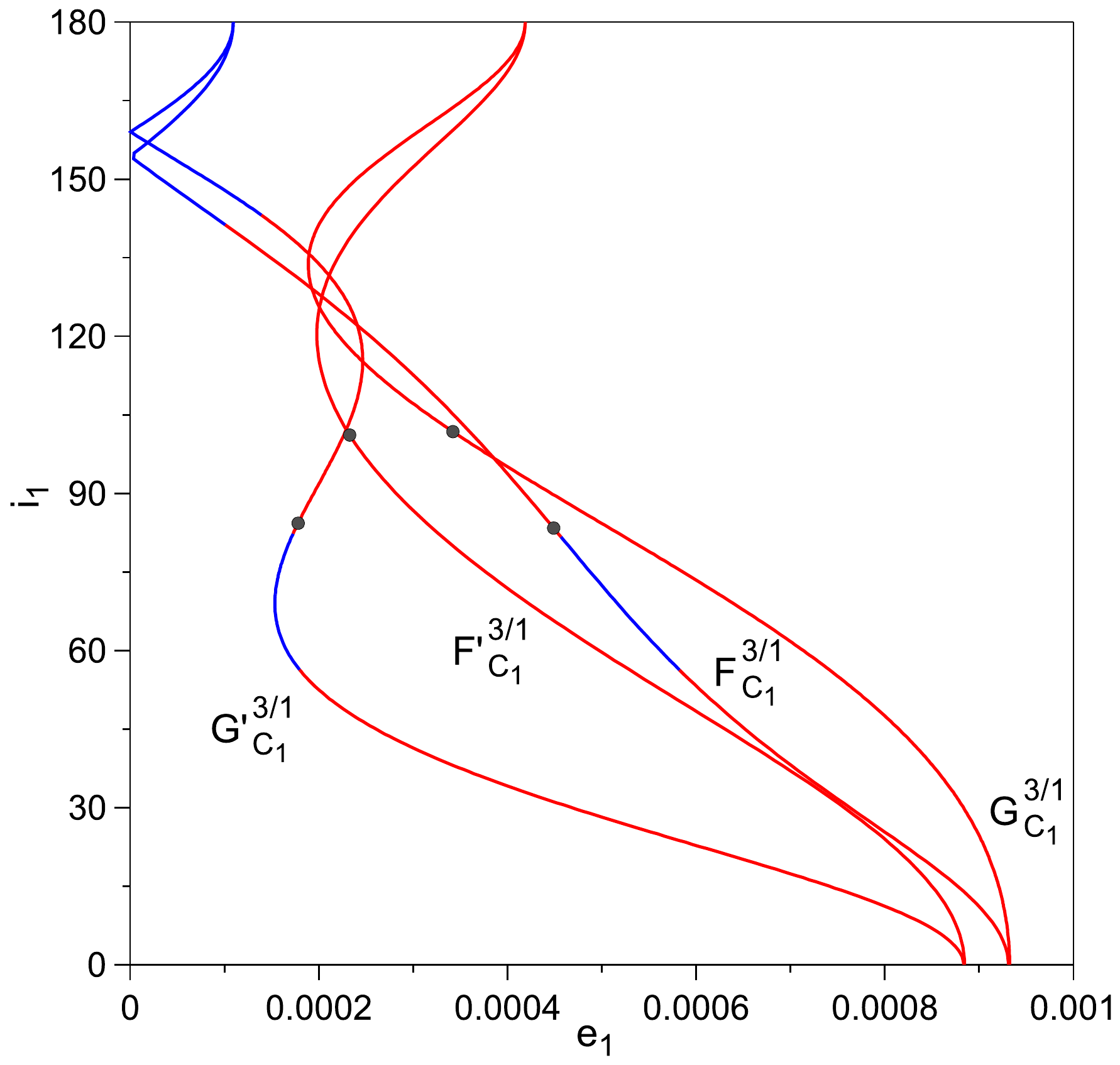}
\end{array} $
\end{center}
\caption{The spatial families $F_{C_1}^{3/1}$, $G_{C_1}^{3/1}$, ${F'}_{C_1}^{3/1}$ and ${G'}_{C_1}^{3/1}$ in the 3/1 MMR bifurcating from the circular family, $C_1$, projected on the plane ($e_1,i_1$). The grey dots at $(e_1,i_1)=(0.000178,84^{\circ})$, $(e_1,i_1)=(0.0002327,101^{\circ})$, $(e_1,i_1)=(0.0003454,101^{\circ})$ and $(e_1,i_1)=(0.0004451,84^{\circ})$ correspond to bifurcation points from the 3D-CRTBP to the 3D-ERTBP. Stability (instability) is showcased by blue (red).}
\label{31circ}
\end{figure}

\subsubsection{3D-CRTBP}

The 3/1 MMR is the only resonance, within the segment studied, which has two v.c.o. even when the multiplicity of the circular periodic orbits is equal to 1 and these v.c.o. are always apparent as the multiplicity increases (see Fig. \ref{circmul} and Table \ref{multab}).

In Fig. \ref{31circ}, we present four spatial families in the 3D-CRTBP emanating from the circular family, $C_1$. $F_{C_1}^{3/1}$ and ${F'}_{C_1}^{3/1}$ consist of $xz$-symmetric periodic orbits, while $G_{C_1}^{3/1}$ and ${G'}_{C_1}^{3/1}$ of $x$-symmetric periodic orbits. $F_{C_1}^{3/1}$ and ${G'}_{C_1}^{3/1}$ have stable periodic orbits when $56^{\circ}<i_1<82^{\circ}$ and $i_1>142^{\circ}$, while ${F'}_{C_1}^{3/1}$ and $G_{C_1}^{3/1}$ are whole unstable. All of them terminate at planar retrograde orbits.

When these families were examined with respect to Method II, we found four bifurcation points to the 3D-ERTBP at $(e_1,i_1)=(0.000178,84^{\circ})$, $(e_1,i_1)=(0.0002327,101^{\circ})$, $(e_1,i_1)=(0.0003454,101^{\circ})$ and $(e_1,i_1)=(0.0004451,84^{\circ})$. Only the prograde orbits will be continued to the 3D-ERTBP in the following.

\begin{figure}
\begin{center}
$\begin{array}{cc}
\includegraphics[width=0.9\columnwidth,keepaspectratio]{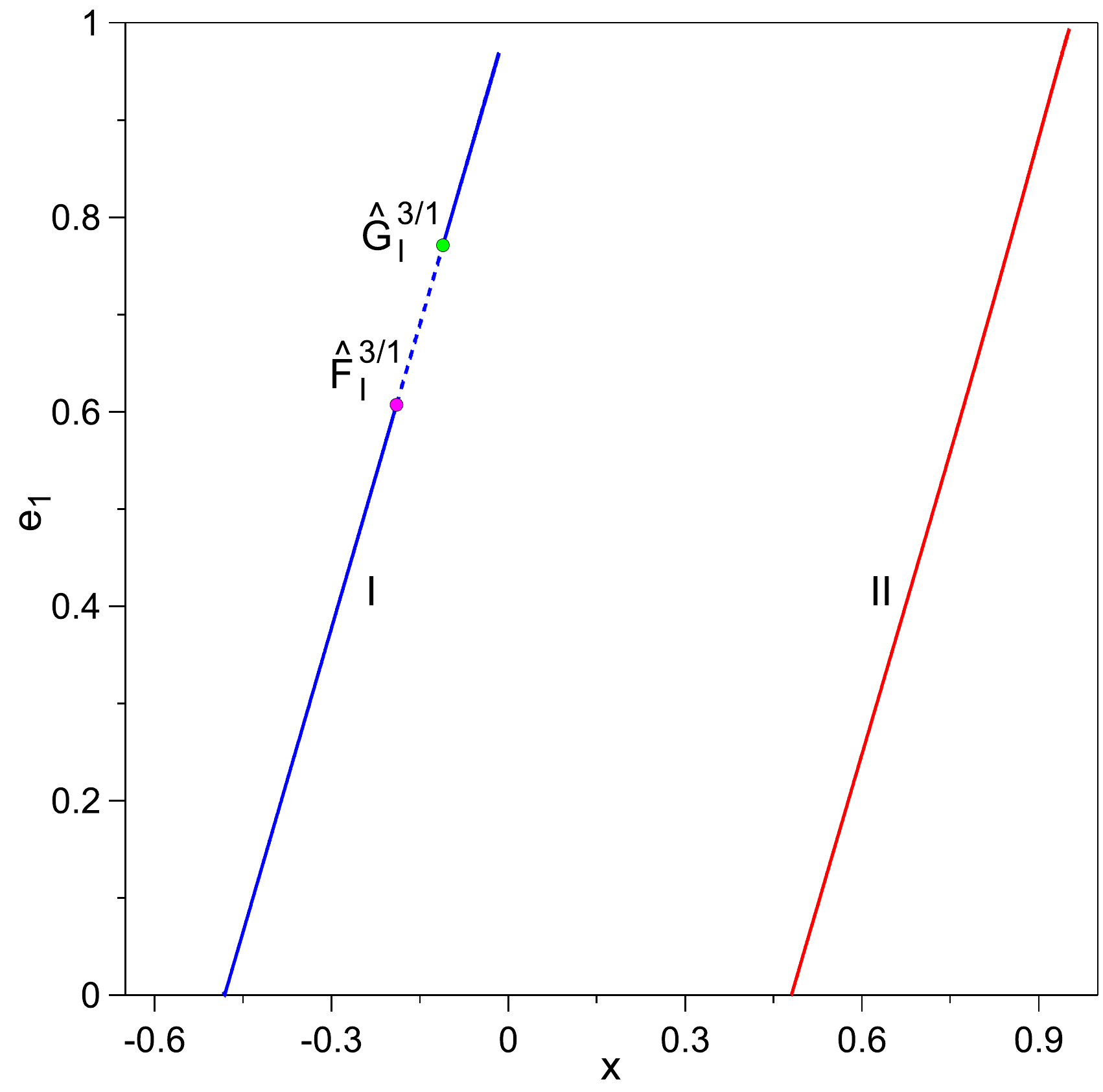} 
\end{array} $
\end{center}
\caption{Families of periodic orbits in the 2D-CRTBP for 3/1 MMR examined with respect to Method I. The  v.c.o. generate spatial symmetric periodic orbits in the 3D-CRTBP. Colours and lines as in Fig. \ref{32cvco}.
}
\label{31cvco}
\end{figure}

In Fig. \ref{31cvco}, we present the families of the 3/1 MMR in the 2D-CRTBP examined with regards to the vertical stability (Method I). The family $I$ exhibits a behaviour similar to the family $I$ of the 3/2 and 2/1 MMRs. Precisely, it possesses a pair of v.c.o, $\hat{F}^{3/1}_I$ and $\hat{G}^{3/1}_I$, located at eccentricity values $e_1=0.61$ and $e_1=0.77$, respectively. Hence, the family is horizontally and vertically stable (blue solid line) and gets only vertically unstable (blue dashed line) between the periodic orbits defining the pair of v.c.o. The family $II$ is horizontally unstable (red), but vertically stable (solid line).

\begin{figure}
\begin{center}
$\begin{array}{c}
\includegraphics[width=0.9\columnwidth]{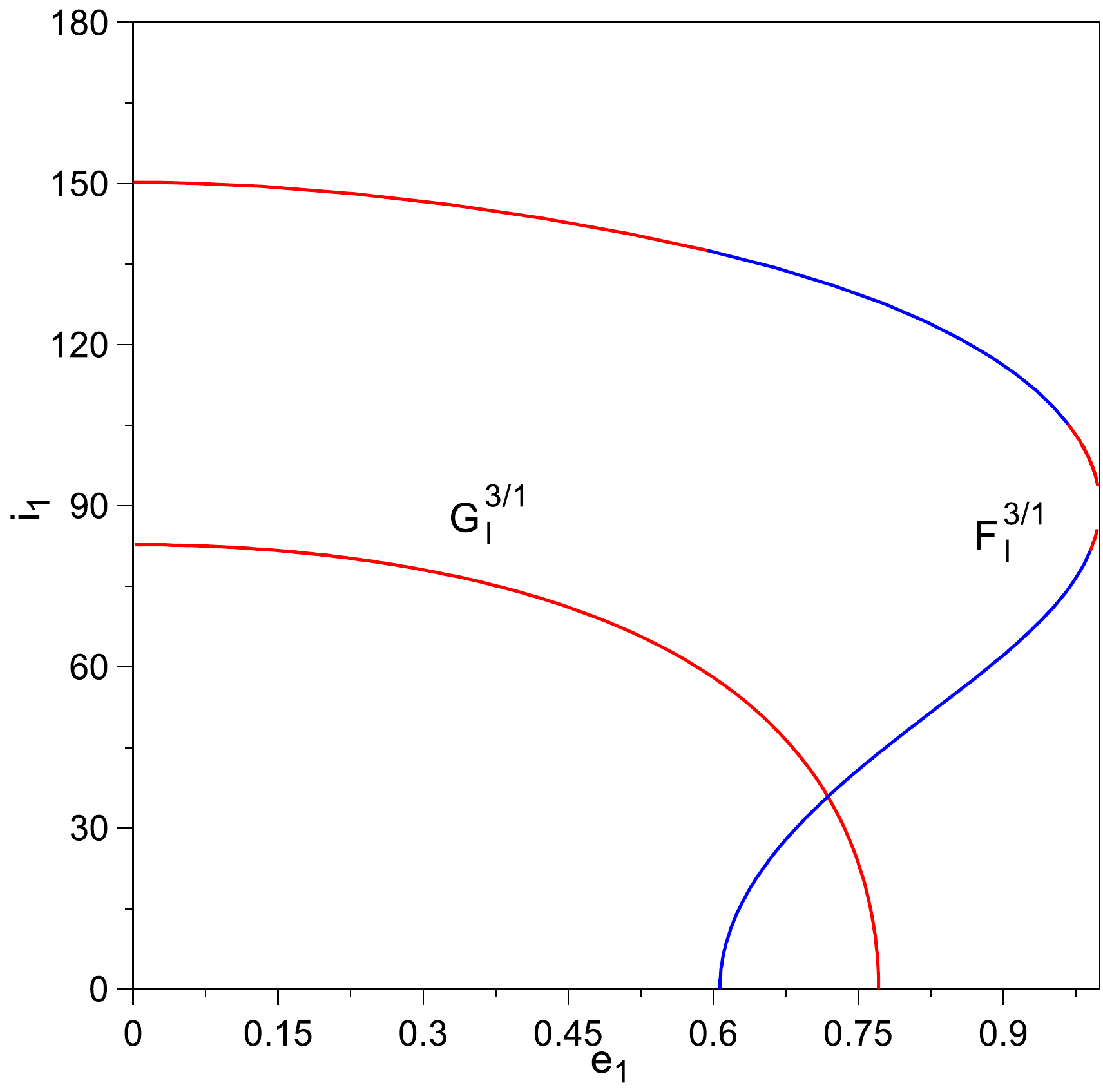}\\
\end{array} $
\end{center}
\caption{Spatial families of periodic orbits in the 3D-CRTBP in the 3/1 MMR bifurcating from the 2D-CRTBP projected on the plane ($e_1,i_1$). $G^{3/1}_I$ and $F^{3/1}_I$ bifurcate from the v.c.o. at $e_1=0.77$ and $e_1=0.61$ of the family $I$. Blue (red) colour stands for stability (instability).}% $56^{\circ}$.
\label{31c3d}
\end{figure}

In Fig. \ref{31c3d},  we present the two spatial families, $G^{3/1}_I$ and $F^{3/1}_I$, in the 3D-CRTBP projected on the plane $(e_1,i_1)$. The spatial family $F^{3/1}_I$ of $xz$-symmetric periodic orbits is stable up to $79^{\circ}$ for prograde orbits and for $104^{\circ}<i_1<138^{\circ}$ for retrograde orbits (the region of stability is interrupted at $90^{\circ}$ as $e_1$ approaches 1), whereas $G^{3/1}_I$, the spatial family of $x$-symmetric periodic orbits, is unstable. $G^{3/1}_I$ and $F^{3/1}_I$ reach a circular periodic orbit at $i_1=82.77^{\circ}$ and $i_1=150.22^{\circ}$, respectively, then they are reflected backwards and terminate at the v.c.o. which generated them. Examination of these spatial families with respect to Method II revealed no bifurcation points leading to the 3D-ERTBP. 

\subsubsection{3D-ERTBP}
\begin{figure}
\begin{center}
$\begin{array}{cc}
\includegraphics[width=0.9\columnwidth,keepaspectratio]{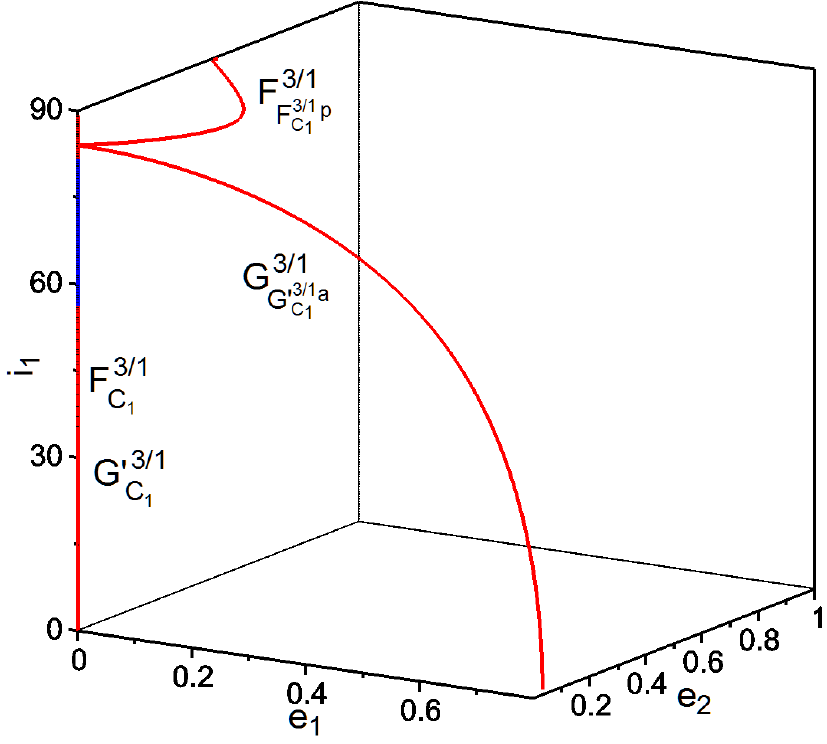} 
\end{array} $
\end{center}
\caption{The unstable families, $F^{3/1}_{F^{3/1}_{C_1}p}$ and $G^{3/1}_{{G'}^{3/1}_{C_1}a}$, of periodic orbits in the 3D-ERTBP for 3/1 MMR, which bifurcate, according to Method II, from the families $F^{3/1}_{C_1}$ and ${G'}^{3/1}_{C_1}$ (shown for $e_2=0$) at the points $(e_1,i_1)=(0.0004451,84^{\circ})$ and $(e_1,i_1)=(0.000179,84^{\circ})$, respectively, of the 3D-CRTBP shown in Fig. \ref{31circ}.}
\label{31el}
\end{figure}

In Fig. \ref{31el}, we present the unstable families, $F^{3/1}_{F^{3/1}_{C_1}p}$ and $G^{3/1}_{{G'}^{3/1}_{C_1}a}$, in the 3D-ERTBP bifurcating from the prograde orbits of the 3D-CRTBP shown in Fig. \ref{31circ}. They correspond to the location of the giant at pericentre and apocentre, respectively. Along $F^{3/1}_{F^{3/1}_{C_1}p}$ there is a change of the configuration of the terrestrial planet at $i_1=90^{\circ}$. The other two families for the location of the giant at the apocentre and the pericentre, respectively, could not be found.

\begin{figure}
\begin{center}
$\begin{array}{cc}
\includegraphics[width=0.9\columnwidth,keepaspectratio]{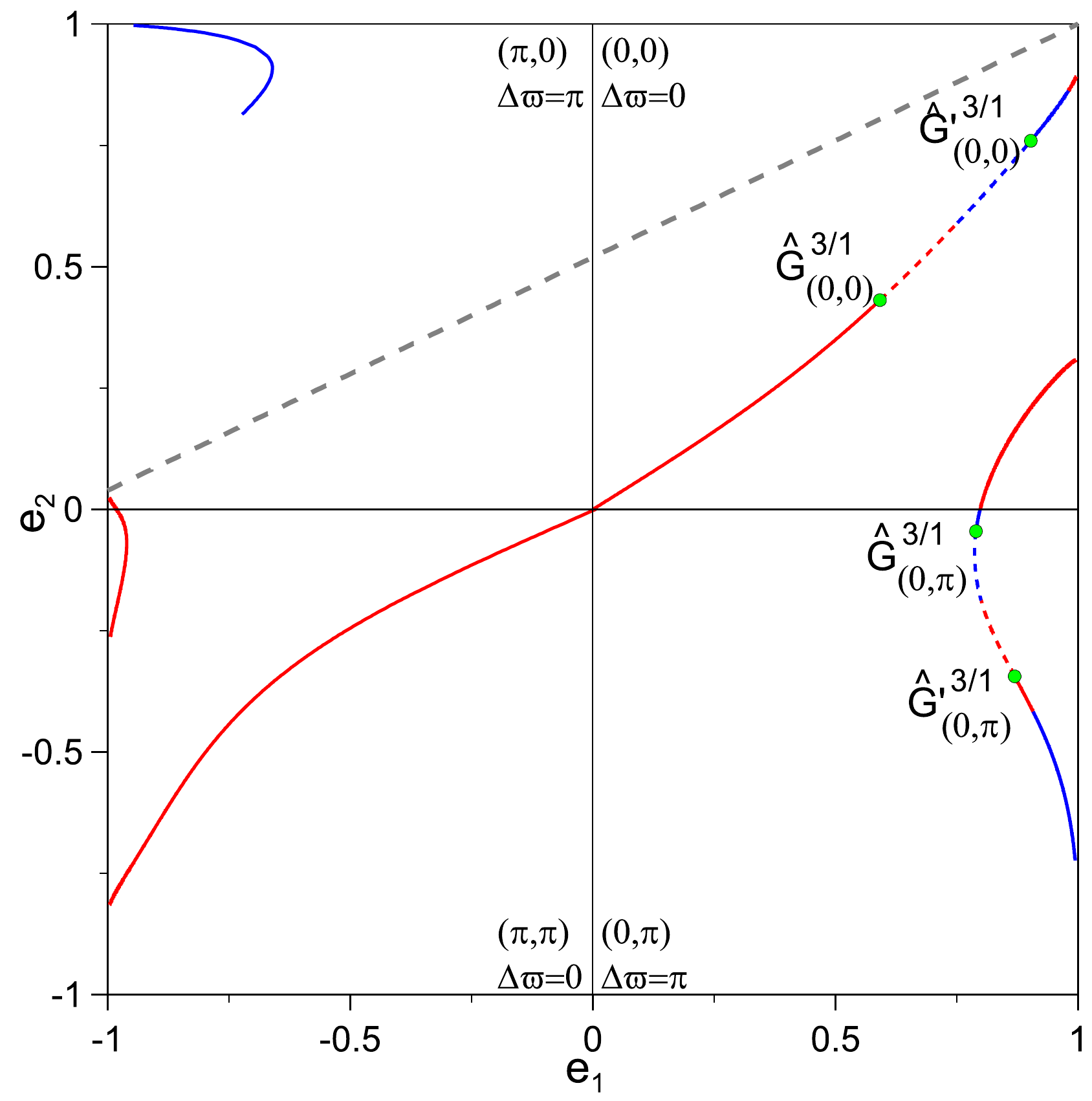} 
\end{array} $
\end{center}
\caption{Families of periodic orbits in the 2D-ERTBP for 3/1 MMR examined with respect to Method I. The v.c.o. generate spatial families in the 3D-ERTBP. Presented as in Fig. \ref{32evco}.}
\label{31evco}
\end{figure}

In Fig. \ref{31evco}, we examine the planar families of the 3/1 MMR in the 2D-ERTBP with respect to their vertical stability. All of them are vertically stable (solid line), apart from two segments between two v.c.o. in each of the configurations $(0,0)$ and $(0,\pi)$, where vertical instability is observed (dashed line). In the configuration $(0,0)$, the v.c.o. $\hat{G}^{3/1}_{(0,0)}$ and $\hat{G}'^{3/1}_{(0,0)}$ are located at $(e_1,e_2)=(0.59,0.43)$ and $(e_1,e_2)=(0.90,0.76)$, respectively. In the configuration $(0,\pi)$, the v.c.o. $\hat{G}^{3/1}_{(0,\pi)}$ and $\hat{G}'^{3/1}_{(0,\pi)}$ are located at $(e_1,e_2)=(0.78,0.05)$ and $(e_1,e_2)=(0.87,0.35)$, respectively.

\begin{figure}
\begin{center}
$\begin{array}{c}
\includegraphics[width=0.8\columnwidth]{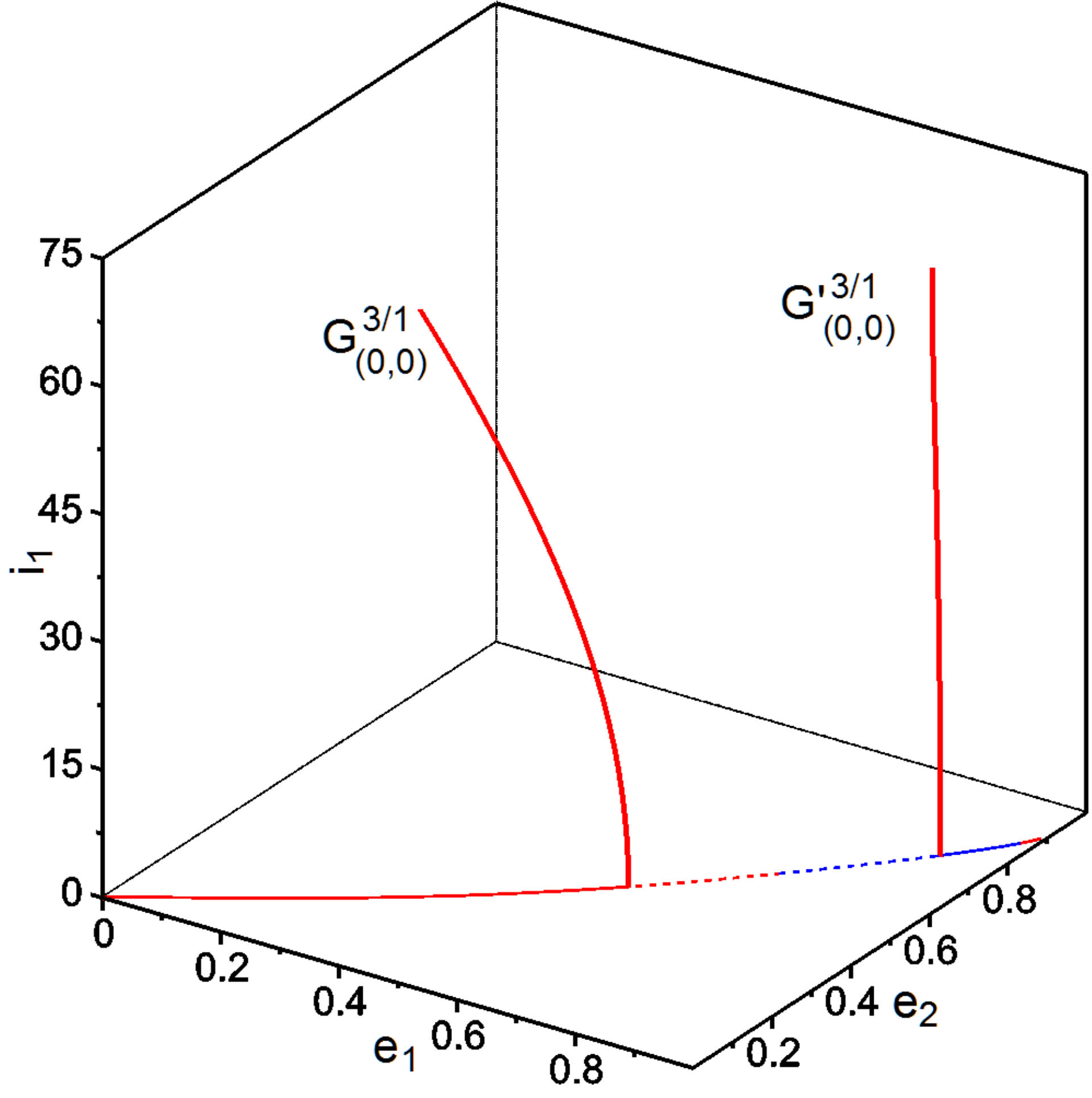}\\ \includegraphics[width=0.8\columnwidth]{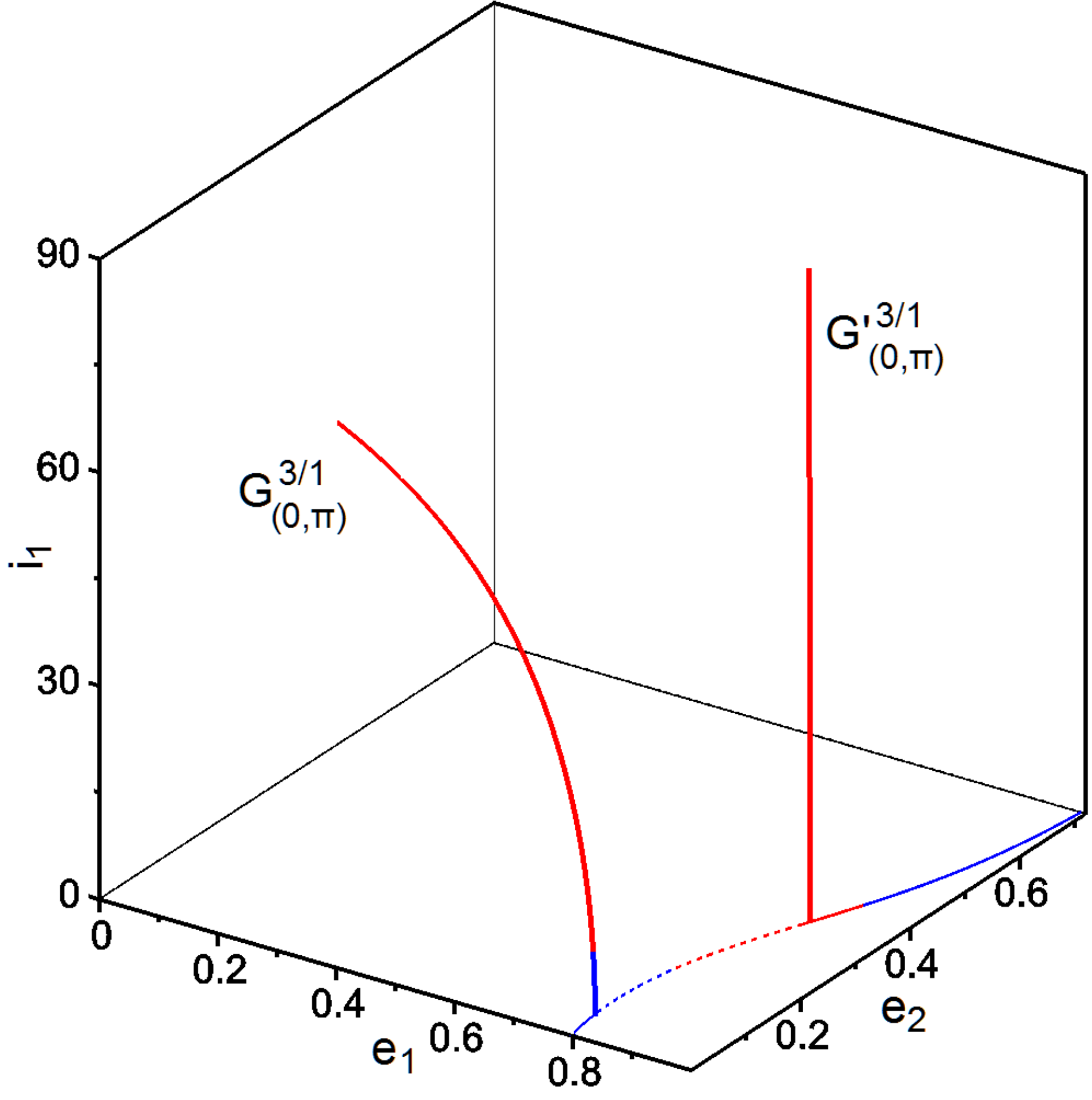}
\end{array} $
\end{center}
\caption{Spatial families in the 3D-ERTBP of the 3/1 MMR. $G^{3/1}_{(0,0)}$ and $G'^{3/1}_{(0,0)}$ bifurcate from the v.c.o. at $(e_1,e_2)=(0.59,0.43)$ and $(e_1,e_2)=(0.90,0.76)$ of the configuration $(0,0)$. $G^{3/1}_{(0,\pi)}$ and $G'^{3/1}_{(0,\pi)}$ bifurcate from the v.c.o. at $(e_1,e_2)=(0.78,0.05)$ and $(e_1,e_2)=(0.87,0.35)$ of the configuration $(0,\pi)$. The planar families encompassing those v.c.o. are also shown.}
\label{31e3d}
\end{figure}

In Fig. \ref{31e3d}, we present the spatial families of $x$-symmetric periodic orbits of the 3/1 MMR in the 3D-ERTBP. The families of the configuration $(0,0)$, $G^{3/1}_{(0,0)}$ and $G'^{3/1}_{(0,0)}$, are totally unstable. In the configuration $(0,\pi)$, the family $G^{3/1}_{(0,\pi)}$ is stable up to an inclination value $i_1=10^{\circ}$, while the family $G'^{3/1}_{(0,\pi)}$ is unstable.

\subsection{4/1 MMR}

\begin{figure}
\begin{center}
$\begin{array}{c}
\includegraphics[width=0.9\columnwidth]{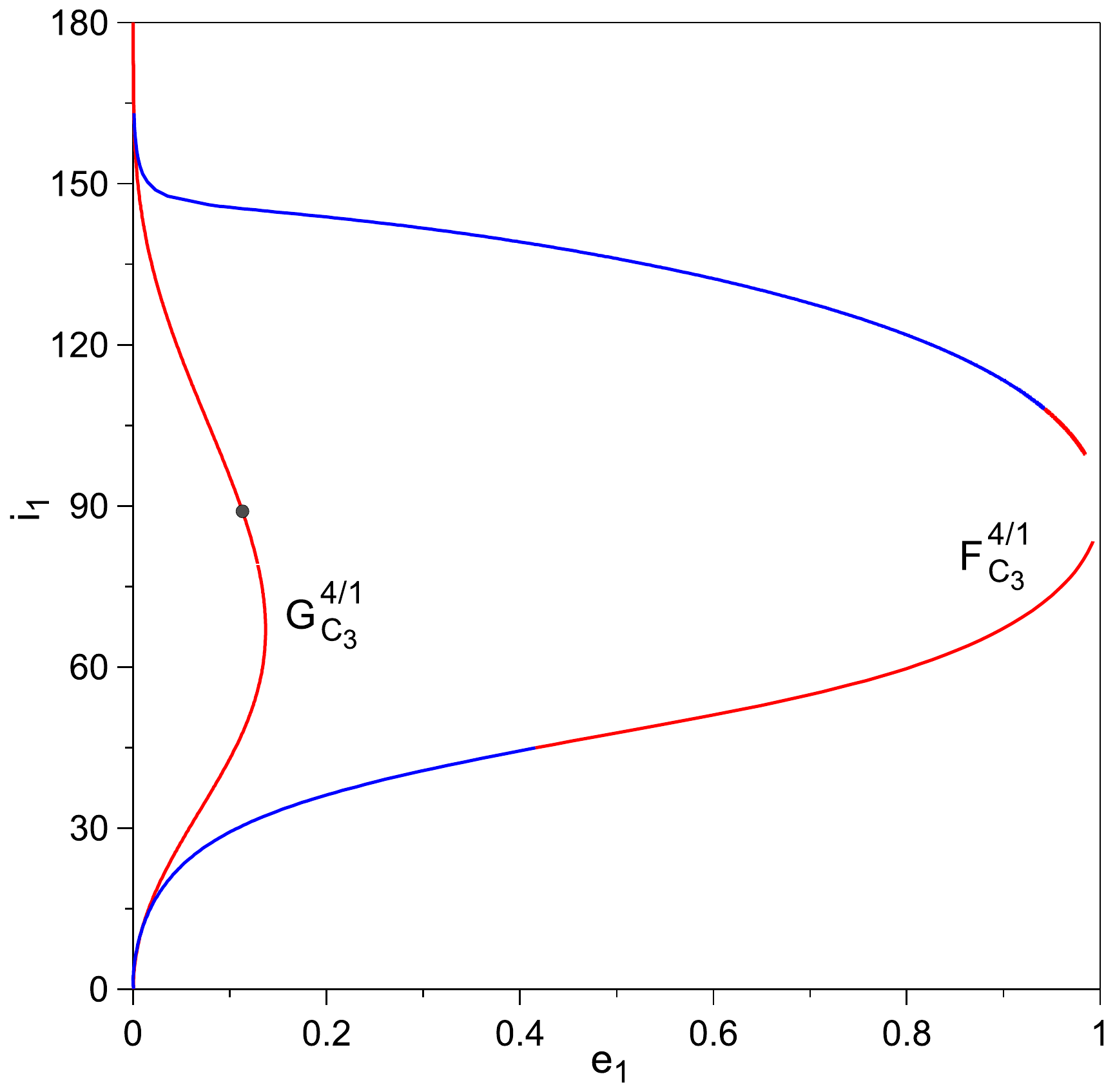}
\end{array} $
\end{center}
\caption{The spatial families, $F_{C_3}^{4/1}$ and $G_{C_3}^{4/1}$, in the 4/1 MMR bifurcating from the circular family, $C_3$, projected on the plane ($e_1,i_1$). The grey dot at $(e_1,i_1)=(0.113,89^{\circ})$ corresponds to a bifurcation point from the 3D-CRTBP to the 3D-ERTBP.  is depicted by blue.}
\label{41circ}
\end{figure}

\subsubsection{3D-CRTBP}
Likewise 5/2 MMR, when the multiplicity of the circular periodic orbits becomes equal to 3, we have the first occurrence of v.c.o. at 4/1 MMR along the circular family (see $C_3$, $C_6$ and $C_9$ in Fig. \ref{circmul}). 

In Fig. \ref{41circ}, we present the families $F_{C_3}^{4/1}$ and $G_{C_3}^{4/1}$. $F_{C_3}^{4/1}$ has $xz$-symmetric periodic orbits which are stable up to $45^{\circ}$ for prograde motion and for the segment $108^{\circ}<i_1<163^{\circ}$ for retrograde orbits. At $90^{\circ}$ this family breaks as $e_1\rightarrow 1$ as the continuation stopped. The family $G_{C_3}^{4/1}$ consists of $x$-symmetric unstable periodic orbits. Both of the families terminate at planar circular retrograde orbits. When examined with respect to Method II,  one bifurcation point to the 3D-ERTBP was found at $(e_1,i_1)=(0.113,89^{\circ})$, which will not be continued to the 3D-ERTBP.

\begin{figure}
\begin{center}
$\begin{array}{cc}
%\textnormal{3/2 MMR}  & \textnormal{2/1 MMR} \\
\includegraphics[width=0.9\columnwidth,keepaspectratio]{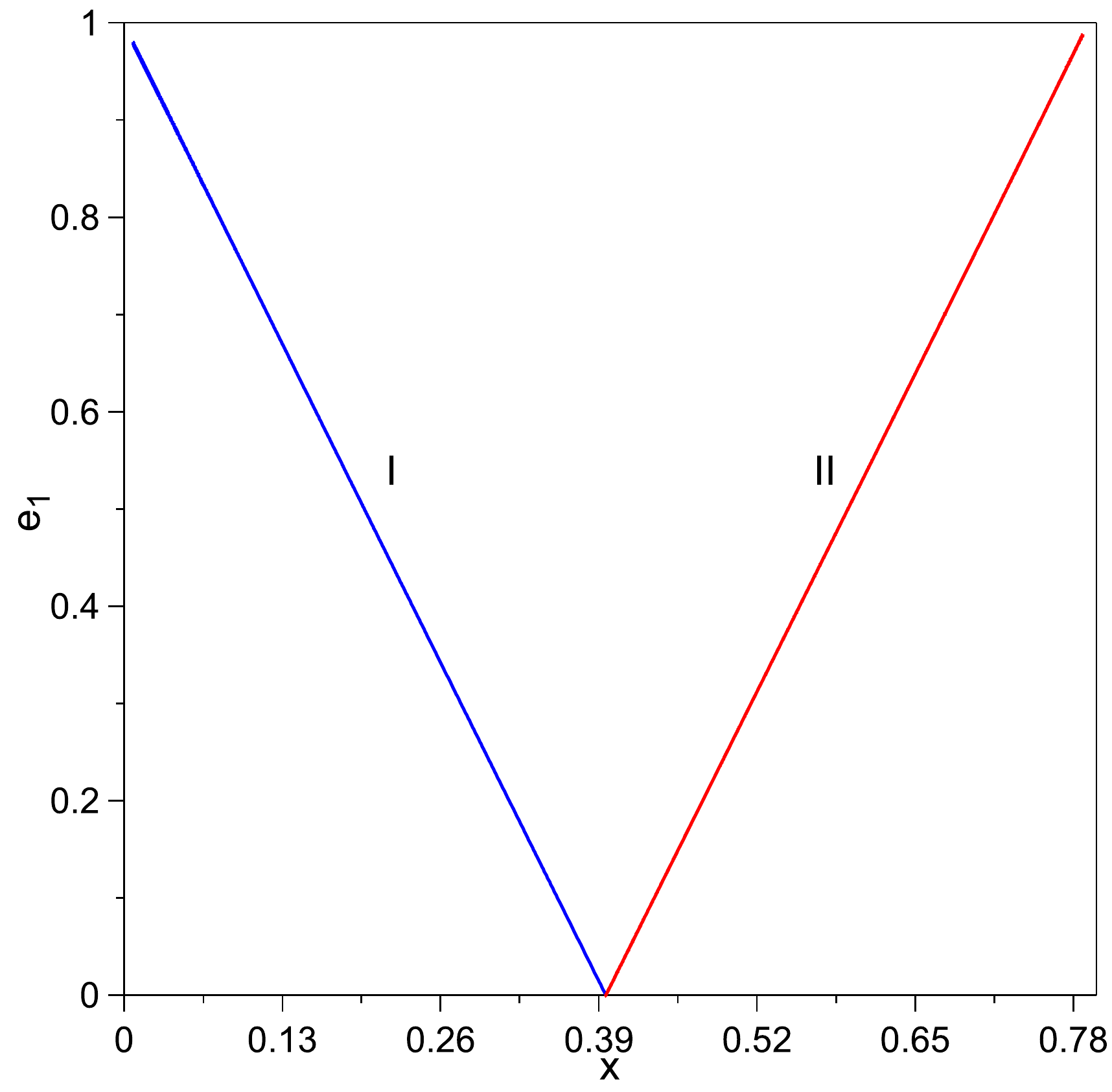} 
\end{array} $
\end{center}
\caption{Families of periodic orbits in the 2D-CRTBP for  4/1 MMR examined with respect to Method I. Colours and lines as in Fig. \ref{32cvco}.
}
\label{41cvco}
\end{figure}

In Fig. \ref{41cvco}, we examine the planar families of the 4/1 MMR in the 2D-CRTBP with respect to the vertical stability. We find that both of the families, $I$ (horizontally stable) and $II$ (horizontally unstable), are vertically stable (solid line). Therefore, no v.c.o. exists and hence, no bifurcation point from the 2D-CRTBP to the 3D-CRTBP. 

\subsubsection{3D-ERTBP}
\begin{figure}
\begin{center}
$\begin{array}{cc}%\textnormal{3/2 MMR}  & \textnormal{2/1 MMR} \\
\includegraphics[width=0.9\columnwidth,keepaspectratio]{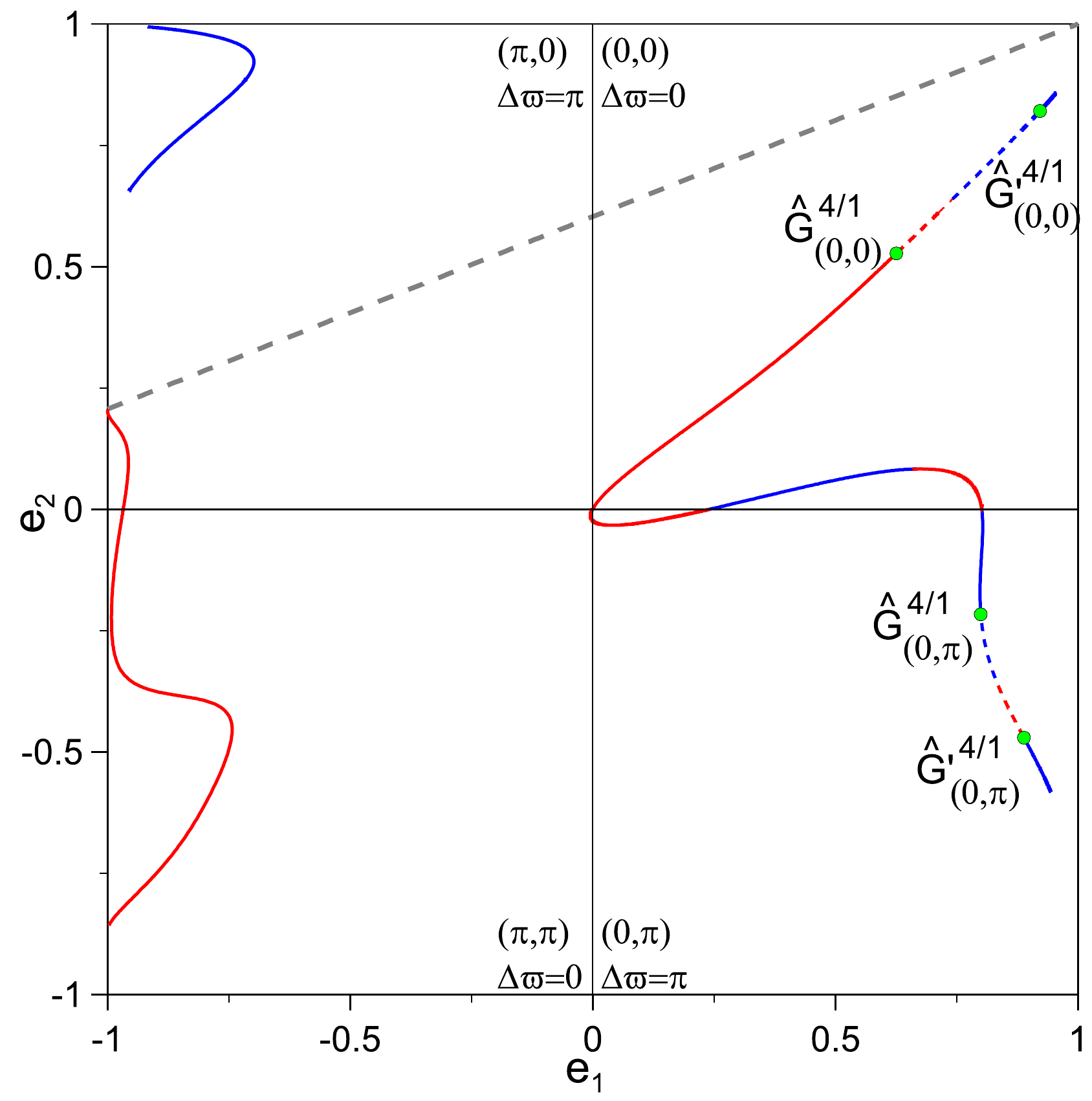} 
\end{array} $
\end{center}
\caption{Families of periodic orbits in the 2D-ERTBP for 4/1 MMR examined with respect to Method I. The v.c.o. generate spatial families in the 3D-ERTBP. Presented as in Fig. \ref{32evco}.}
\label{41evco}
\end{figure}

In Fig. \ref{41evco}, we examine the families of the 4/1 MMR in the 2D-ERTBP with respect to Method I. The results are qualitatively similar to the ones of 3/1 MMR in the 2D-ERTBP. Particularly, all of the families are vertically stable (solid line), apart from two segments between two v.c.o. in each of the configurations $(0,0)$ and $(0,\pi)$ (dashed line). In the configuration $(0,0)$, the v.c.o. $\hat{G}^{4/1}_{(0,0)}$ and $\hat{G}'^{4/1}_{(0,0)}$ are located at $(e_1,e_2)=(0.62,0.52)$ and $(e_1,e_2)=(0.92,0.82)$, respectively. The v.c.o. $\hat{G}^{4/1}_{(0,\pi)}$ and $\hat{G}'^{4/1}_{(0,\pi)}$ of the configuration $(0,\pi)$ are located at $(e_1,e_2)=(0.79,0.21)$ and $(e_1,e_2)=(0.88,0.47)$, respectively.

\begin{figure}
\begin{center}
$\begin{array}{c}
\includegraphics[width=0.8\columnwidth]{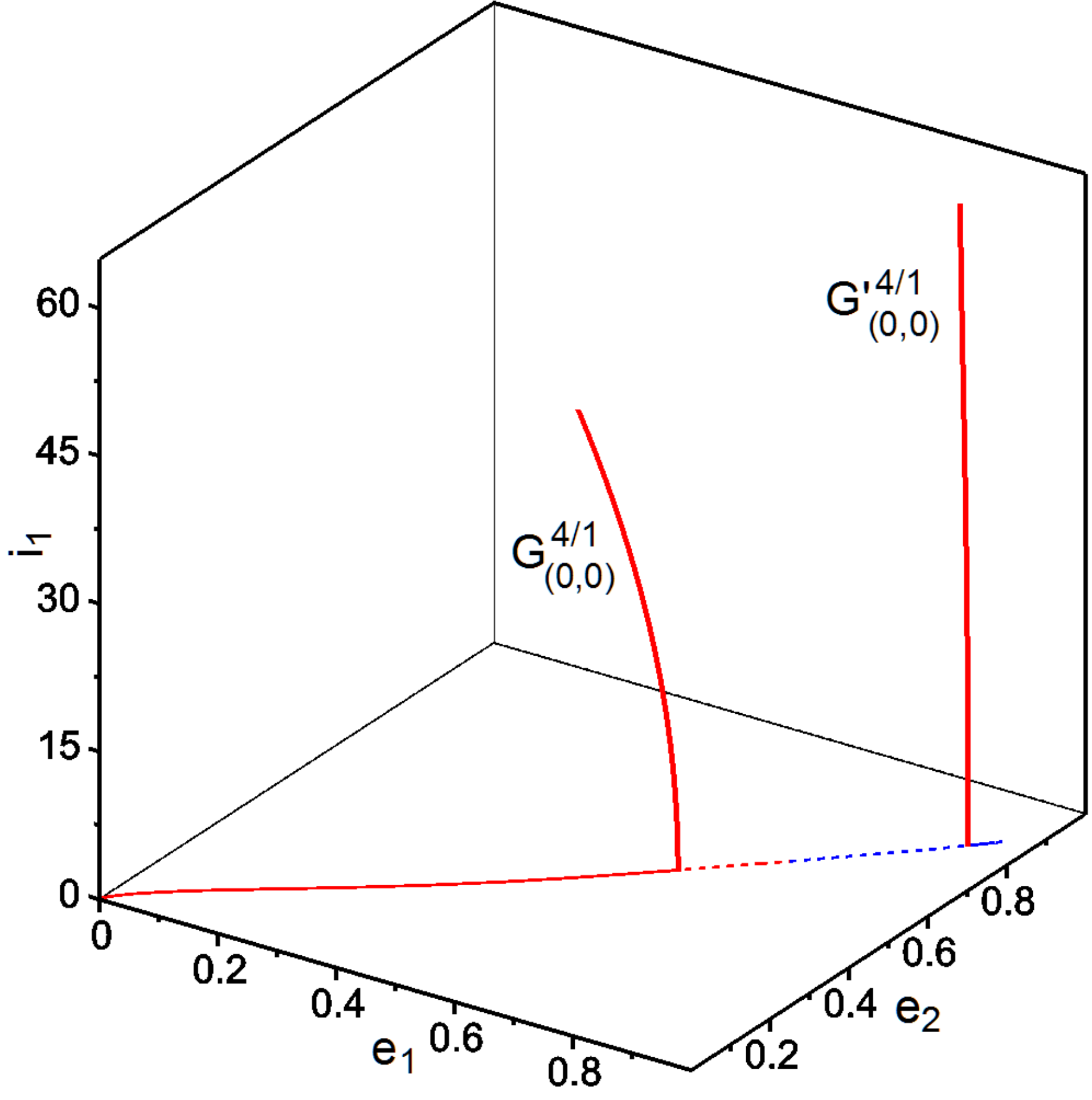}\\ \includegraphics[width=0.8\columnwidth]{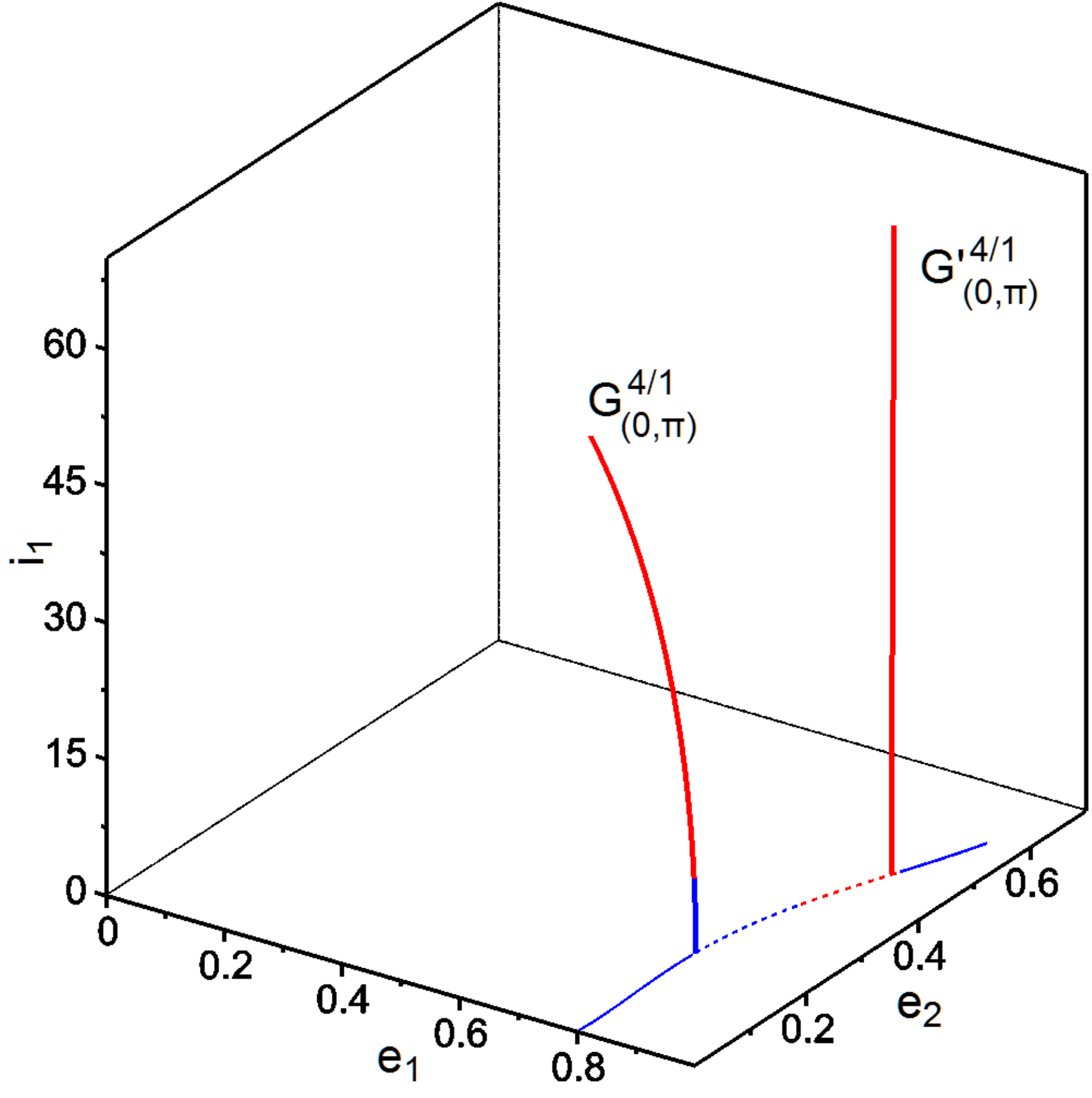}
\end{array} $
\end{center}
\caption{Spatial families in the 3D-ERTBP in the 4/1 MMR along with the bifurcation points of the planar ones. $G^{4/1}_{(0,0)}$ and $G'^{4/1}_{(0,0)}$ bifurcate from the v.c.o. at $(e_1,e_2)=(0.62,0.52)$ and $(e_1,e_2)=(0.92,0.82)$, respectively, of the configuration $(0,0)$. $G^{4/1}_{(0,\pi)}$ and $G'^{4/1}_{(0,\pi)}$ bifurcate from the v.c.o. at $(e_1,e_2)=(0.79,0.21)$ and $(e_1,e_2)=(0.88,0.47)$, respectively, of the configuration $(0,\pi)$.}
\label{41e3d}
\end{figure}

In Fig. \ref{41e3d}, we present the spatial families of $x$-symmetric periodic orbits in the 3D-ERTBP of the 4/1 MMR. Their linear stability is qualitatively similar to the one observed in 3/1 MMR (Fig. \ref{31e3d}). Particularly, the families $G^{4/1}_{(0,0)}$ and $G'^{4/1}_{(0,0)}$ are totally unstable, the family $G^{4/1}_{(0,\pi)}$ is stable up to an inclination value $i_1=8^{\circ}$, and the family $G'^{4/1}_{(0,\pi)}$ is unstable.

\subsection{5/1 MMR}

\begin{figure}
\begin{center}
$\begin{array}{c}
\includegraphics[width=0.9\columnwidth]{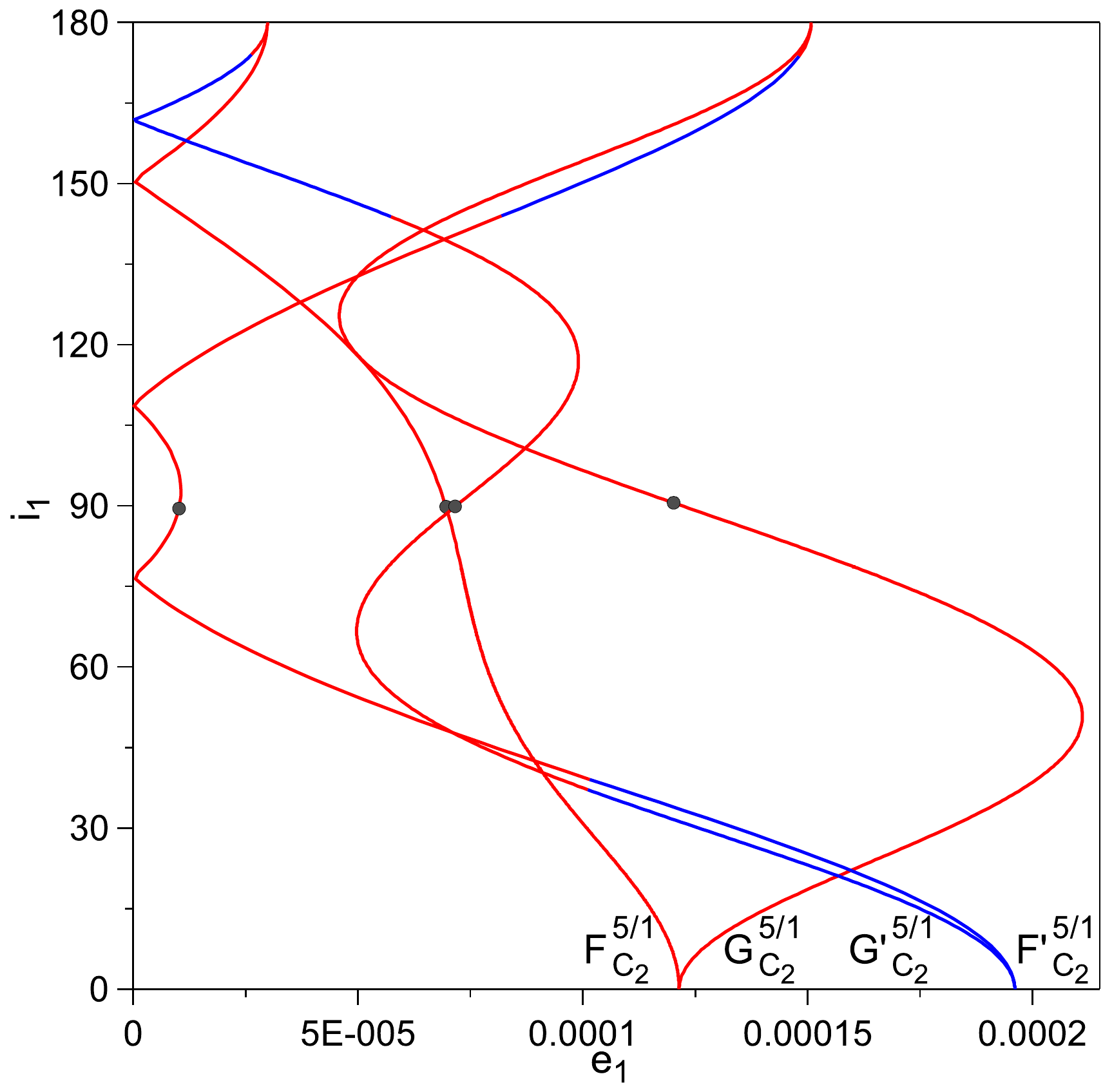}
\end{array} $
\end{center}
\caption{The spatial families, $F_{C_2}^{5/1}$, $G_{C_2}^{5/1}$, ${F'}_{C_2}^{5/1}$ and ${G'}_{C_2}^{5/1}$, in the 5/1 MMR bifurcating from the circular family, $C_2$, projected on the plane ($e_1,i_1$). The grey dots at $(e_1,i_1)=(0.0000104,90^{\circ})$, $(e_1,i_1)=(0.0000696,90^{\circ})$, $(e_1,i_1)=(0.0000716,90^{\circ})$ and $(e_1,i_1)=(0.0001225,90^{\circ})$ correspond to bifurcation points from the 3D-CRTBP to the 3D-ERTBP. Stable periodic orbits are blue coloured.}
\label{51circ}
\end{figure}

\subsubsection{3D-CRTBP}
When the multiplicity of the circular periodic orbits becomes equal to 2, we have the first occurrence of v.c.o. at 5/1 MMR along the circular family (see $C_2$, $C_4$, $C_6$ and $C_8$ in Fig. \ref{circmul} and Table \ref{multab}).

In Fig. \ref{51circ}, we present four families, $F_{C_2}^{5/1}$ and $G_{C_2}^{5/1}$ which are whole unstable, and ${F'}_{C_2}^{5/1}$ and ${G'}_{C_2}^{5/1}$ which are stable when $i_1<37^{\circ}$ for prograde orbits and when $144^{\circ}<i_1<174^{\circ}$ for retrograde orbits. All of them terminate at planar retrograde orbits.

When these families were examined with respect to Method II, four bifurcation points (one at each family) to the 3D-ERTBP were revealed (grey dots in Fig. \ref{51circ}) at $(e_1,i_1)=(0.0000104,90^{\circ})$, $(e_1,i_1)=(0.0000696,90^{\circ})$, $(e_1,i_1)=(0.0000716,90^{\circ})$ and $(e_1,i_1)=(0.0001225,90^{\circ})$. The families emanating from these bifurcation points will not be computed here.

\begin{figure}
\begin{center}
$\begin{array}{cc}
%\textnormal{3/2 MMR}  & \textnormal{2/1 MMR} \\
\includegraphics[width=0.9\columnwidth,keepaspectratio]{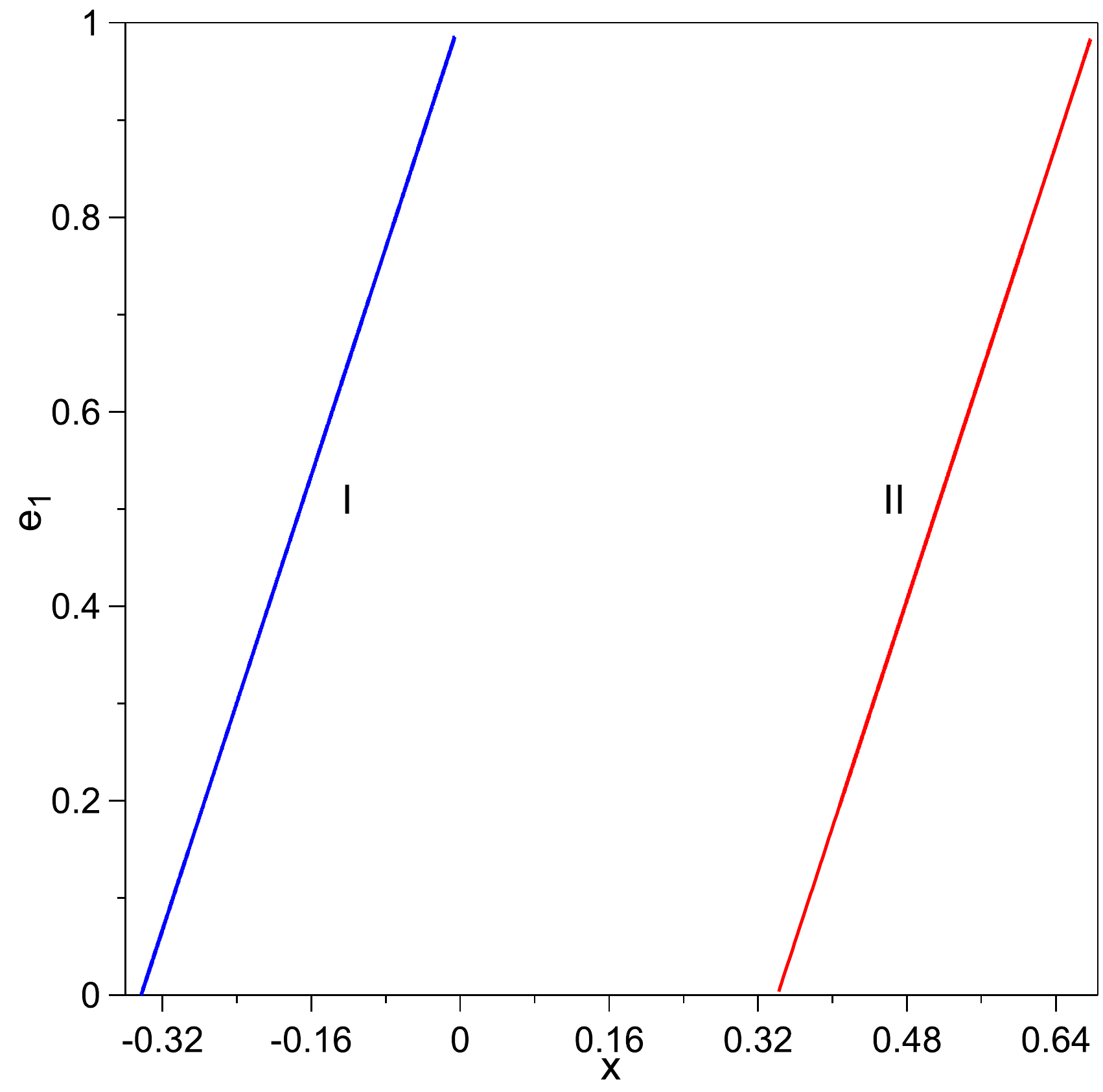} 
\end{array} $
\end{center}
\caption{Families of periodic orbits in the 2D-CRTBP for  5/1 MMR examined with respect to Method I. Colours and lines as in Fig. \ref{32cvco}.
}
\label{51cvco}
\end{figure}

In Fig. \ref{51cvco}, we examine the  families of the 5/1 MMR in the 2D-CRTBP with respect to Method I. Likewise 4/1 MMR, no v.c.o. exists along both of the families $I$ (horizontally stable) and $II$ (horizontally unstable), which are vertically stable (solid line). 

\subsubsection{3D-ERTBP}
\begin{figure}
\begin{center}
$\begin{array}{cc}%\textnormal{3/2 MMR}  & \textnormal{2/1 MMR} \\
\includegraphics[width=0.9\columnwidth,keepaspectratio]{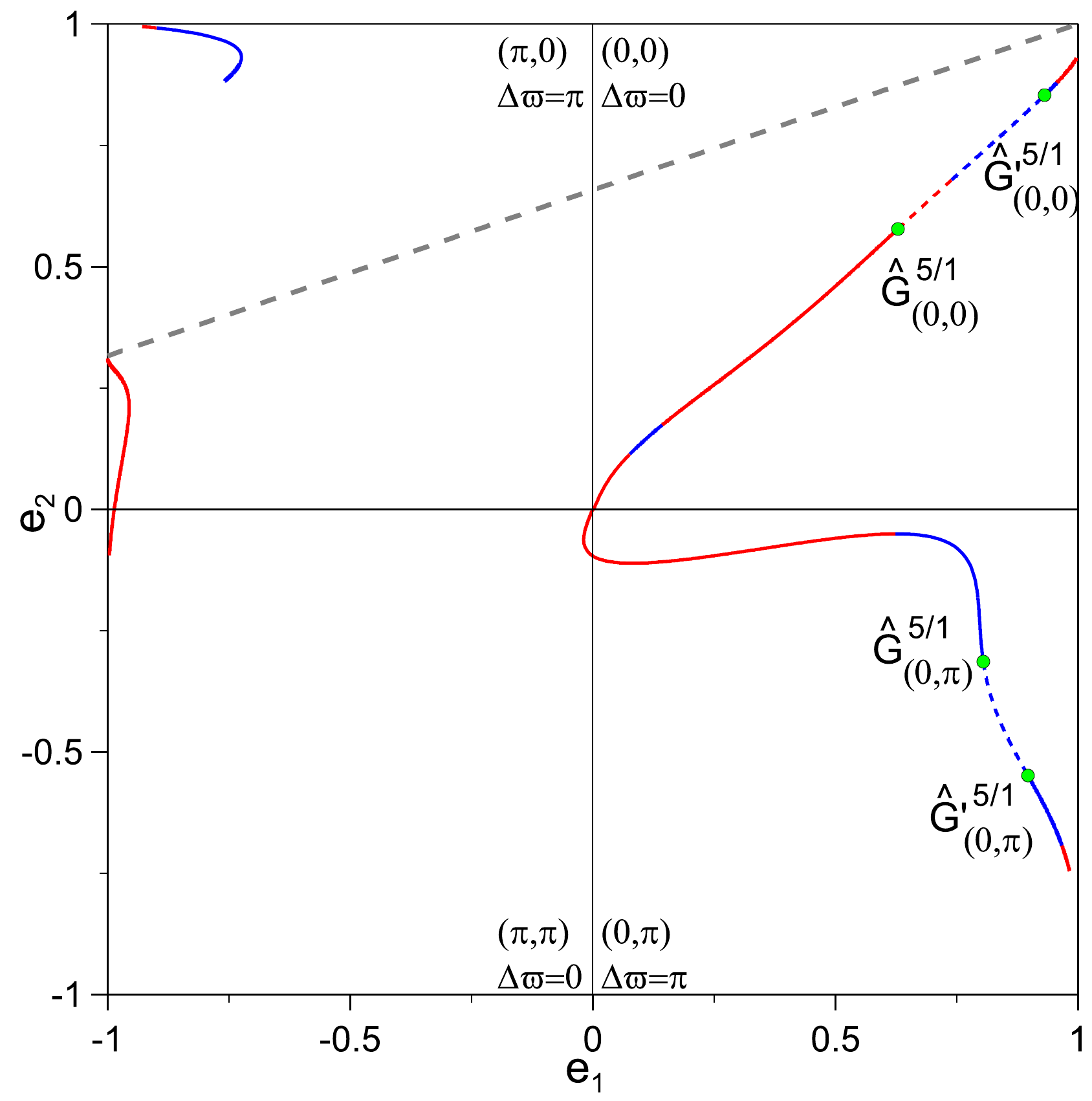} 
\end{array} $
\end{center}
\caption{Families of periodic orbits in the 2D-ERTBP for 5/1 MMR examined with respect to Method I and presented as in Fig. \ref{32evco}.}
\label{51evco}
\end{figure}

In Fig. \ref{51evco}, we examine the families of the 5/1 MMR in the 2D-ERTBP with respect to Method I. The results are qualitatively similar to the ones of 3/1 and 4/1 MMRs in the 2D-ERTBP. Particularly, all the families are vertically stable (solid line), except two segments bounded by two v.c.o. in each of the configurations $(0,0)$ and $(0,\pi)$ (dashed line). In the configuration $(0,0)$, the v.c.o. $\hat{G}^{5/1}_{(0,0)}$ and $\hat{G}'^{5/1}_{(0,0)}$ are located at $(e_1,e_2)=(0.62,0.57)$ and $(e_1,e_2)=(0.93,0.85)$, respectively, whereas the v.c.o. $\hat{G}^{5/1}_{(0,\pi)}$ and $\hat{G}'^{5/1}_{(0,\pi)}$ of the configuration $(0,\pi)$ are located at $(e_1,e_2)=(0.80,0.31)$ and $(e_1,e_2)=(0.89,0.54)$, respectively.

\begin{figure}
\begin{center}
$\begin{array}{c}
\includegraphics[width=0.8\columnwidth]{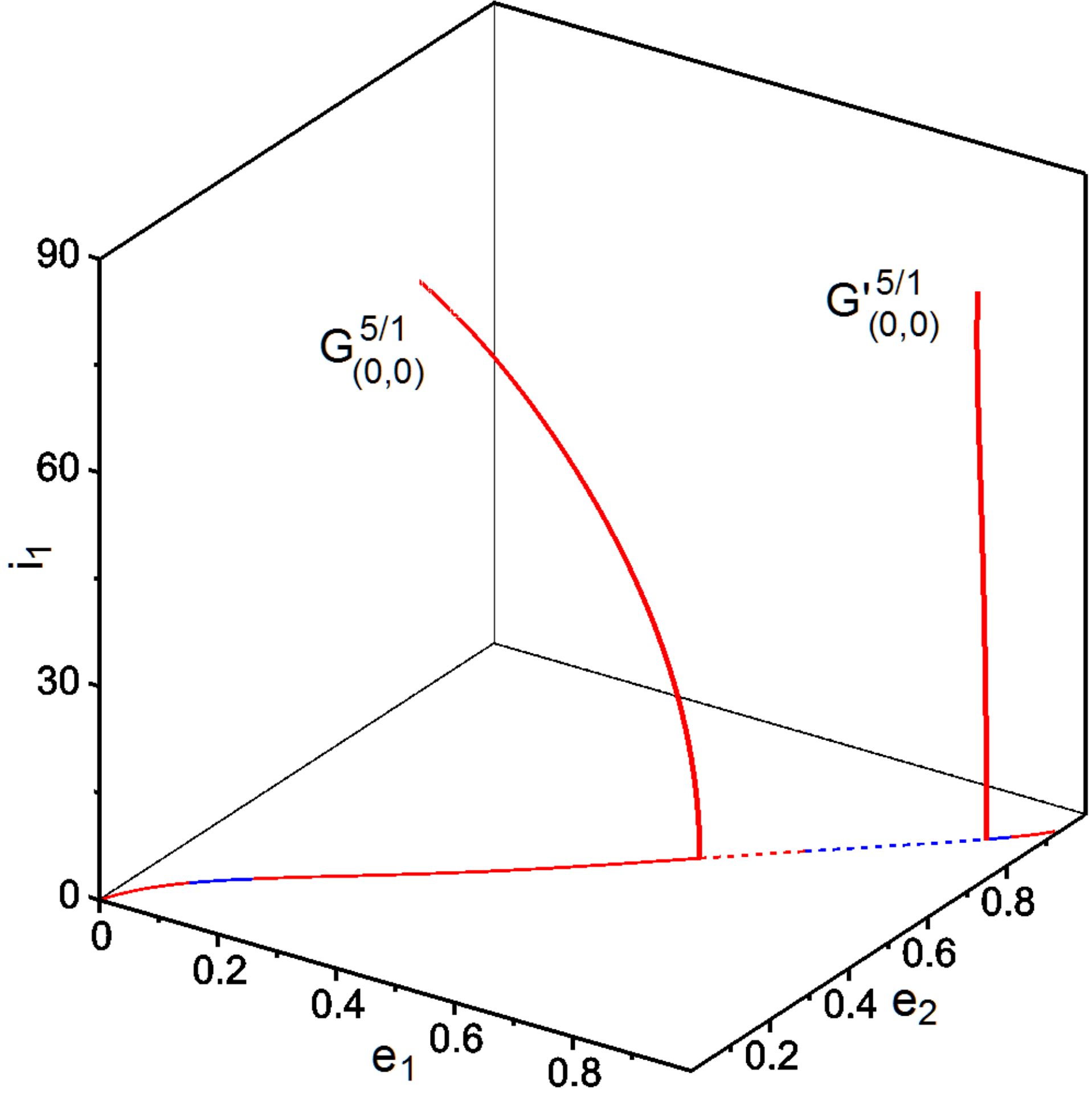}\\\includegraphics[width=0.8\columnwidth]{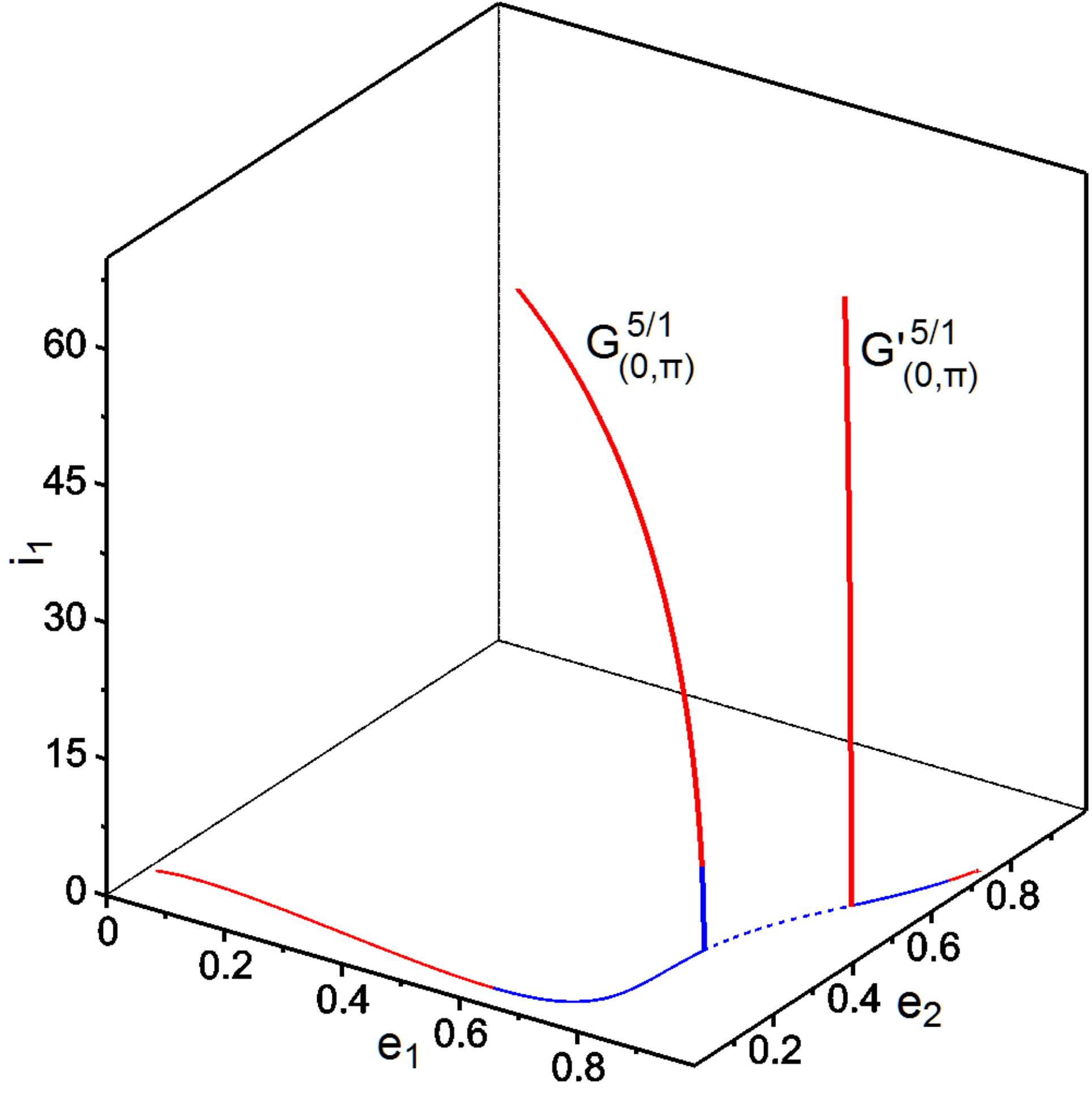}\\
\end{array} $
\end{center}
\caption{Spatial families of periodic orbits in the 3D-ERTBP in the 5/1 MMR generated by the planar ones. $G^{5/1}_{(0,0)}$ and $G'^{5/1}_{(0,0)}$ bifurcate from the v.c.o. at $(e_1,e_2)=(0.62,0.57)$ and $(e_1,e_2)=(0.93,0.85)$, respectively, of the configuration $(0,0)$. $G^{5/1}_{(0,\pi)}$ and $G'^{5/1}_{(0,\pi)}$ bifurcate from the v.c.o. at $(e_1,e_2)=(0.80,0.31)$ and $(e_1,e_2)=(0.89,0.54)$, respectively, of the configuration $(0,\pi)$.}
\label{51e3d}
\end{figure}

In Fig. \ref{51e3d}, we present the spatial families of $x$-symmetric periodic orbits in the 3D-ERTBP of the 5/1 MMR. Their linear stability is qualitatively similar to the one observed in 3/1 and 4/1 MMRs (Fig. \ref{31e3d} and Fig. \ref{41e3d}). Particularly, the families $G^{5/1}_{(0,0)}$ and $G'^{5/1}_{(0,0)}$ are totally unstable, the family $G^{5/1}_{(0,\pi)}$ is stable up to an inclination value $i_1=9^{\circ}$, and the family $G'^{5/1}_{(0,\pi)}$ is unstable.

\section{Possible existence of inclined resonant configurations}\label{appi}
\subsection{Planetary dynamics}\label{app}
\citet{kiaasl} performed an exhaustive study to identify the planar dynamical neighbourhoods that could host an inner massless body (called terrestrial planet in that work) in MMR or not with an outer giant planet. The 2/1 MMR was discussed therein, while the 3/2, 5/2, 3/1, 4/1 and 5/1 MMRs were accordingly discussed in \citet{spis}. We presently illustrate the spatial case, where the inner massless body is inclined, and provide clues of potential coexistence and survival. We use DS-maps, in order to visualise the phase space of spatial configurations in the neighbourhood of horizontally and vertically stable planar elliptic periodic orbits. We also showcase the dynamical vicinity of spatial circular periodic orbits. We should note that along the families of the ERTBP the mean-motion ratio is maintained constant, while the semi-major axes vary slightly. On the DS-maps the semi-major axes are constant, equal to value of the periodic orbit we select each time.

\begin{figure}
\begin{center}
\includegraphics[width=0.85\columnwidth]{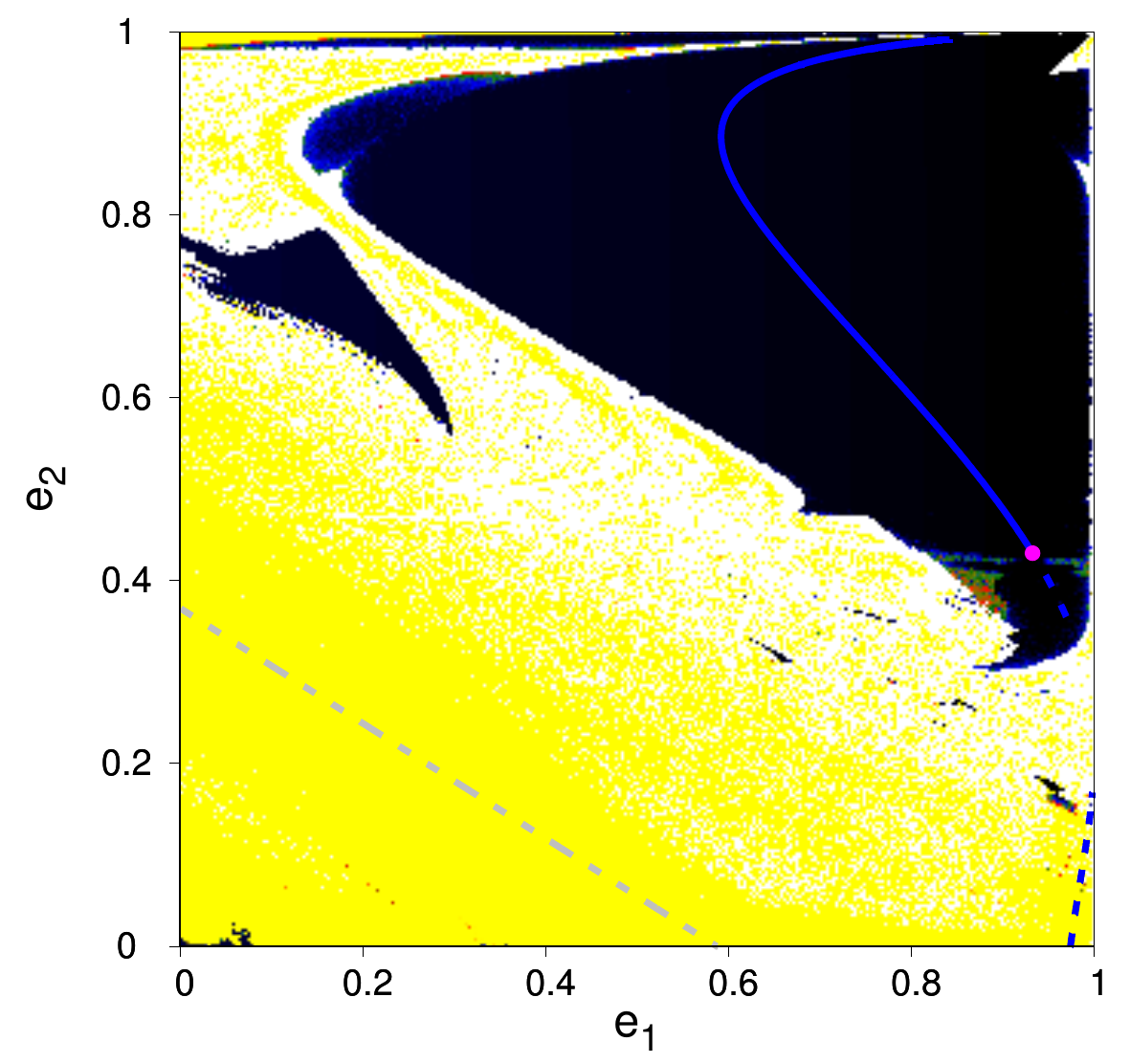}\\ 
\includegraphics[width=0.85\columnwidth]{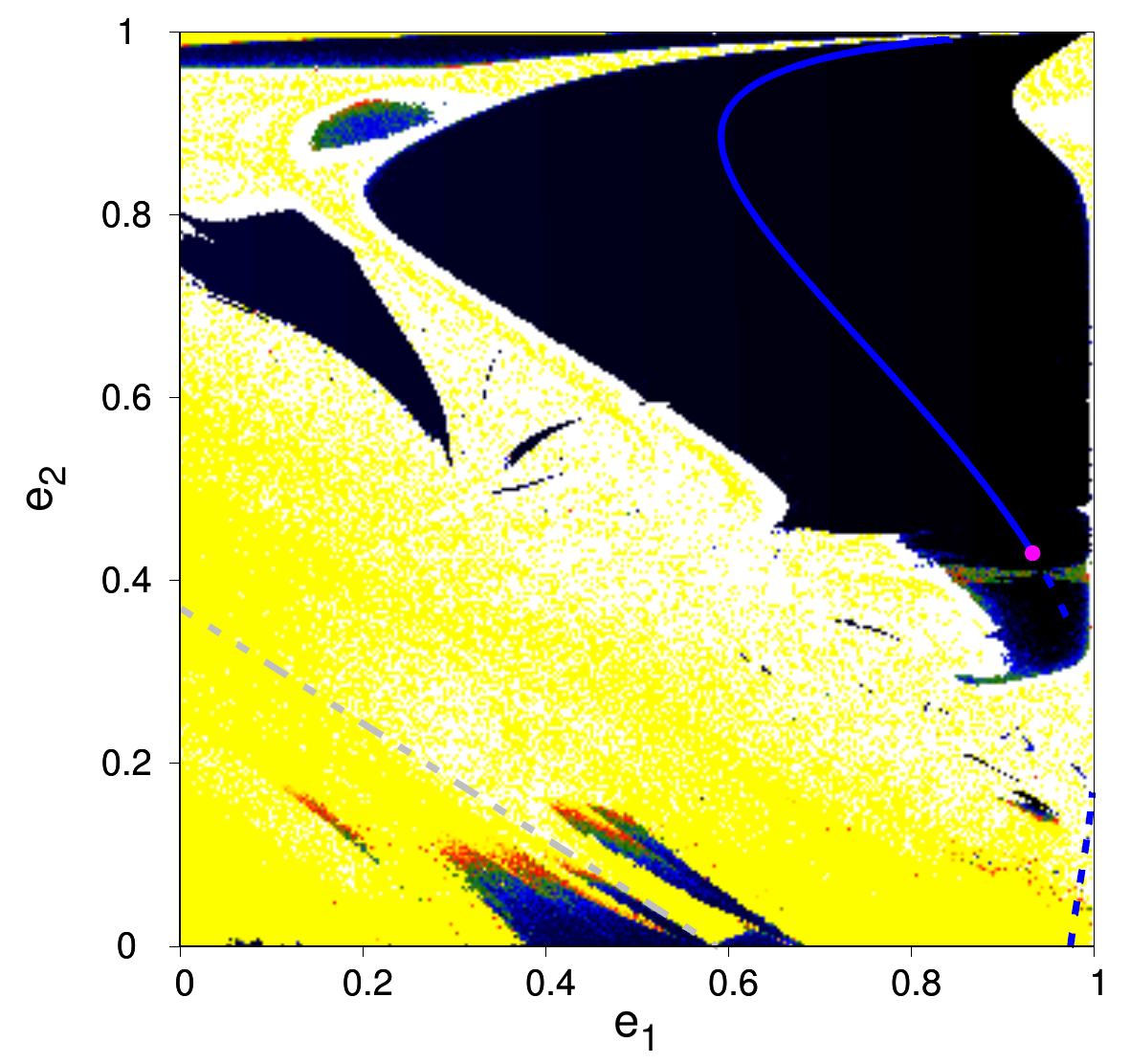}\\ 
\includegraphics[width=0.85\columnwidth]{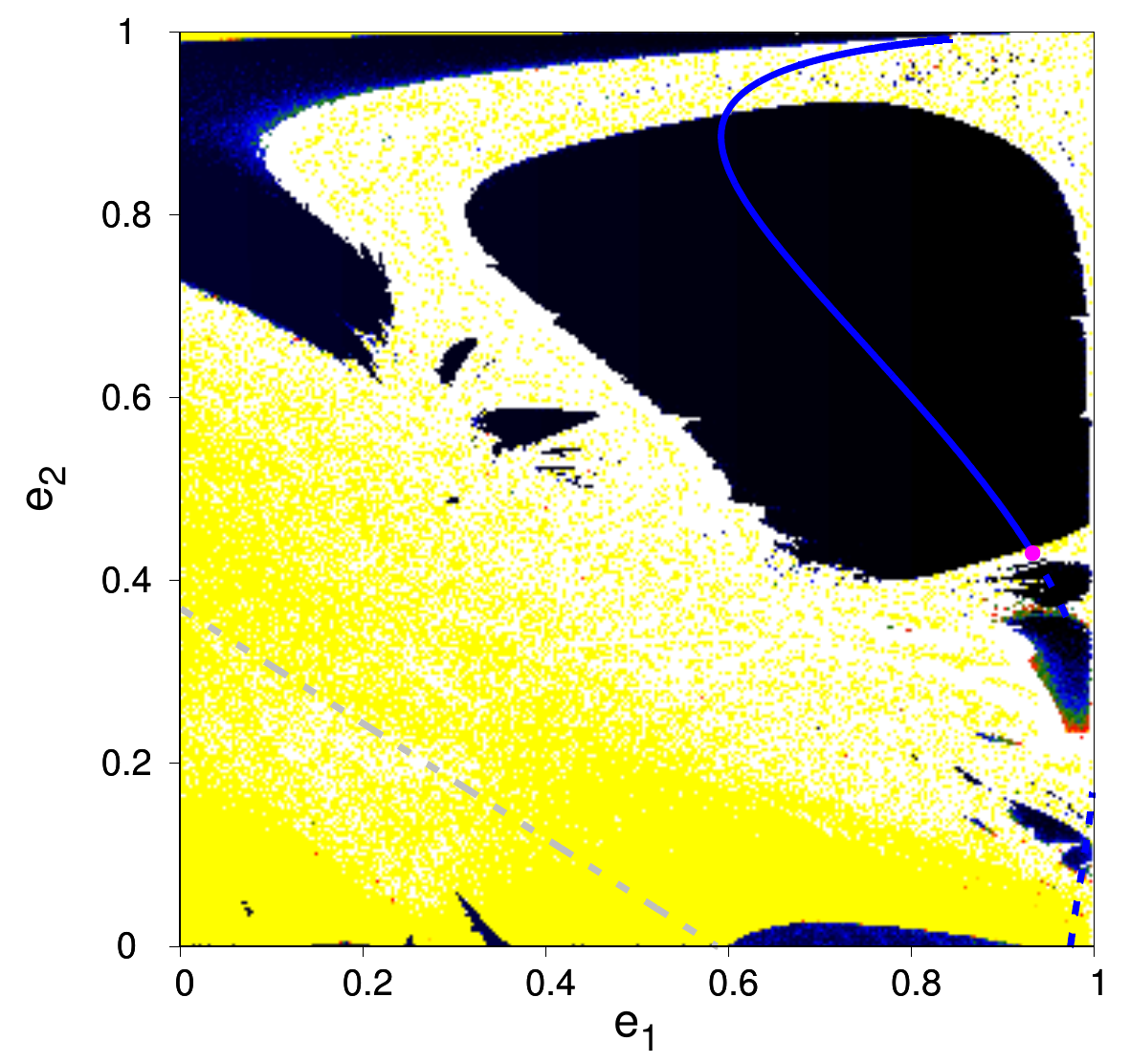}\\ 
\hspace{1cm}\includegraphics[width=.65\columnwidth]{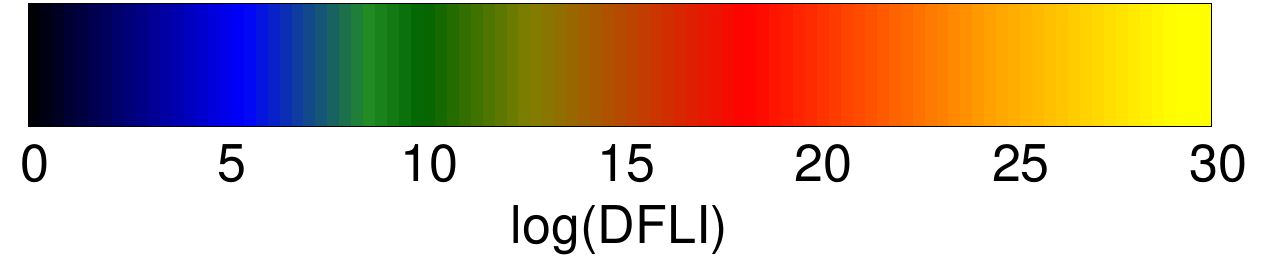}\\
\end{center}
\caption{DS-maps on the plane $(e_1,e_2)$ of the configuration $(\pi,0)$ of the 2/1 MMR in the 2D-ERTBP (Fig. \ref{21evco}) with $i_1=20^{\circ}$ (top), $i_1=30^{\circ}$ (middle) and $i_1=50^{\circ}$ (bottom). The orbital elements that remained constant for the computation are $a_2/a_1=0.6312$, $\omega_1=\Omega_1=270^{\circ}$, $M_1=180^{\circ}$, $i_2=0^{\circ}$, $\omega_2=270^{\circ}$, $\Omega_2=90^{\circ}$ and $M_2=0^{\circ}$.}
\label{21ds}
\end{figure}

In Fig. \ref{21ds}, we study the planar family of the configuration $(\pi,0)$ of the 2/1 MMR in the 2D-ERTBP (Fig. \ref{21evco}) with regards to the horizontal and vertical stability. Particularly, we used a mutual inclination of $20^{\circ}$, $30^{\circ}$ and $50^{\circ}$, in order to showcase the extent of such a region of concurrent horizontal and vertical stability. The main region of stability around the horizontally and vertically stable periodic orbits spans a very broad domain even when $i_1=50^{\circ}$. However, we observe that at the v.c.o. (magenta dot) the region of stability breaks and afterwards, it is populated by chaotic orbits, which become strongly chaotic as the inclination increases. 

\begin{figure}
\begin{center}
$\begin{array}{c}
\includegraphics[width=0.85\columnwidth]{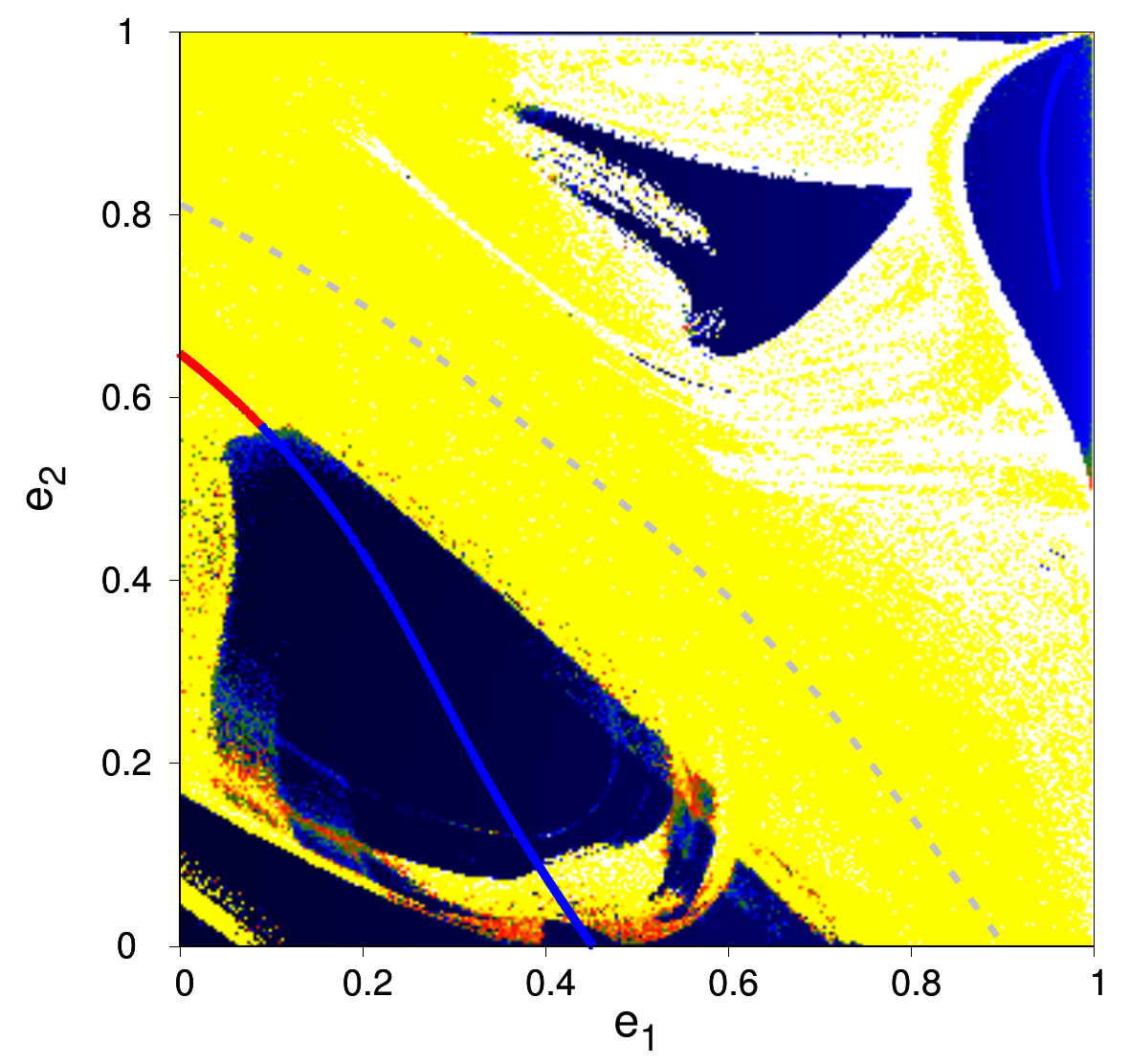}\\ 
\includegraphics[width=0.85\columnwidth]{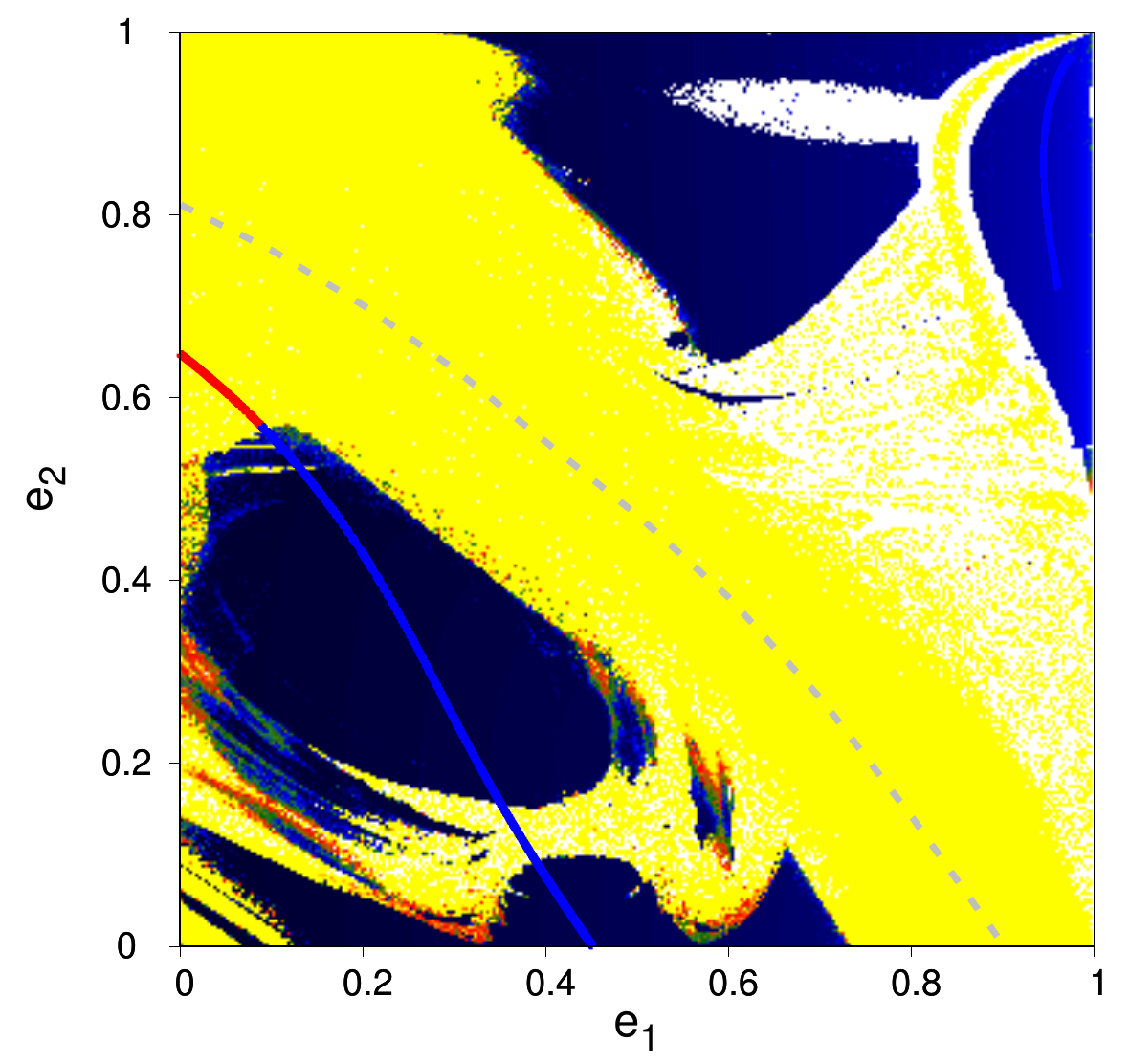}\\
\includegraphics[width=0.85\columnwidth]{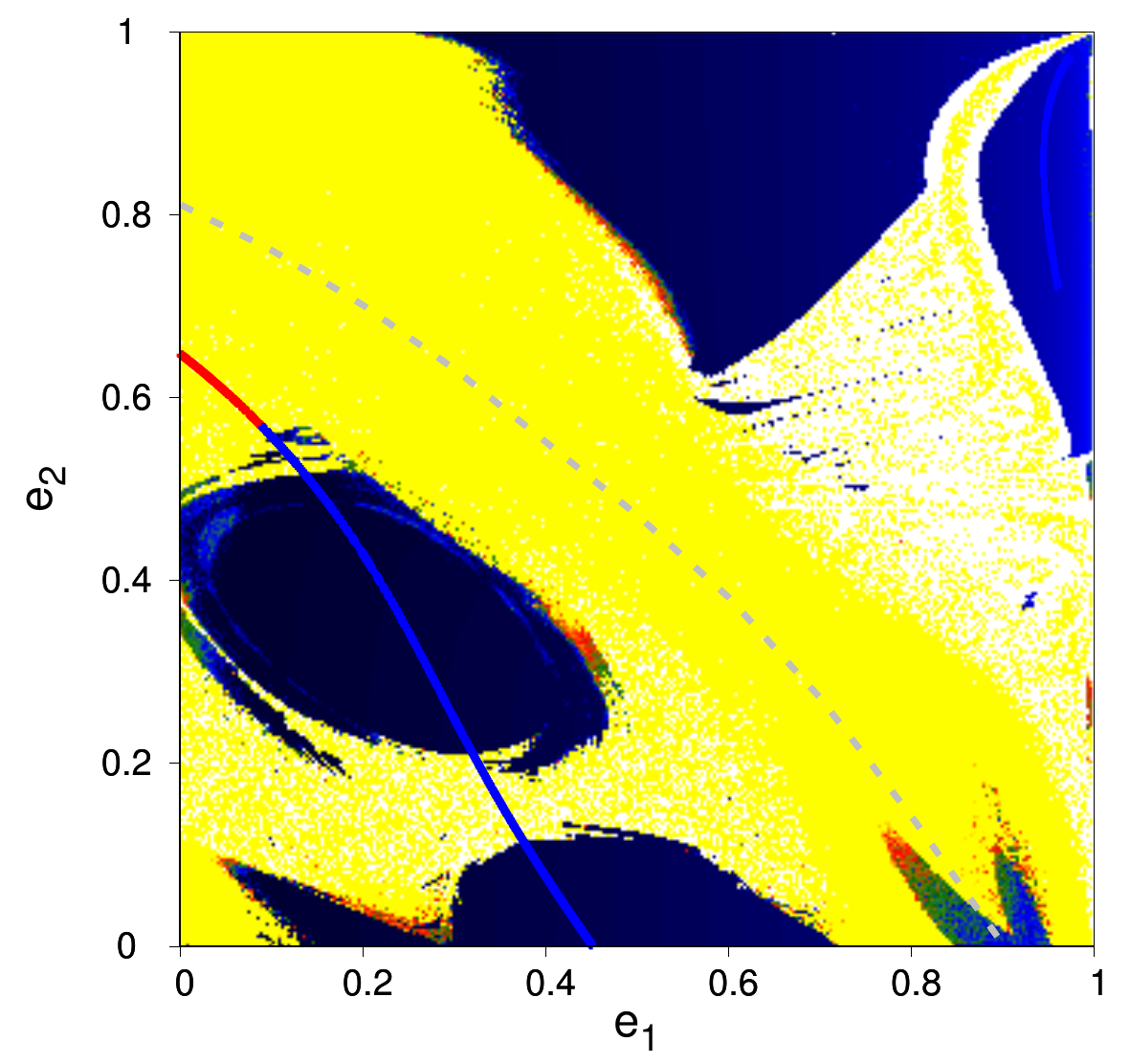}\\
\end{array} $
\end{center}
\caption{DS-maps on the plane $(e_1,e_2)$ of the configuration $(0,\pi)$ of the 3/2 MMR in the 2D-ERTBP (Fig. \ref{32evco}) with $i_1=10^{\circ}$ (top), $i_1=20^{\circ}$ (middle) and $i_1=30^{\circ}$ (bottom). The orbital elements that remained constant for the computation are $a_2/a_1=0.7595$, $\omega_1=90^{\circ}$, $\Omega_1=270^{\circ}$, $M_1=0^{\circ}$, $i_2=0^{\circ}$, $\omega_2=90^{\circ}$, $\Omega_2=90^{\circ}$ and $M_2=180^{\circ}$.}
\label{32ds}
\end{figure}

Accordingly, in Fig. \ref{32ds}, we study the planar family of the configuration $(0,\pi)$ of the 3/2 MMR in the 2D-ERTBP existing for $e_1<0.5$ (Fig. \ref{32evco}) with regards to the horizontal and vertical stability and used a mutual inclination of $10^{\circ}$, $20^{\circ}$ and $30^{\circ}$. We observe that the region of stability extends up to the point where the family becomes horizontally unstable. Moreover, we observe another region of regular motion around the other horizontally and vertically stable family of this configuration at high values of both eccentricities. This domain is maintained even when $i_1=50^{\circ}$.

\begin{figure}
\begin{center}
$\begin{array}{c}
\includegraphics[width=0.85\columnwidth]{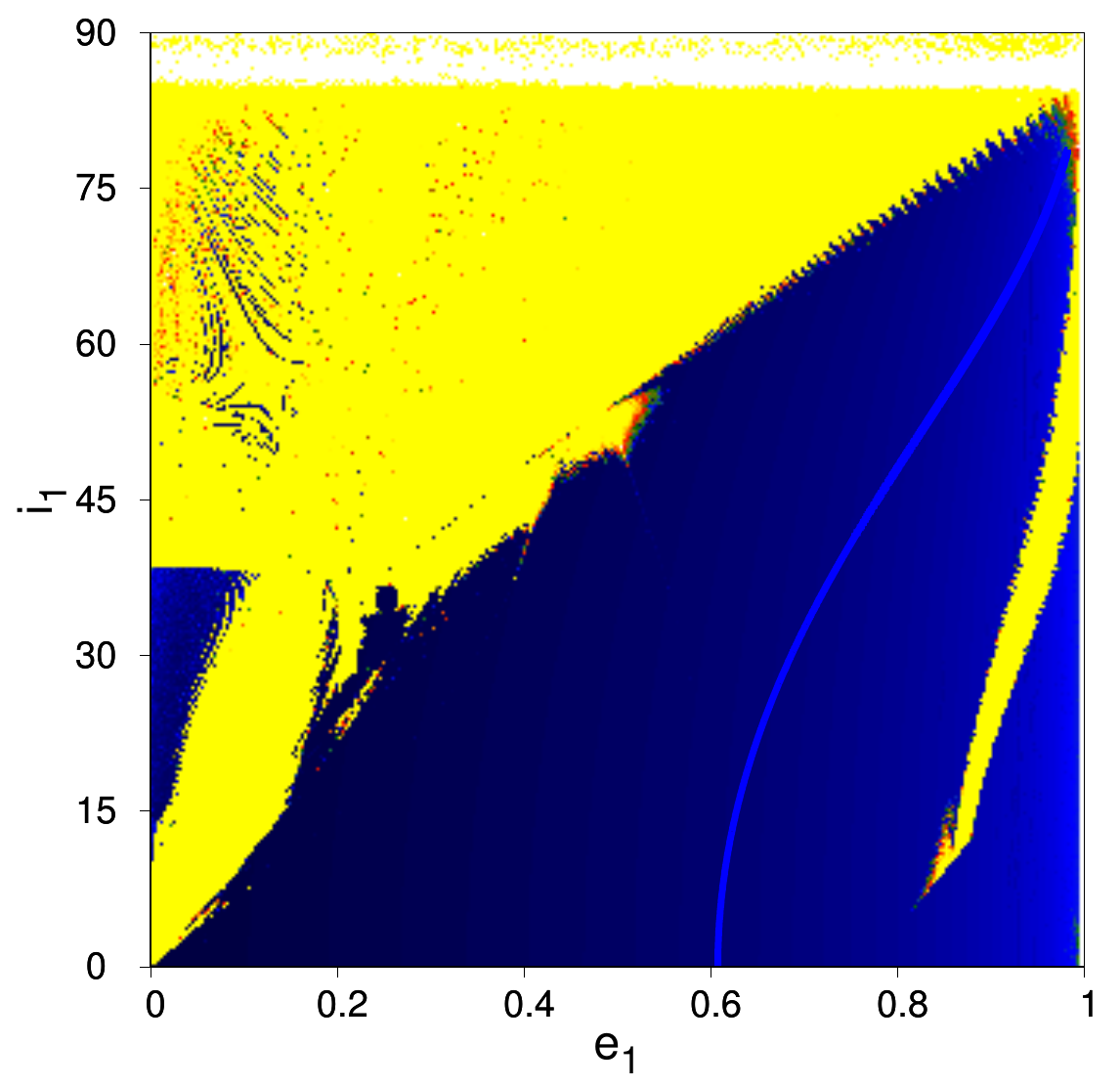}\\
\end{array} $
\end{center}
\caption{DS-map around the family $F^{3/1}_I$ (Fig. \ref{31c3d}) of the 3/1 MMR in the 3D-CRTBP on the plane $(e_1,i_1)$. The orbital elements of the $xz$-symmetric stable periodic orbit that remained constant for the computation are $a_2/a_1=0.4806$, $\omega_1=\Omega_1=90^{\circ}$, $M_1=0^{\circ}$ and $e_2=i_2=\omega_2=\Omega_2=M_2=0^{\circ}$.}
\label{31ds}
\end{figure}

In Fig. \ref{31ds}, we performed an exploration in the neighbourhood of the spatial family $F^{3/1}_I$ of the 3D-CRTBP (Fig. \ref{31c3d}), in order to exhibit the extent of the regular domain along the family. Within this region the inclination resonance takes place, and the inclination resonant angles librate as illustrated in Fig. \ref{rang}b. We observe that the region of stability is particularly broad even when the inner massless body becomes highly eccentric and reaches $80^{\circ}$.
 
To sum up, we have illustrated two cases that can guarantee the long-term stability of spatial resonant configurations in the restricted three-body problem. The systems should either reside in the neighbourhood of both horizontally and vertically stable planar periodic orbits, or in the vicinity of linearly stable spatial periodic orbits. Interestingly, very high mutual inclinations of the massless body were manifested in these regions.

\subsection{Asteroid dynamics}\label{aster}
The present work can also be applied to asteroid dynamics and space mission designs \citep[see e.g.][]{per,rick}. For instance, it gives an explanation to many resonance captures of massless particles (asteroids) at arbitrary inclination observed in \citet{mn15,mn18}. In particular, when the giant planet is on circular orbit, captures in 3/2 MMR can be explained by Fig. \ref{32c3d}, in 2/1 MMR by Fig. \ref{21c3d}, in 5/2 MMR by Fig. \ref{52circ}, in 3/1 MMR by Figs. \ref{31circ} and \ref{31c3d} and in 4/1 MMR by Fig. \ref{41circ}. However, some features cannot, for many reasons. Firstly, additional spatial families can exist, such as spatial isolated families of symmetric periodic orbits \citep[see e.g.][for an illustration of spatial isolated families in the 3D-GTBP in Fig. 22]{av12} or families of spatial asymmetric periodic orbits. Secondly, these features could also be artefacts of the numerical simulations, as resonance capture is dependent on the initial location of the migrating body, the drift chosen for the migration and the integration time (see for instance the parametric study of resonance capture of \citet{litsi09b}). Also, the features reported could consist of temporal captures, whereas the proximity to periodic orbits assures long-term stability. Thirdly, there exist additional mechanisms that ensure the stability of planetary systems without the existence of stable periodic orbits in their dynamical vicinity, namely secondary resonance, apsidal resonance, nodal resonance, Kozai resonance and apsidal difference oscillation.

\section{Conclusions}\label{con}
Our study focused on spatial planetary systems consisting of a star, an inclined inner terrestrial (massless) planet and an outer giant (massive) planet. However, our results can be applied to any celestial architectures, which can be modelled by the 3D-CRTBP or the 3D-ERTBP, in particular, to asteroid dynamics, as shown in Sect. \ref{aster}. 
More specifically, we examined the circular family computed for various multiplicity values of the circular periodic orbits and the planar families of elliptic periodic orbits of the 2D-CRTBP and the 2D-ERTBP for the 3/2, 2/1, 5/2, 3/1, 4/1 and 5/1 interior MMRs with respect to the vertical stability. We identified the vertical critical orbits that exist in each case for each MMR. Starting from the v.c.o. of the circular family we generated and computed spatial families of symmetric periodic orbits. Accordingly, beginning from the v.c.o. of the 2D-CRTBP and the 2D-ERTBP we derived spatial families of symmetric periodic orbits in the 3D-CRTBP and the 3D-ERTBP. Moreover, we identified the bifurcation points of the 3D-CRTBP to the 3D-ERTBP and computed the families of prograde spatial periodic orbits.

Whether interested in locating potential exo-Earths, or generally massless inner bodies locked in MMR with outer massive ones, the spatial families of stable periodic orbits provide the regions of stability in phase space, where inclined planetary configurations can be hosted and survive for long-time spans.  Additionally, the neighbourhood of horizontally and vertically stable planar periodic orbits can provide such spatial regular domains, too. The only difference being the fact that inclination resonance takes place only in the former case, where the evolution is about a spatial periodic orbit.

Our results regarding the stability of the spatial families can be summarised as follows.\\
In the 3D-CRTBP: 
\begin{itemize}
\item For the direct or prograde orbits, the stable periodic orbits exist in the 3/2 MMR when $i_1<55^{\circ}$, in the 2/1 MMR for $i_1<90^{\circ}$, in the 5/2 MMR for $i_1<40^{\circ}$, in the 3/1 MMR for $i_1<79^{\circ}$ and when $56^{\circ}<i_1<82^{\circ}$ (the different ranges correspond to different families), in the 4/1 MMR for $i_1<45^{\circ}$ and in the 5/1 MMR for $i_1<37^{\circ}$. 
\item For the retrograde orbits, stability exists in the 3/2 MMR when $103^{\circ}<i_1<165^{\circ}$ and when $162^{\circ}<i_1<178^{\circ}$ (the different ranges correspond to different symmetries), in the 2/1 MMR for $95^{\circ}<i_1<180^{\circ}$, in the 5/2 MMR for $109^{\circ}<i_1<164^{\circ}$, in the 3/1 MMR for $104^{\circ}<i_1<138^{\circ}$ and when $i_1>142^{\circ}$ (the different ranges correspond to different families), in the 4/1 MMR for $108^{\circ}<i_1<163^{\circ}$ and in the 5/1 MMR for $144^{\circ}<i_1<174^{\circ}$.
\end{itemize}
In the 3D-ERTBP, where we only continued prograde orbits: 
\begin{itemize}
\item Starting from the v.c.o. of the 2D-ERTBP, we found that in the 2/1 MMR the family of periodic orbits in the configuration $(\pi,0)$ is whole stable and reaches $70^{\circ}$. Additionally, segments of stability (up to $\sim10^{\circ}$) exist for the 3/1, 4/1 and 5/1 MMRs in the configuration $(0,\pi)$.
\item Starting from a bifurcation point of the 3D-CRTBP, we found that all of the families were unstable.
\end{itemize}
A straightforward deduction could be that all the MMRs studied are more likely to host an inner inclined terrestrial planet in their dynamical neighbourhood as long as the orbit of the outer giant planet is circular. For an elliptic orbit of the giant planet, the 2/1 MMR can host an inclined terrestrial planet when the giant is highly eccentric, while for the 3/1, 4/1 and 5/1 MMRs in the 3D-ERTBP, the inclined massless body should have a low inclination, in order to survive.

However, there are additional suitable stable domains that can host the system of interest, when the giant's orbit is elliptic. This can be realised via the notion of the horizontal stability exhibited together with the vertical stability.  As shown through the DS-maps, the planar families of periodic orbits of the 2D-ERTBP, which are both horizontally and vertically stable (solid blue lines), can provide regular domains for inclinations up to high values. Same holds for the horizontally and vertically stable periodic orbits of the 2D-CRTBP.

Following the Methods I and II, we computed spatial families of symmetric periodic orbits. As changes of stability along the families were observed, it will be interesting to explore whether these points would generate asymmetric or symmetric (of double period) periodic orbits in the future.

In this work, we  examined the stable inclined configurations for the main MMRs, when the giant's orbit is circular (3D-CRTBP) and when the giant's orbit is elliptic (3D-ERTBP). We based our results on the computation of the spatial families of symmetric periodic orbits bifurcating from the circular family, the 2D-CRTBP, the 3D-CRTBP and the 2D-ERTBP and the exploration of the stable domains around the planar families of horizontally and vertically stable periodic orbits. This study is of interest to exo-Earths hunters, since potential inclined, inner, terrestrial planets in MMR with a detected giant planet can be dynamically constrained within the regular domains provided by these families. Finally, the results can also contribute to the field of the dynamics of high inclination retrograde orbits in asteroid dynamics and space mission designs. 

\section*{Acknowledgements}
Computational resources have been provided by the Consortium des \'Equipements de Calcul Intensif (C\'ECI), funded by the Fonds de la Recherche Scientifique de Belgique (F.R.S.-FNRS) under Grant No. 2.5020.11.

%%%%%%%%%%%%%%%%%%%%%%%%%%%%%%%%%%%%%%%%%%%%%%%%%%



%\bibliographystyle{mnras}
%\bibliography{nbib}

% Don't change these lines
\bsp	% typesetting comment
\label{lastpage}

%\begin{landscape}
\begin{appendix}
\section{The circular family} \begin{landscape}
\label{appendix}
\begin{figure}
\begin{center}
$\begin{array}{c}
\includegraphics[height=13cm,keepaspectratio]{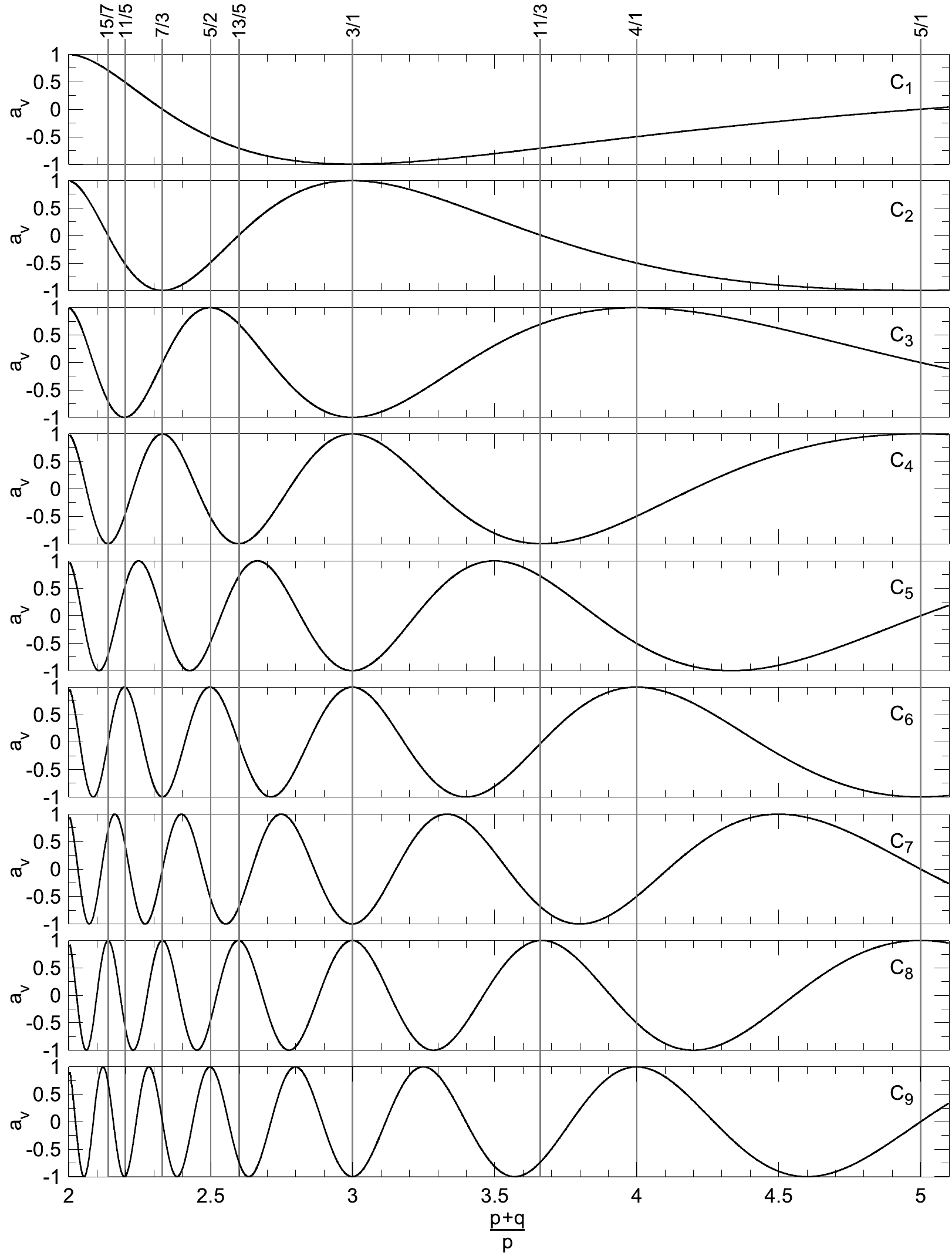}\vspace{-0.5cm}\\
\end{array} $
\end{center}
\caption{The vertical stability index, $a_v$, is computed along the circular family, $C_i$,  as the multiplicity, $i$, of the periodic orbits increases. The appearance of v.c.o., when $a_v=\pm 1$, at specific MMRs is showcased.}\vspace{-0.5cm}
\label{circmul}
\end{figure}
%\end{landscape}\begin{landscape}
\begin{table}
\begin{center}
\caption{The systematic emergence of MMRs possessing a v.c.o. as the multiplicity, $i$, of the periodic orbits of the circular family increases.}
\label{multab}
\resizebox{1.35\textwidth}{!}{
\begin{tabular}[b]{@{}cc@{}c@{}c@{}c@{}c@{}ccccc@{}c@{}c@{}ccc@{}c@{}c@{}ccccc@{}c@{}c@{}c@{}c@{}ccc@{}c@{}c@{}c@{}c@{}ccccc@{}c@{}c@{}cc@{}}
\hline%{@{}c@{}}
\small
$i$ & \multicolumn{42}{c}{MMR}\\
\hline
%\hline
1&&&&&&&&&&&&&&&&&&&&&&&&&&&&3/1&&&&&&&&&&&&&&\\
%\hline
2&&&&&&&&&&&&&&7/3&&&&&&&&&&&&&&3/1&&&&&&&&&&&&&&5/1\\
%\hline
3&&&&&&&&&11/5&&&&&&&&&&5/2&&&&&&&&&3/1&&&&&&&&&4/1&&&&&\\
%\hline
4&&&&&&&15/7&&&&&&&7/3&&&&&&&13/5&&&&&&&3/1&&&&&&&11/3&&&&&&&5/1\\
5&&&&&19/9&&&&&&9/4&&&&&&17/7&&&&&&8/3&&&&&3/1&&&&&7/2&&&&&&13/3&&&\\
%\hline
6&&&&23/11&&&&&11/5&&&&&7/3&&&&&5/2&&&&&19/7&&&&3/1&&&&17/5&&&&&4/1&&&&&5/1\\
7&&&27/13&&&&&13/6&&&&25/11&&&&12/5&&&&23/9&&&&&11/4&&&3/1&&&10/3&&&&&19/5&&&&9/2&&\\
8&&31/15&&&&&15/7&&&29/13&&&&7/3&&&&27/11&&&13/5&&&&&25/9&&3/1&&23/7&&&&&11/3&&&21/5&&&&5/1\\
%\hline
9&35/17&&&&&17/8&&&11/5&&&&16/7&&31/13&&&&5/2&&&29/11&&&&&14/5&3/1&13/4&&&&&25/7&&&4/1&&&&23/5&\\
\hline
\end{tabular}
}
	\end{center}
 \end{table}\end{landscape}
\end{appendix}
%\end{landscape}
% Don't change these lines
%\bsp	% typesetting comment
%\label{lastpage}
\end{document}